\documentclass[a4paper]{article} 

\providecommand{\esymbol}{e}
\providecommand{\esymbolv}{\ev}

\usepackage{setspace} 
\usepackage{xr}
\usepackage{wrapfig}
\usepackage{enumitem}
\usepackage{color}
\usepackage{amsmath,amsfonts,amssymb}
\usepackage{epsfig}
\usepackage{graphicx}
\usepackage{colortbl}
\usepackage{url}
\usepackage{lineno}

\definecolor{darkblue}{rgb}{0.05,0.,0.65}
\definecolor{grey}{rgb}{0.9, 0.9, 0.9}

\newcommand{\myparagraph}[1]{\vspace{-3mm}\paragraph{#1}}

\newcommand{\mybox}[1]{\fbox{\parbox{16.5cm}{\begin{center}\parbox{15.5cm}{#1}\end{center}}}}

\newcommand{\coout}[1]{[{\color{magenta} #1}]}
\newcommand{\co}   [1]{[{\it \color{red} #1}]}
\newcommand{\todo} [1]{[{\color{blue} #1}]}

\usepackage{amsmath,amsfonts,amssymb}
\usepackage{relsize}
\usepackage{ marvosym }

\newcommand{\sign}     {\mbox{sign}}

\newcommand{\onevector}{\hat 1}
\newcommand{\onevec}{\mathbf{1}}

\newcommand{\const}{\mbox{\rm const.}}

\newcommand{\inv}{^{-1}}

\newcommand{\md}{{\rm d}}

\newcommand{\diag}{\mbox{\rm Dg}}

\newcommand{\e}   {\mbox{\rm e}}

\newcommand{\hv}{{\bf h}}
\newcommand{\cv}{{\bf c}}
\newcommand{\vv}{{\bf v}}

\newcommand{\uv}{{\bf u}}

\newcommand{\bv}{{\bf b}}
\newcommand{\gv}{{\bf g}}
\newcommand{\ev}{{\bf e}}
\newcommand{\xv}{{\bf x}}
\newcommand{\dv}{{\bf d}}

\newcommand{\nv}{{\bf n}}

\newcommand{\mv}{{\bf m}}
\newcommand{\av}{{\bf a}}

\newcommand{\tauv}{{\boldsymbol \tau}}
\newcommand{\sigmav}{{\boldsymbol \sigma}}
\newcommand{\alphav}{{\boldsymbol \alpha}}

\newcommand{\gammav}{{\boldsymbol \gamma}}

\newcommand{\etav}{{\boldsymbol \eta}}

\newcommand{\zetav}{{\boldsymbol \zeta}}

\newcommand{\nuv}{{\boldsymbol \nu}}

\newcommand{\Nmat}{\mathbf{N}}

\newcommand{\Mmat}{\mathbf{M}}

\newcommand{\Hmat}{\mathbf{H}}

\newcommand{\Amat} {\mathbf{A}}

\newcommand{\Bmat} {\mathbf{B}}

\newcommand{\trans}{^{\top}}
\newcommand{\half}{\frac{1}{2}}

\newtheorem{theorem}{Theorem}[section]

\newtheorem{definition}[theorem]{Definition}

\newtheorem{lemma}{Lemma}
\newtheorem{proposition}{Proposition}

\DeclareMathAlphabet{\mathpzc}{OT1}{pzc}{m}{it}
\DeclareMathAlphabet{\mathcalligra}{T1}{calligra}{m}{n}

\providecommand{\esymbol}{e}

\newcommand{\kcat}     {k_{\rm cat}}

\usepackage{etoolbox}

\newtoggle{bookversion}      \togglefalse{bookversion}
\newtoggle{shortbookversion} \togglefalse{shortbookversion}
\newtoggle{anhang}           \togglefalse{anhang}
\newtoggle{hidecomments}      \togglefalse{hidecomments}

\newcommand{\bookco}[1]{}

\definecolor{brown}{rgb}{0.9,0.69,0.34}
\definecolor{samoabrownlight}{rgb}{0.89,0.69,0.4}
\definecolor{samoabrowndark} {rgb}{0.5,0.3,0.15}
\definecolor{cbasamoabrown1}{rgb}{0.87,0.6,0.23}
\definecolor{cbasamoabrown2}{rgb}{0.87,0.6,0.23}
\definecolor{cbabrown1}{rgb}{0.87,0.6,0.23}
\definecolor{cbabrown2}{rgb}{0.87,0.6,0.23}
\definecolor{cbabrown3}{rgb}{0.87,0.6,0.23}
\definecolor{cbabrown4}{rgb}{0.87,0.6,0.23}
\definecolor{cbaecoblue1}{rgb}{0.8,0.8, 1.0}
\definecolor{cbaecoblue2}{rgb}{0.7,0.7, 1.0}
\definecolor{cbaecoblue3}{rgb}{0.87,0.6,0.23}
\definecolor{cbaecoblue4}{rgb}{0.87,0.6,0.23}
\definecolor{cbablue2}{rgb}{0.87,0.6,0.23}
\definecolor{cbapink}{rgb}{.99,0.92,0.75}
\definecolor{cbabeige1}{rgb}{0.86, 0.797, 0.625} 
\definecolor{cbabeige2}{rgb}{0.93, 0.812, 0.56}  
\definecolor{cbabeige3}{rgb}{1.0, 0.97, 0.88}  
\definecolor{cbahelleslila}{rgb}{1.0, 0.99, 1.0}  
\definecolor{cbalightgrey}{rgb}{0.95,0.95,0.95}
\definecolor{cbatablecolor1}{rgb}{0.86, 0.797, 0.625} 
\definecolor{cbatablecolor2}{rgb}{1,1,1}         

\newcommand{\myvalue}      {value}

\newcommand{\gain}         {gain}

\newcommand{\target}      {target}

\newcommand{{\fluxvalue}}  {flux \myvalue}
\newcommand{{\fluxgain}}   {flux \gain}

\newcommand{{\valueflow}}    {value flow}
\newcommand{{\Valueflow}}    {Value flow}

\newcommand{ {\flow}}        {flux profile}
\newcommand{ {\Flow}}        {Flux profile}

\newcommand{\favour}     {favour}
\newcommand{\favoured}    {favoured}

\newcommand{\fluxpattern}{flux pattern}

\newcommand{\mfFBA}{FBA with minimal fluxes}
\newcommand{\MfFBA}{FBA with minimal fluxes}
\newcommand{\mcFBA}{FBA with molecular crowding}

\newcommand{\mwfFBA}{linear FCM}

\newcommand{\stateequal}{kinetically equal}
\newcommand{\statedistinct}{kinetically distinct}
\newcommand{\Statedistinct}{Kinetically distinct}
\newcommand{\compromisecost}{compromise cost}
\newcommand{\Compromisecost}{Compromise cost}

\newcommand{\basicflow}{basic {\flow}}
\newcommand{\combinedflow}{combined {\flow}}
\newcommand{\Combinedflow}{Combined {\flow}}

\providecommand{\esymbol}   {e}
\providecommand{\esymbolv}  {\mathbf{e}}

\providecommand{\prodrate}  {r}

\newcommand{\enzymev}   {\mathbf{\esymbol}}
\newcommand{\intprod}   {\prodrate^{\rm int}}

\newcommand{\ratev}{\nuv}
\newcommand{\ratelaw}{k}

\newcommand{\ffit}        {{\mathcal F}} 
\newcommand{\fluxbene}    {b}
\newcommand{\fluxcost}    {a}
\newcommand{\metcost}     {g}

\newcommand{\hminus}      {h}

\newcommand{\wsymbol}     {w}

\newcommand{\partialder}{^{\centerdot}} 

\newcommand{\Nint}    {\Nmat^{\rm int}}
\newcommand{\Ntot}    {\Nmat^{\rm tot}}

\newcommand{\lnc}{m}
\newcommand{\lncv}{\mv}

\newcommand{\fes}    {\fes}
\newcommand{\fel}    {\fel}
\newcommand{\fevs}   {\fevs}

\newcommand{\hul}     {\hminus_{\esymbol_l}}

\newcommand{\hus}     {\hminus_{\rm \esymbol}}

\newcommand{\metcostkin}{\metcost^{\rm kin}}

\newcommand{\enzymemetcost}{\metcost^{\rm enz}}

\newcommand{\fluxbenev}     {{\bf \fluxbene}}

\newcommand{\bvdir}        {{\fluxbenev_{\rm v}^{\rm int}}}

\newcommand{\hmet}   {\metcost}

\newcommand{\acost}     {\fluxcost}  
\newcommand{\acostenz}  {\fluxcost^{\rm enz}}  
\newcommand{\acostkin}  {\fluxcost^{\rm kin}}  
\newcommand{\acostv}    {{\bf \acost}_{\rm v}}

\newcommand{\hatacostvl}{\hat{\acost}_{v_l}}

\newcommand{\apointcostvl}   {\acost_{v_l}\partialder}
\newcommand{\hatapointcostvl}{\hat{\acost}_{v_l}\partialder}
\newcommand{\asumpointcostvl}{\acost_{v}\partialder}

\newcommand{\sPolytope}   {{\mathcal P}_{\rm F}}       
\newcommand{\pPolytope}   {{\mathcal P}_{\rm P}}       
\newcommand{\bPolytope}   {{\mathcal P}_{\rm B}}       
\newcommand{\mPolytope}   {{\mathcal P}_{\rm M}}

\newcommand{\fluxenzymecostl}  {\Delta \wsymbol_{\intprod_l:}}

\newcommand{\kM}{K_{\rm M}}

\newcommand{\us}{\esymbol}
\newcommand{\ul}{\esymbol_{l}}

\newcommand{\he}{\hminus_e}

\newcommand{\vvA}{\vv_{\rm A}}
\newcommand{\vvB}{\vv_{\rm B}}
\newcommand{\vvC}{\vv_{\rm C}}
\newcommand{\lncvA}{\lncv_{\rm A}}
\newcommand{\lncvB}{\lncv_{\rm B}}
\newcommand{\lncvC}{\lncv_{\rm C}}

\newcommand{\aenz}{\acost^{\rm enz}}
\newcommand{\akin}{\acost^{\rm kin}}

\renewcommand{\hv}{{\bf h}}

\renewcommand{\Nmat} {{\bf N}}

\renewcommand{\coout}[1]{}

\definecolor{brown}{rgb}{0.9,0.69,0.34}
\definecolor{lightyellow}{rgb}{1,0.99,0.85}

\newcommand{\psfilesfluxcostfunctions}{ps-files}

\definecolor{orange}{rgb}{1,0.75,0.3}
\definecolor{brown}{rgb}{0.95,0.6,0.3}
\definecolor{pink}{rgb}{1,0.7,0.7} 
\definecolor{purple}{rgb}{0.8,0.7,1}
\definecolor{lightblue}{rgb}{0.9,0.9,1}

\renewcommand{\co}[1]{}
\renewcommand{\todo}[1]{#1}
\renewcommand{\myparagraph}[1]{}

\begin{document}

\title{Flux cost functions and optimal metabolic states}

\author{Wolfram Liebermeister\\[3mm] 
Universit\'e Paris-Saclay, INRAE, MaIAGE, 78350 Jouy-en-Josas, France}

\maketitle

\begin{abstract}
  The metabolic fluxes in cells follow physical, biochemical, and
  economic principles. Some flux balance analysis (FBA) methods trade
  flux benefit against flux cost. However, if flux cost functions are
  linear and meant to describe underlying enzyme costs, this entails
  that enzyme efficiencies are constant and ignores the interplay
  between fluxes, metabolite concentrations and enzyme levels in
  cells. Here I introduce realistic flux cost functions that describe
  an ``overhead cost'', namely the minimum enzyme and metabolite cost
  associated with the fluxes in a kinetic model. These flux cost
  functions have general mathematical properties. Enzymatic flux cost
  functions, which represent enzyme costs, scale proportionally with
  the flux profile and are concave on the flux polytope.  Kinetic flux
  cost functions represent the sum of enzyme and metabolite costs. If
  two flux profiles are superimposed, their different demands for
  metabolite concentrations cause an extra compromise cost, which
  makes flux cost functions strictly concave in almost all cases.
  When fluxes change their direction, the enzymatic cost jumps
  abruptly.  Based on general flux cost functions, I propose two
  methods for flux modelling: Flux Cost Minimisation, a nonlinear
  variant of FBA with flux minimisation, and Flux Benefit
  Optimisation, a nonlinear variant of FBA with molecular
  crowding. The optimal flux profiles, at a given flux benefit, are
  vertices of the flux polytope.  In models without other flux
  constraints, these vertices are elementary flux modes.  Linear
  approximations of enzymatic flux cost can be used in FBA. In
  contrast to flux costs chosen ad hoc, these functions reflect the
  enzyme kinetics and extracellular concentrations in realistic
  kinetic models.  Based on enzymatic flux costs, we can describe the
  cell growth rate as a convex function on the flux polytope and
  derive growth-optimal metabolic states and statistical distributions
  for the fluxes in cell populations. Therefore, flux cost functions
  unify kinetic models and flux analysis, provide parameters for FBA
  models, and shows that these models are physically justified.
\end{abstract}

\textbf{Keywords:} Flux balance analysis, enzyme cost, metabolite
cost, concave function, elementary flux mode.

\textbf{Abbreviations:} ECM: Enzyme Cost Minimisation; FCM: Flux Cost
Minimisation; FBA: Flux Balance Analysis; CM: Common Modular rate law;
mfFBA: FBA with minimal fluxes; mwfFBA: FBA with minimal weighted
fluxes; mcFBA: FBA with molecular crowding; CAFBA: constrained
allocation FBA; RBA: Resource Balance Analysis

\co{meike: alternative definition of EFMs (Meike): if you know one
  flux, you know all the fluxes!  stimmt das? beweise wo?}

\co{WO in paper? oder in FCM paper? (und hier nur in tabelle bzw als kommentar fuer ECM
  angeben?) are ECM scores strongly convex?  (with regularisation for
  sure!) - and does this provide advantages?} \co{aus wikipedia
  (``convex function''), applied to FCM: q(ln c) is strongly convex if
  (and only if) q(ln c) is differentiable twice and if there exists an
  $\alpha >0$ s.t. the lowest hessian eigenvalue of q is $\ge \alpha$
  for any ln c. (if it is strongly convex, it is also strictly
  convex). (strictly convex means that the lowest hessian eigenvalue
  in each point is positive). With curved regularisation, strong
  convexity holds for sure! allerdings nicht klar, was ``stronly
  convex'' in der praxis bringt.}

\co{allg: struktur des artikels ist gut, nur schlecht geschrieben; an leser denken, was er schon weiß und was ihn interessiert}

\co{elementary modes = minimal supports of stationary flux
  distributions}

 \co{"rate" in FBA is often yield, multiplied with an uptake rate.  
   ref scpf:08 Is maximization of molar yield in metabolic networks favoured by evolution?}
 
  \co{FCM; lesen / cite beeliveau: fundamental limits on bacterial
    growth (datei auf desktop)} \co{FCM (wo sonst?): Develop a CBA für
    David Tourignys approach in which costs do not belong to
    individual reactions, but to elementary modes. This probably
    requires constant concentrations (to allow for an enzyme cost
    function that is additive between modes). David assumes this
    linearisation anyway, and does it for short time intervals.}
  \co{FCM: nonlinear h(e); for q(x) to be convex, h(e) needs to be
    (not only convex but) convex and non-increasing!!} 
 \co{(kommt auch in CBA opt):
    "maximum-entropy-augmented objective function”: refer to david
    tourigny's entropy principle (boltzmann distribution for
    prefactors of EFMS) in CBA optimality? optimality function on flux
    polytope defines, through boltzmann distribution (with a certain
    temperature), a "mean" flux distribution that is the predicted
    actual flux distribution (the averaging could be on the level of a
    single cell, as a mixture; or on the level of the population, with
    each cell choosing one optimal corner) in any case, cite david t
    in cba optimality! Note: a de martino: resource allocation models
    cannot explain sub-optimal growth rates; maximum entropy can!}
  \co{note: flux polytope (by definition bounded) vs flux polyhedron
    (can also be unbounded)?}  \co{notes from Stefans talk (1. März
    2021) (SIEHE LATEX SOURCECODE):
%
%
%
%
%
%
}

\co{message (for website?):\\
  1. Given a flux distribution, the enzyme cost (as a function of
  (thermodynamically) feasible metabolite log-concentrations) is
  convex. This holds for a large class of kinetic rate laws (including
  allosteric regulation)\\ 2. The enzymatic flux cost, in the space
  feasible flux distributions is optimized
  on a polytope vertex (except for rare cases of ``singular models'', as defined below), i.e.~typically an  elementary flux mode.\\
  3. Non-elementary flux modes can be as good as (but not better than)
  the best elementary flux mode.}

  \co{JA! auch ins google doc! use "thermodynamically inconsistent" for
  rate laws, "..infeasible" for flux modes} \co{FCM: "enzyme/growth
  relation" (note that enzyme cost, here, means total mass
  concentration)} \co{JA! abstract and intro: "enzymatic flux cost"
  means: "a flux cost that represents an enzyme cost"; or "that is
  actually an overhead enzyme cost"} \co{JA! F-polytope statt
  T-polytope (FCM; CBA fluxes)} \co{JA! reciprocal statt inverse, also
  in CBA optimality usw} \co{auch CBA flux: elementary modes = minimal
  supports of stationary flux distributions} \co{JA! present kinetic
  optimization, mueller et al 2015 as a basic version of many
  optimization problems (also in CBA theory)} \co{JA! einmal FN! avoid
  confusion with the term ``linear'' (proportional or affine)?  uea}
\co{JA! in text klarer zwischen flux cost and scaled flux cost
  ${\hat a}^{\rm enz} = a^{\rm enz}/b$ unterscheiden; \co{waere es
    besser, allgemein nicht von B-polytope, sondern von
    \co{S-polytope} (fuer scaled fluxes) zu sprechen, und erst bei
    FBA-anwendung zu sagen, dass wir jetzt benefit-skalierte fluesse
    betrachten?}  \co{generate better figure 1 from branch point
    example} wie scaled v's, auch benefit-scaled flux cost hataenz}
\co{enzyme cost (and possibly metabolite cost) are conceptualised as
  effective (enzymatic or kinetic) flux cost functions}

\co{NICHT SO WICHTIG} \co{schauen + einbauen! steffen klamt paper rate
  vs yield, linear-fractions :
  https://www.sciencedirect.com/science/article/pii/S1096717617303853}
\co{Mention elementarmoden typ 3} \co{``flux modelling'' statt ``flux
  analysis''} \co{notion of optimality: definition in mathematik:
  $\ge$.}  \co{fix F-polytope (vs cone?)} \co{fix flux pattern vs flux
  template?}  \co{CBM (constraint-based modeling)} \co{ueberal term
  "thermo-physiological"} \co{hier und sonst definiert? auch andere
  artikel: note: ``flux profile'' is stationary by definition
  (otherwise they are called ``non-stationatry flux distribution'')}
\co{Manu'a-algorithmus programmieren} \co{mention ``proxy objective''
  and ``overhead cost''} \co{use magma colour scale for cost!  auch
  magma-aehnliche skalen in schematischen bildern?}  \co{gute
  beige-farbe: eed499} \co{``enzyme demand function'' e(v,c)} \co{some
  good graphics are in additional material!  recover them} \co{wort
  fuer kinetic M-cost?  ``effective'' sollte ein allgemeinwort fuer
  alle solchen kosten bleiben // use ``indirect cost'' instead of
  overhead cost?  das hier erwaehnen!} \co{joosts beweis: gibts eine
  referenz?}  \co{in Fig1: Show actual model results -- one running
  example for all figures?  Im prinzip ist das das branch point
  example!}  \co{F-polytop als oberbegriff fuer S- und B-polytop
  einfuehren // in diesem abschnitt und im rest des artikels}
\co{Kriterium fuer lokal optimale fluesse Programmieren!}  \co{what
  about predefined conserved moiety concentrations?  nonlinear
  constraints on the M-polytope? wo erwaehnen?}  \co{discuss this
  article with Meike+Stefan/Ralf to make sure I'm doing justice to
  their work}

\iftoggle{bookversion}
{\section{Flux cost functions}}
{\section{Introduction}}

\myparagraph{\ \\Metabolic fluxes in cells} The metabolic state of a
cell, consisting of metabolic fluxes, metabolite concentrations, and
enzyme levels, is constantly adapted to external conditions and
cellular demands. Since the aim of metabolism is substance conversion,
the metabolic fluxes are the main variables to be controlled.  Which
metabolic pathways should a cell use in a given environment, and how
should pathway fluxes be adjusted to varying nutrient supply?  When
should pathways be switched on or off?  Which enzyme and metabolite
concentrations \todo{belong to} a flux profile, and more generally,
how do enzyme and metabolite costs affect the choice between flux
profiles?  \co{A well-studied example is why and when cells switch
  (deliberately) from respiration to (less yield-efficient)
  respiro-fermentation? See fig 1. kurz fakten (text von unten im
  artikel), dann erklaerungen; auch substrate efficiency (sustainable)
  and enzyme efficiency (fast growth); compare to r and k strategists;
  as we will see below, ..}

To predict metabolic strategies, we may first ask what fluxes are
physically possible.  In Flux Balance Analysis (FBA), {\flow}s are
required to respect flux bounds, be stationary (i.e.~mass-balanced
internal metabolites), and possibly show thermodynamically feasible
flux directions.  But which of the possible {\flow}s will be realised
by cells, depending on enzyme parameters and environmental conditions?
Kinetic models explain metabolic fluxes by rate laws, enzyme levels,
and metabolite concentrations. But even if all rate laws were known,
these predictions would be uncertain because enzyme and metabolite
concentrations are highly variable.  Flux analysis models work
differently: they assume that cells choose, among all physically
feasible {\flow}s, the most profitable one, e.g.~the one that provides
the best compromise between biomass production rate (benefit) and
total enzyme demand (cost). The idea is simple: just like the
efficiency of a single enzyme (the flux per enzyme level), the overall
catalytic rate of metabolism (the biomass production per total
metabolic enzyme) may be optimised.  A cost-efficient metabolism can
be useful both for fast growth and for allocating protein resources to
other tasks, at a given growth rate or even in non-growing cells.

\begin{figure*}[t!]
  \begin{center} 
\includegraphics[width=14.5cm]{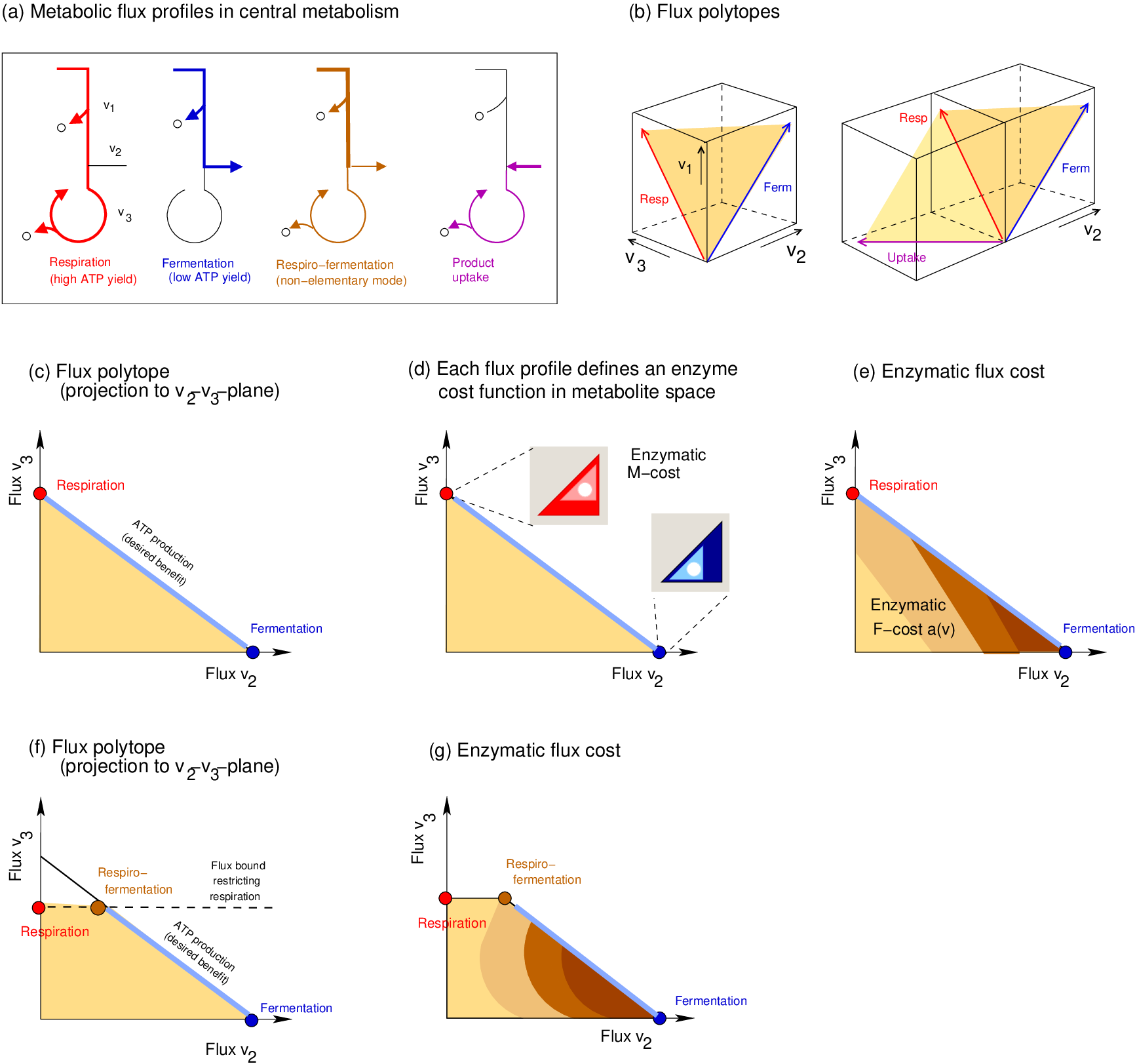}
\caption{Metabolic flux profiles, flux cost, and  optimal
  fluxes.  (a) Fluxes in central metabolism. Three basic {\flow}s
  (for pure fermentation, respiration, and respiration on ethanol) and a
  convex combination of them (respiro-fermentation) are shown.  With 
  ATP production (small circles) as a benefit function,  all
  {\flow}s can be scaled to a unit benefit value (i.e.~unit ATP production
  rate). For simplicity, ethanol uptake is described as a reverted
  ethanol overflow (using the same enzyme).  (b) {\Flow}s as points in
  flux space,  spanned by the reaction fluxes $v_{1}$,
  $v_{2}$ and $v_{3}$. The plane of stationary fluxes is defined by
  the constraint $v_{1}=v_{2}+v_{3}$.  The feasible {\flow}s with
  given flux directions form a convex polytope $\sPolytope$. On the right, two such
  polytopes (for varying directions of the ethanol excretion flux) are
  shown. (c) {\Flow}s as points in flux space (projection on the
  $v_2/v_3$ plane). Flux profiles with unit  ATP
  production lie on a diagonal line (called B-polytope and shown
  in light blue). (d) Each flux profile defines a set of feasible
  metabolite profiles, forming a convex polytope in log-metabolite
  space (called M-polytope).  Given a desired {\flow}, each metabolite
  profile requires a particular enzyme profile with a particular
  cost, which can be  an effective metabolite cost function
  on the M-polytope. Minimising this function yields an optimal
  metabolite profile, an optimal enzyme profile, and a minimal cost
  value called enzymatic flux cost. (e) Enzymatic flux cost as a
  function on the flux polytope. The enzymatic cost $\acostenz(\vv)$ is defined as the
  minimum enzyme cost at which a {\flow} can be realised in a given
  kinetic model.  The {\flow} with the smallest cost on the B-polytope
  is assumed to be optimal.  (f) A bound on  respiration (flux
  bound shown by dashed line) gives rise to a new vertex point,
  describing respiro-fermentation.  (g) In the (schematic) example
  show, respiro-fermentation is the optimal strategy.}
 \label{fig:definition} \end{center}
\end{figure*}

\myparagraph{Flux prediction by cost and benefit} Cost-benefit
optimisation is a common approach in economics and also in flux
prediction.  Constraint-based modelling (CBM) methods such as Flux
Balance Analysis \cite{Orth2010} or Resource Balance Analysis
\cite{gofs:11} constrain metabolic fluxes by assuming a steady state
and imposing linear flux bounds.  \todo{Then, a flux benefit can be
  maximised or a flux cost is minimised at a fixed benefit.} Under
these constraints, flux costs can then be constrained or minimised.
In classical FBA, there are separate bounds on single fluxes (an
assumption without a good physical justification, unless individual
enzyme levels are known). \co{[auch in CBA II:] Accordingly, "rate" in
  FBA is often yield, multiplied with an uptake rate. and since the
  medium-dependent uptake rates are usually unknown (they depend not
  only on transporter kinetics, but also transporter expression, which
  in turn depends on the entire metabolic strategy), these growth rate
  predictions are typically unreliable (cite papers by Martin
  lercher). what remains is the safe prediction of biomass yield on
  substratem which may be a fitness objective in itself - however,
  relevant only in certain cases. \co{ref scpf:08 Is maximization of
    molar yield in metabolic networks favoured by evolution?}}
Internal cycle fluxes, which have no effect on the benefit, remain
undetermined and can lead to underdetermined, unrealistic flux
solutions. To avoid this, other FBA methods constrain or penalise
fluxes more generally. In FBA with molecular crowding
(mcFBA\co{UEA!!}) \cite{bvem:07}, each {\flow} is associated with an
enzyme demand, written as a weighted sum of the absolute fluxes (or
the fluxes themselves, if fluxes are known to be positive).  This
enzyme demand is then bounded (e.g.~to account for limited space in
the cell; hence the name ``FBA with molecular crowding'').  FBA with
flux minimisation (mfFBA\co{uea JA!!})  \cite{holz:04} does exactly
the opposite: flux costs (described by a possibly weighted sum of
absolute fluxes) are not bounded during optimisation, but are
themselves minimised (at a given flux benefit). Both methods can predict sparse {\flow}s in which only
the most profitable pathways are active.  To justify linear flux cost
functions or linear flux constraints, it is assumed that fluxes are
proportional to enzyme levels with constant proportionality factors
(called catalytic rates, apparent catalytic constants, or enzyme
efficiencies) \cite{khma:16}.

\myparagraph{Fluxes, enzyme demands, and enzyme efficiencies}
\co{sagen: um eFBA realistisch zu machen, muss man die enzyme
  efficiencies verstehen. Since they vary with the fluxes (SEE ECM!),
  and since the fluxes depend on them (see FBA)!) FBA etc assume
  constant enzyme efficiencies, in order to translate fluxes into
  enzyme levels; either for costs or for upper bounds: in any case a
  function of the fluxes that represents a total enzyme amound (flux
  cost function). such functions (representing variable enzyme
  efficiencies) are the topic of this paper}
\co{FCM: wort "metabolic productivity"\\
  Biomass/substrate productivity (“yield”)\\
  Biomass/enzyme productivity (“rate”, “specific enzyme cost”) = biomass flux /enzyme level\\
  Biomass/space productivity (“specific kinetic cost”) = biomass flux
  / space demand} \co{biomass/catalytic rate or ``enzyme
  productivity''}, \co{use in FCM und uea!} \co{The methods above
  optimise either biomass yield or (enzyme-specific) production rate;
  or in other words, substrate or enzyme productivity. while
  output/substrate productivity depends only on the shape of the flux
  distribution itslef, output/enzyme productivity depends on enzyme
  kinetics or, simply stated, on the state dependent enzyme catalytic
  rates (and therefore on metabolic concentrations).} In reality, the
enzyme efficiencies are not constant, but depend on metabolite
concentrations and vary between metabolic
states of the cell. Enzyme levels and stationary fluxes  are
not simply proportional!   Proportionality holds if metabolite
concentrations are constant, but this is usually not the case: when
enzyme levels are changing, this changes both steady-state fluxes,
metabolite concentrations, and enzyme efficiencies. The enzyme catalytic
rates play a key role in determining metabolic strategies. In
{\mcFBA}, they serve as the conversion factors between enzyme levels
and fluxes. A linear conversion from fluxes to enzyme levels assumes
that enzymes work at a constant catalytic rate (approximated by the
$\kcat$ value or by a smaller ``catalytic rate'' $k_{\rm app}$), thus
ignoring enzyme kinetics. In kinetic models, enzyme efficiencies \co{cat rates uea?} 
depend  on metabolite concentrations, as described by the rate
laws -- and since metabolite concentrations vary, the efficiencies
vary as well!  \co{JA! refs, erklaerungen, compare
  to kcat values, mention factorized rate laws usw?}  In FBA models
based on resource allocation (like {\mfFBA}, {\mcFBA}, or CAFBA
\co{CITE!!  auch sonst}), the enzyme efficiencies are important
parameters: they determine the choice of fluxes, metabolic strategies,
and growth rates.  The biomass/enzyme productivity (i.e.~the biomass
production rate per amount of metabolic enzyme) also plays an
important role in protein allocation models (e.g.~used to explain
bacterial growth laws \cite{sgmz:10}), which can be used to relate the
biomass/enzyme productivity to cell growth rates (as shown in
\cite{wnfb:18} and below in this article). Thus, enzyme efficiencies
are important.  If they depend on metabolite concentrations, if
metabolite concentrations depend on the metabolic state, and if the
enzyme levels in optimal states depend again on enzyme effiencies, all
these variables need to be found at the same time and the relation
between fluxes and enzyme levels will be complicated.  The complex
interplay between all these variables and its consequences for flux
prediction are addressed in this paper.

\myparagraph{Computing fluxes, enzyme demands, and enzyme efficiencies
  based on optimality principles} So, let us open Pandora's box and
predict fluxes, metabolite concentrations, and enzyme levels at the
same time.  \co{sagen: gemeinsam waere moeglich, aber praktisch mit
  brute force oft nicht zu loesen (oder man wuesste nicht, ob man das
  optimum gefunden hat; und wuerde in jedem fall das problem nicht
  richtig verstehen!)}  To optimise all of them, we combine flux
modelling with underlying kinetic models and perform a layered
optimisation: first, given a predefined flux distribution, we optimise
enzyme and metabolite concentrations; by solving this problem, we can
associate our flux profile with an effective (enzyme and metabolite)
cost, summarising all details of the kinetic model (both optimality
and kinetics) in a nonlinear flux cost function. By associating each
flux profile with a cost, we can find an optimal flux profile by
minimising this cost function in flux space.  The result is an optimal
{\flow} together with its optimal enzyme and metabolite concentrations
and the corresponding enzyme efficiencies. Altogether, this method
resembles a minimal-flux FBA with a flux cost function obtained by
solving an optimality problem for enzyme and metabolite
concentrations. This latter problem, a search for optimal enzyme
levels that minimise the overall enzyme demand \cite{brow:91,klhe:99}
is called enzyme cost minimisation (ECM) \cite{nfbd:16} and can be
reformulated as a convex optimisation over metabolite profiles.  In
ECM , different flux modes favour different metabolite concentrations
and therefore different enzyme efficiencies. This leads to nonlinear
relationships between fluxes and enzyme cost.  In this article, we
consider  flux cost functions $\fluxcost(\vv)$ more generally and  study
their mathematical properties. We describe nonlinear variants of FBA
based on these functions (FCM and FBM), and draw conclusions for
cellular growth rates and the probabilities of metabolic states in
cell populations.  \co{clearly introdue ``flux cost function'' in
  general, and those representing known other costs // auch in CBA
  opt, lag, kin usw!!  ``flux cost'' describes simply the requirements
  (and incentives!!)  associated with fluxes}

\myparagraph{Flux cost minimisation} \co{wirklich ernstnehmen, dass
  kosten PRO FLUSS das wichtige sind. das gut diskutieren! (kommt
  schon in CBA fluxes und CBA labour)} \co{use word ``layered'' also
  to describe FCM approach} In FBA models, flux cost can either be
minimised (as in minimal-flux FBA) or constrained (as in
{\mcFBA}). Assuming nonlinear flux cost functions, I propose two
methods for flux prediction called Flux Cost Minimisation (FCM) and
Flux Benefit Maximisation (FBM). \co{WO? metabolic ``targets''
  (e.g. flux costs and benefit) can appear as objective or constraint
  - by changing their roles, we obtain different (but closely related)
  optimality problems (see CBA opt); we discuss this here, for flux
  cost minimisation ve flux benefit maximisation (mention also
  pareto?)} Both methods lead to concave optimality problems
\cite{wpht:14,wnfb:18}.  In FCM, we search for flux distributions that
minimise flux cost at a given flux benefit (e.g.~a given rate of
biomass production). \co{WO? FCM challenges the assumption of linear
  enzyme-flux relationships, but shows how they can be dervied as
  practical approximations.} Applied to metabolism as a whole, it
yields the ratio of biomass production and enzyme cost. This
biomass/catalytic rate (or ``enzyme productivity'') is an important
quantity in protein allocation models \cite{sgmz:10}, where it
determines the achievable growth rate of cells.  This approach --
minimising the enzyme cost per biomass production rate, computing the
resulting cost/benefit ratio, and translating it into cellular growth
rates -- has previously been applied to a model of central metabolism
\cite{wnfb:18}.

 \co{FCM with kinetics
  derived flux cost functions is pratisch nur EINZELN in jedem flux
  space segment (wir sehen unten, dass zwischen den segmenten spruenge
  sind), dh downside: flux directions must be known; or abs function
  must be used} 

\co{new paragraph, ref to EFM papers, dann daans paper. then say thet
  here the basic logic is simple. we argue that flux cost functions
  are concave and that therefore optimal flux modes are polytope
  corners (which are EFM or convex combinations of EFM).} \co{von daans resultat mit EFM-kombis ausgehen, hier erklaeren, dass das einfach heißt, dass die ecken so sind + hier zeigen warum (bzw unter welchen bedingungen) (nur) ecken optimal sind}

Importantly, under certain assumptions the optimal flux solutions are
elementary flux modes. \co{kurz el-modes-dings erklaeren; sagen: hier
  3. beweis!}  \co{hier mehr betonen, dass ich hier auf die arbeiten
  von meike und ralf aufbaue und mehrtheorie dazu mache} This had been
proven for similar optimality problems (e.g.~maximising a linear flux
benefit at a given total enzyme amount) \cite{wpht:14,murs:14}: in
these problems, the enzyme cost per flux is a concave function in flux
space, and it was shown that optimal {\flow}s must be elementary flux
modes (EFMs), no matter which underlying kinetic model is assumed.
\co{say that FCM has been exemplified in EFM paper}

\co{one aim of this article: clarify the exact conditions under which optimal flux modes (in the FCM problem) must be polytope vertices (which in models without flux bounds also means, EFMs); another aim: show how linear Flux cost functions for FBA can be derived from kinetic models, and how they reflect details like kinetic constants and external concentrations; relate this to simpler methods like mcFBA with transporter efficiency (satFBA) and related ways to include kinetics in RBA.}

\myparagraph{Overview of this article} Here I discuss how flux cost
functions can be defined, what are their mathematical properties, and
how they can be used in modelling.  \todo{The functions allow us to
  represent different types of biological costs as costs in flux
  space, in order to, then, simply optimise over the flux
  distributions.} I first show how enzyme and metabolite costs can be
written as flux cost functions, and how such functions reflect the
underlying kinetic models and optimality principles.  Then I study the
mathematical properties. Unlike flux cost functions FBA, they need not
be linear: when two flux modes are superimposed, their costs may not
add up, but there may be an extra {\compromisecost}, that makes the
function strictly concave. \co{and as a consequence, optimum points
  can only be polytope vertices} \co{in fact, I argue here that --
  while counterexamples can be constructed in theory -- this is the
  case in almost all realistic models; and that reasons for non-vertex
  solutions fluxes of a different nature, eg side objectives, no
  strict optimality etc (rather say this in discussion than here?)}
\co{folgt konkav direkt aus konvex?  JA!  das sagen, hier und unten!}
Based on these mathematical properties, I discuss how enzymatic and
kinetic flux cost functions can be used for flux prediction and to
define linearised cost functions for FBA, how predicted enzyme demands
can be converted into cell growth rates \cite{sgmz:10}, and how we can
compute probability distributions of metabolic states in cell
populations. This connects metabolic models to cell or population
models The running example in this article is a simple branch point
model like in figure \todo{1}, but all concepts apply to metabolic
networks of any size.

\section{Flux costs scoring  enzyme and metabolite concentrations}

\subsection{Enzymatic and kinetic flux cost functions}

\myparagraph{\ \\Flux cost function defined by enzyme demand} A cell
can realise a {\flow} $\vv$ by different combinations of enzyme and
metabolite concentrations. High concentrations are costly: in
experiments, overexpressing an idle protein reduces cell growth. In
the case of enzymes, the same costs may exist and may
\todo{counterbalance} the effects of the higher catalysed fluxes
\cite{deal:05,szad:10}. An explanation for this growth deficit is that
material resources and cell space are limited and that, therefore,
higher enzyme or metabolite concentrations in one pathway imply lower
protein concentrations, and therefore a lower performance, in other
pathways, which compromises cell growth. To model this, fluxes may be
penalised by effective flux cost functions or may be constrained by
$\kcat \cdot \us$, the product of catalytic constant $\kcat$ and
enzyme level $\us$ to be bounded or minimised.  \co{hugo! \co{(text
    aus CBA fluxes hierher?)}  also refer to enzyme + metabolite
  crowding?  refer to combination of the two objectives in tepper et
  al.} In flux modelling, enzyme costs can be taken into account in
several ways.  Molecular Crowing FBA \cite{bvem:07} assumes that each
{\flow} requires some overall enzyme amount, which is bounded due to
limited space in cells. To translate fluxes into enzyme amounts, a
simple linear relationship is used. FBA with molecular crowding
(mcFBA) \co{uea statt {\MfFBA}} \cite{holz:04} works similarly: after
predefining a required flux benefit ($b=\sum_l b_{v_l}'\,v_l$), one
minimises a sum $\acost=\sum_l |v_l|$ or a weighted sum of fluxes
$\acost=\sum_l a_{v_l} |v_l|$ with cost weights $a_{v_l}$ (where flux
signs are ignored).  Again, these flux costs may represent enzyme
costs.  In classical FBA, metabolic {\flow}s are predicted by assuming
a linear benefit function $b=\bv_{v}' \cdot\,\vv$ (e.g.~the biomass
production rate) and maximising this benefit on the flux polytope.
More generally, in FBA methods, maximisation of flux benefit
(e.g.~biomass production) and minimisation of flux cost (e.g.~the sum
of absolute fluxes) can be combined in different ways, e.g.~by layered \co{statt nested uea!}
optimisation \cite{holz:04} or multi-criteria optimisation
\cite{scks:07,szzh:12}.

\myparagraph{Enzymatic and kinetic flux cost} In all these methods, we
assume that fluxes and enzyme levels are strictly proportional, which
leads to linear flux cost functions. Moreover, enzymes are sometimes
assumed to operate at their maximal rate (called $\kcat$ value), and
deviations from this ideal assumption have been dubbed ``unused enzyme
capacity'' \cite{obup:16}. However, part of this ``unused capacity''is
due to the fact that the enzymes within a larger network cannot
operate at their maximal speed, so we need to beware of idealised
models that ignore the trade-offs between enzymes and underpredict the
enzyme demands!  If flux cost weights for FBA are computed from
$\kcat$ values, these are lower bounds on the actual cost weights and
the actual values remain unclear. \co{FN about schaetzung in wortel
  SI.}  All this is of course unrealistic. To address these problems,
we now define realistic flux cost functions which reflect the minimal
enzyme and metabolite costs that come with the fluxes in a given
kinetic model (Figure \ref{fig:definition}).  Both types of cost have
been considered before (enzyme costs in
\cite{reic:83,brow:91,fnbl:13}, enzyme plus metabolite costs in
\cite{sche:91, tnah:13}).  To write them as flux costs, we first need
to score enzyme levels, metabolite concentrations, and fluxes by cost
and benefit functions: enzyme cost is a linear function of enzyme
levels (i.e.~penalising high enzyme levels), metabolite cost is a
convex function of the logarithmic metabolite concentrations
(e.g.~penalising deviations from some ideal metabolite profile), and
flux benefit is a linear function of the fluxes.  Given these
metabolic targets, we may formulate different optimality problems: we
may assume that cells maximise either the benefit-cost difference, the
benefit/cost ratio, or the benefit at a fixed cost, or that they
minimise cost at a fixed benefit. Below, we mostly consider this
latter case, where ``cost'' can refer to enzyme cost or the sum of
enzyme and metabolite costs. having all this in place, we can finally
assign to each flux profile the metabolite and enzyme profiles that
can realise this flux profile in the most beneficial way, quantify
these profile by their cost functions and assign this cost as an
effective ``flux cost'' to our metabolic {\flow}.

\begin{figure*}[t!]
  \fbox{\parbox{16.5cm}{
      \begin{center} 
        \parbox{15.5cm}{
       
        \textbf{Box 1: Cost functions for enzymes, metabolites, and fluxes in cells}\\

        \coout{kurz einfuehren, wie in CBA theory?``manifold''/''leaf/sheet''/''space''/''state''}

        \small
        
        The fluxes $v$, metabolite concentrations $c$, and enzyme levels
        $\esymbol$ in cells depend on each other. An enzyme cost
        $h(\esymbolv)$, a function of the enzyme levels, can be
        translated into an effective metabolite cost or a flux
        cost. We obtain three different variants of our cost function
        enzyme cost function, with different cell variables as
        function arguments\footnote{There is an analogy to this in
          classical thermodynamics.  Heat uptake can be considered at
          constant volume or at constant pressure,
          with different functional relationships.}\\

        \includegraphics[width=15.cm]{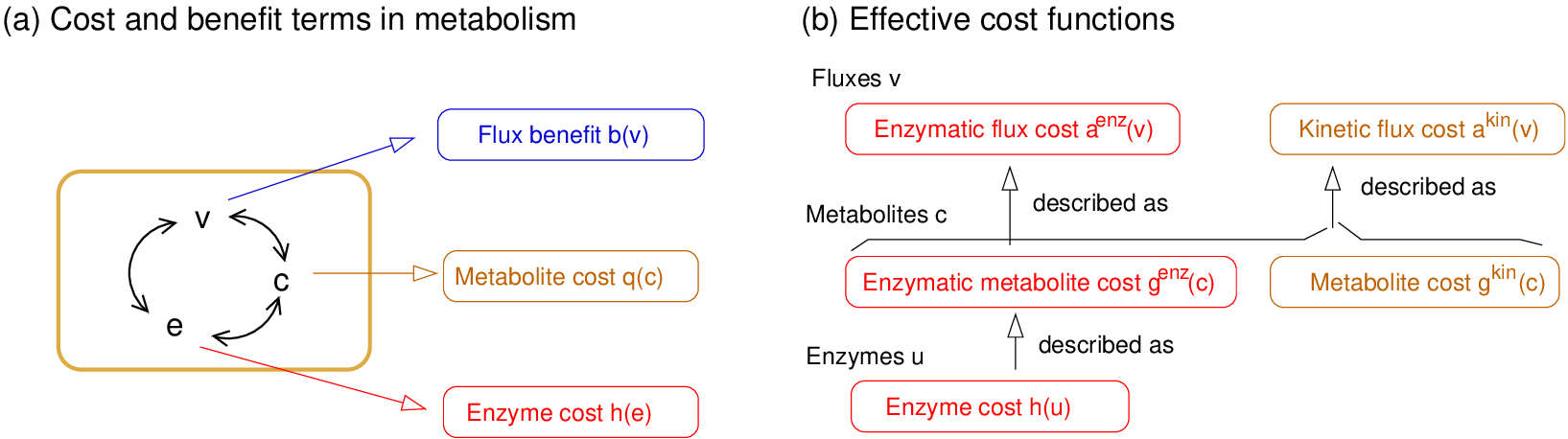}\\
        \label{fig:generalCostFunction}

        \co{chnage colors! metab orangerot, enz gold}
        
          \textbf{1. Enzyme cost} An \emph{enzyme cost} function 
          $\hminus(\esymbol_1, \esymbol_2, ..)$ is assumed to score 
          enzyme levels directly.   If each enzyme molecule has a fixed
          cost (which may differ between  
          enzymes), we obtain a linear enzyme cost
          function. The meaning and physical units of enzyme cost functions
          may vary from model to model. A cost  may, for example,
          refer to enzyme amounts (e.g.~enzyme mass per cell dry
          weight), e.g.~when predicting the enzyme demand of
          engineered pathways, or to overall growth
          defects\footnote{The reasons for such growth defects can be
            diverse, including demands for material, energy, and
            macromolecules for protein production and maintenance;
            space restrictions in cells and on membranes; other
            catalytic activities of the enzymes.} in units of 1/h (for
          absolute growth rate changes) or
          unitless (for relative growth rate changes).\\
          
          \textbf{2.  Metabolite cost} (or ``flux-restricted enzymatic
          M-cost'') A given  flux profile defines thermodynamic constraints on the
          possible metabolite concentrations. The feasible  
          metabolite profiles form a convex polytope in
          log-concentration space, called metabolite polytope or
          M-polytope. Its shape depends on
          network structure and flux directions, equilibrium
          constants, and external metabolite concentrations (see
          \cite{nfbd:16}).  Metabolite cost is described by functions
          on this polytope.  The \emph{enzymatic M-cost}
          \begin{eqnarray}
            \label{EnzymaticMetCost}
            \enzymemetcost(\lncv;\vv) = \hminus(\enzymev(\vv,\lncv))
          \end{eqnarray}           
          is an effective (or ``overhead'') cost, describing the
          enzyme cost $\hminus(\enzymev)$ needed to realise a {\flow}
          $\vv$ at the metabolite profile $\lncv = \ln \cv$, with
          given external metabolite concentrations and within allowed ranges
          for internal metabolite concentrations.  Given a {\flow} $\vv$, the
          enzyme levels $\enzymev(\vv,\lncv)$ follow directly from the
          kinetic rate laws. The cost (\ref{EnzymaticMetCost}) is a
          convex function of log-concentrations \cite{nfbd:16}, and
          the optimal metabolite profile can be computed by convex
          optimisation\footnote{In enzymatic metabolite cost
            minimisation (ECM), we consider a kinetic model (e.g.~with
            common modular (CM) rate law \cite{liuk:10}), a linear
            enzyme cost function, and fixed external metabolite
            levels, and compute the optimal enzyme and metabolite
            profile to realise the predefined {\flow}.}  (see Figure
          \ref{fig:definition}), defining the optimal enzyme profile
          and enzyme cost. By adding a direct metabolite cost $\metcost(\lncv)$, we obtain
          the kinetic metabolite cost
          \begin{eqnarray}
            \metcostkin(\lncv) = \metcost(\lncv) + \enzymemetcost(\lncv).
          \end{eqnarray}
          The metabolite cost g(m) should be convex (or even strictly
          convex) of mathematical convenience. It may describe, for
          example, the total metabolite mass density or a quadratic
          deviation from an (empiricall or biologically) preferred
          metabolite profile.  If $\metcost(\lncv)$ is strictly
          convex, the kinetic metabolite cost has be strictly convex
          and will have a unique
          optimum point, uniquely defining  optimal enzyme and metabolite concentrations.\\

          \textbf{3. Flux cost} (or ``metabolite-optimised enzymatic
          F-cost'') 
          The \emph{enzymatic flux
            cost} \begin{eqnarray} \label{eq:enzymaticFluxCost}
            \aenz(\vv) = \mbox{min}_\lncv\, \enzymemetcost(\lncv;\vv)
            = \mbox{min}_\lncv\, \hminus(\enzymev(\vv,
            \lncv)) \end{eqnarray} is the minimal enzyme cost at which
          we can realise the {\flow} $\vv$ in a given kinetic model
          (see Figure \ref{fig:definition}).  If we add a direct
          metabolite cost, we obtain the \emph{kinetic flux cost}
          \begin{eqnarray}
            \acostkin(\vv) = \mbox{min}_\lncv\, \metcostkin(\lncv|\vv),
          \end{eqnarray}
          i.e.~the minimal sum of enzyme and
          metabolite costs at which $\vv$ can be realised in our
          model. Flux cost functions $\acost(\vv)$ can also be  defined
          ad hoc (like the linear cost functions in FBA), In any case, they should
          increase with the absolute flux for plausibility reasons
          (satisfying $\sign(\partial a/\partial v_l) = \sign(v_l)$
          whenever $v_l \ne 0$), and they may show a jump where a
          reaction flux $v_l$ switches its sign, i.e.~where $v_l=0$.

          \coout{hier kurz auch fluss-optimierte kosten im metabolitraum erwaehnen? // Alle vier Funktionen NUR KURZ erwaehnen; ref to CBA theorie \cite{lieb:18theory}}
      } \end{center} } }
\end{figure*}

\myparagraph{Principle of minimal flux cost} \co{coordinate with
  paragraph before. thread is chaotic!} How can we find an optimal
metabolic state, comprising metabolite concentrations, enzyme levels,
and fluxes?  While each flux profile requires certain metabolite and
enzyme profiles for its realisation, this choice is not unique.  Each
flux profile can be realised by different combinations of enzyme and
metabolite concentrations. To choose the best concentration profiles,
we assume that cells minimise their enzyme cost\footnote{I assume that
  reactions are catalysed by specific enzymes (excluding non-enzymatic
  reactions, isoenzymes, and promiscuous enzyme activities from our
  models). Without this assumption, the problems would be much harder
  (see discussion).}  \cite{fnbl:13,nfbd:16} or their ``kinetic''
(i.e.~enzyme plus metabolite) cost, where costs may represent, for
example, occupied cell volume or molecular mass. By
  conceptualising enzyme and metabolite costs as flux costs, we can
  translate kinetic optimality problems into a simple search for
  optimal fluxes.  Enzyme and metabolite concentrations are not
  mentioned explicitly, but their costs are implicitly taken into
  account.  \co{JA! FN: kinetic cost with sum enz + sum met als
  spezialfall (sum met ist konvex (aber nicht strikt convex) in ln c!
  // come back to this later, when discussing density constraints =
  constraining this ``cost objective'' // also ref to hugo's work,
  minimising sum of BM space occupied (kam oben schon kurz!)} With
this optimality principle, each flux mode defines an optimal
metabolite and enzyme profile.  \co{JA! FN: Briefly say: can be ssolved by
  ECM; def metab polytope (see box 3) convex problem! \co{metabolite
    polytope // difference between external and fixed metabolites}}
Importantly, these profiles are not \emph{required} by the flux
profile (because other profiles would yield the same fluxes), but
\emph{\favoured} by it, because they are less costly than other
(physically and biochemically) possible choices.  The cost of these
profiles can be seen as the ``overhead flux cost'', i.e.~a
cost that is not caused by the fluxes themselves, but are inevitable  if
the fluxes are to be physically realised!  \co{FN Overhead costs can be derived from an ``optimistic''
  optimisation procedure. briefly explain,
  ref to CBA opt!  In the case of fluxes, every possible flux profile we optimise the enzyme and metabolite
  concentrations (``optimistic overhead cost'' uea), resulting in a
  ``layered modelling'' of fluxes and the underlying metabolic states,
  and effective flux costs representing metabolite and enzyme costs as
  overhead costs (kommt auch kurz weiter unten)} By minimising this
flux cost function, we optimise the entire metabolic
state\footnote{Optimising the flux-dependent fitness
  $\fluxbene(\vv)-\acostkin(\vv)$ with a kinetic flux cost
  $\acostkin(\vv)$ is equivalent, by construction, to optimising a
  fitness function
  $\ffit(\vv,\cv,\esymbolv)=\fluxbene(\vv)-\metcost(\cv)-\hminus(\enzymev)$
  \cite{lieb:18theory}, with metabolite and enzyme costs $\metcost$
  and $\hminus$, under physical and physiological constraints for
  fluxes, metabolite concentrations, and enzyme levels.}, including
fluxes, metabolite concentrations and enzyme levels.

\begin{figure*}[t!]
  \begin{center}
 \begin{tabular}{lll}
(a) Branch point model & (b) Enzyme cost in metabolite space & (c) Enzyme cost in flux space \\[1mm]
    \parbox{5cm}{\includegraphics[width=5cm]{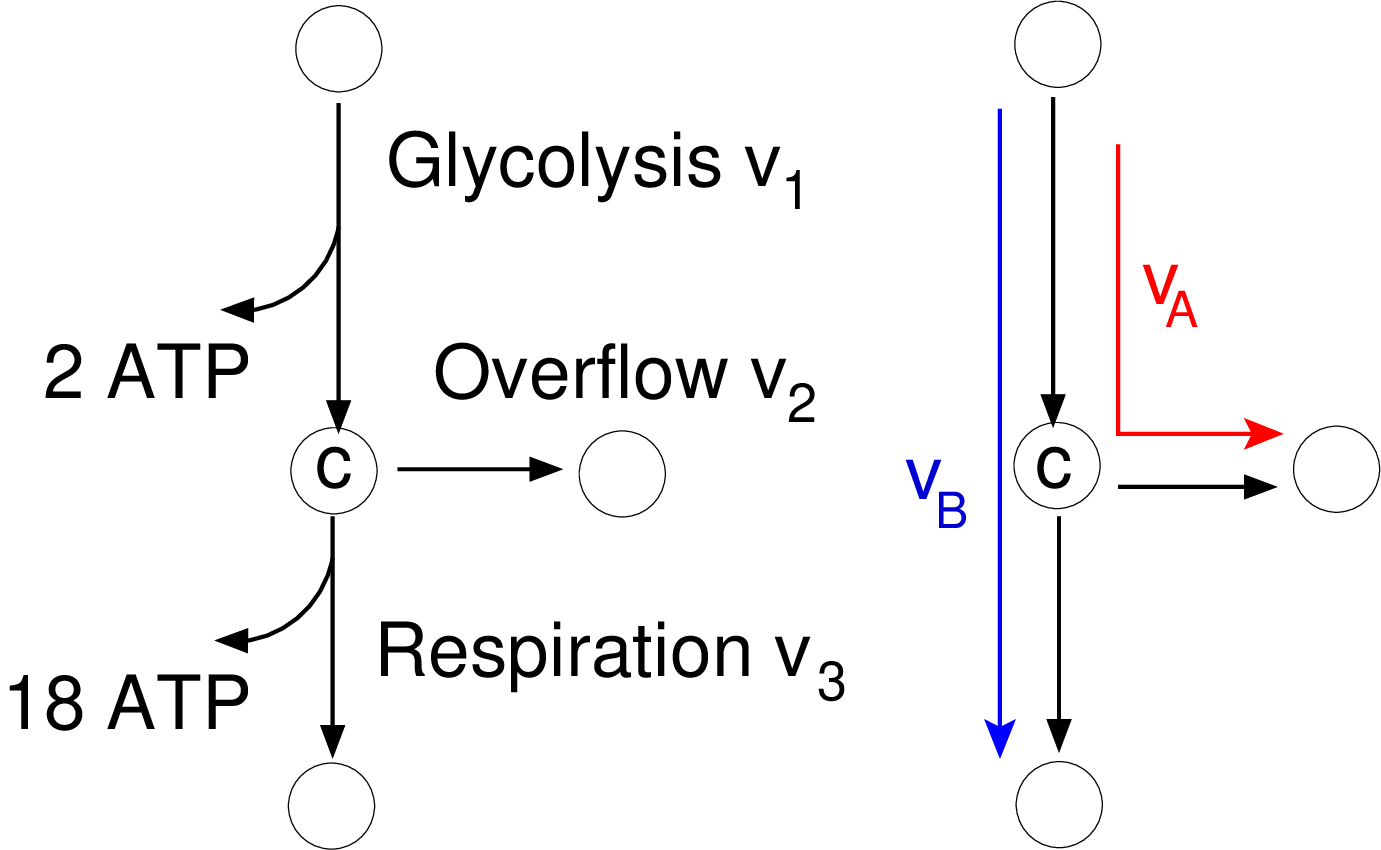}\\[1mm]} & 
    \parbox{5cm}{\includegraphics[width=5cm]{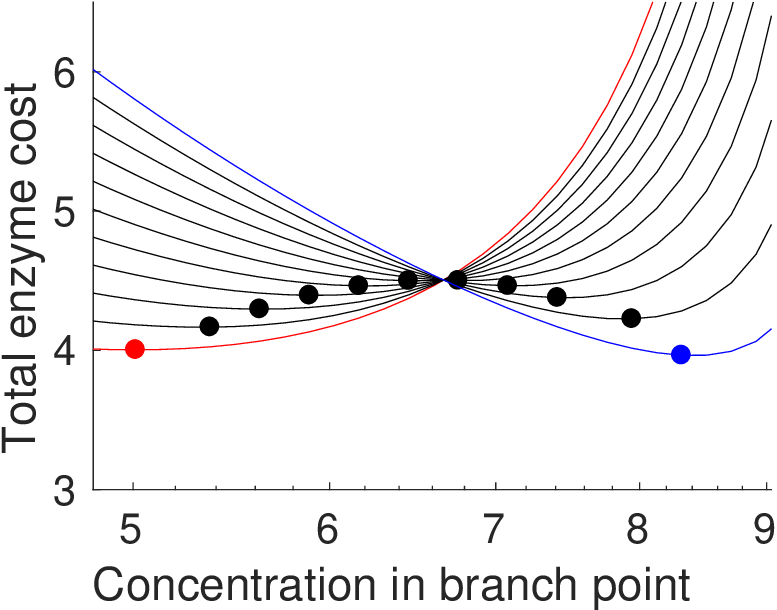}} & 
    \parbox{5cm}{\includegraphics[width=5cm]{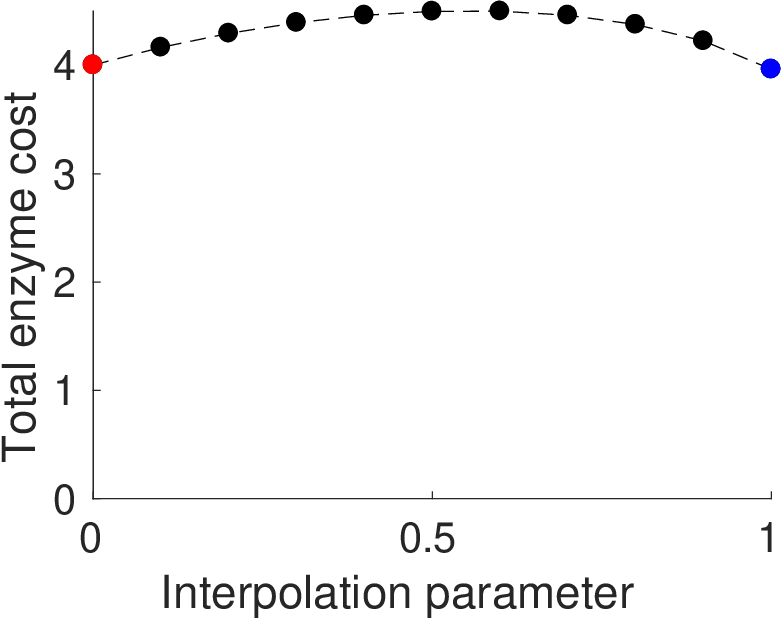}}\\
\end{tabular}
\end{center} \caption{Enzymatic flux cost in a  branch point
  model of fermentation and respiration.  (a) Model of central
  metabolism (see Figure \ref{fig:definition}) with reactions
  representing glycolysis ($v_1$), overflow ($v_2$), and respiration
  ($v_3$).  The internal  metabolite C (e.g.~pyruvate) has
  a variable concentration $c$, while all other concentrations are
  fixed. We consider a kinetic model with substrate-saturated,
  reversible rate laws and a linear benefit function
  $\fluxbene(v_1,v_2,v_3)$ scoring the ATP production rate. All
  stationary {\flow}s are convex combinations of two basic profiles $\vvA$
  (respiration, red) and $\vvB$ (fermentation, blue).
  (b) Enzymatic  cost in metabolite space.   Cost
  functions for the {\basicflow}s $\vvA$ and $\vvB$ (scaled to
  equal ATP production rates) are shown by red and blue lines. Black
  lines show  cost functions for {\combinedflow}s
  $\sigma \,\vvA + (1-\sigma)\,\vvB$, interpolating between $\vvA$ and
  $\vvB$.   For each {\flow}, the optimum point is shown by a dot.  (c)
  Enzyme cost in flux space. The optimal costs from (b) are plotted as
  a function of the interpolation parameter $\sigma$, which varies between 0
  (for {\flow} $\vvA$) to 1 (for {\flow} $\vvB$). The resulting
  cost function $\aenz(\vv)$ is negatively curved and
  therefore strictly concave.} \label{fig:interpolation}
\end{figure*}

\myparagraph{Example of enzymatic flux costs} Let us see an
example. Figure \ref{fig:interpolation} shows a simple model of
central metabolism, describing the choice between overflow metabolism
and respiration (compare Figure \ref{fig:definition} (a)).  We assume
reversible mass-action rate laws and fluxes in forward direction.
Given the fluxes $v_l$, the enzyme cost can be written as a function
of the logarithmic metabolite concentration $\lnc= \ln c$.  We obtain
the formula
$\metcost(\lnc) = \frac{a_0}{b_0-e^\lnc} v_0 + \frac{a_1\,
  e^\lnc}{e^\lnc-b_1} v_1 + \frac{a_2\, e^\lnc}{e^\lnc- b_2} v_2$.  In
our model, any stationary flux distribution is a combination of two
{\basicflow}s\footnote{To compare the two {\flow}s, we fix a desired
  flux benefit of 20 ATP molecules per time unit. Assuming that
  reaction $v_0$ yields 2 ATP molecules (per unit of flux) and $v_2$
  yields 18 ATP molecules (per unit of flux); this is exactly the
  benefit of a respiration flux profile with unit flux. thus, we
  interpolate between the modes $v_A=(10, 10, 0)\trans$ und
  $v_B=(1,0,1)\trans$. We then consider a range of possible
  concentrations $c$ in the branch point.  For the three reactions, we
  consider mass-action laws $r_0 = 10 -\lnc$; $r_1 = \lnc-0.01$,
  $r_2 = \lnc-0.01$ and cost weights $h_0=1$, $h_1=1$, and
  $h_2 = 28$.}. In the first profile, the entire flux goes through the
overflow reaction; in the second one, the entire flux goes through
respiration.  Since the flux directions are fixed, any other {\flow}s,
with different flux ratios, must be convex combinations of these basic
profiles.  Figure \ref{fig:interpolation} (b) and (c) show enzyme cost
as a function in metabolite space and in flux space. In metabolite
space, the blue and red lines denote the costs for the basic {\flow}s,
and the minimum cost for a {\combinedflow} must lie in between these
two lines. The overall minimum cost is always  achieved by one of the
basic {\flow}s (blue dot for
$\vvB$ in the example shown). \co{comment on fig c!}

\subsection{Shape of the enzymatic cost flux function}


Enzymatic flux costs can be computed by numerical optimisation, but
there is no closed formula to describe them. So what are their
mathematical properties?  The enzymatic flux cost $\acostenz(\vv)$ is
a nonlinear function defined on the set of thermodynamically feasible
{\flow}s.  Its precise shape depends on rate laws, parameters,
external conditions, and constraints of the kinetic model. In the
following I consider models with biochemically plausible rate laws
such as the common modular (CM) rate law, which has convenient
mathematical properties. More specifically, we consider factorised
rate laws \cite{nflb:13}, which comprise all plausible reversible rate
laws and ensure thermodynamically correct reaction
directions\footnote{In flux modelling (e.g.~\cite{wpht:14}), chemical
  reactions are often split into separate forward and backward
  reactions.  Afterwards, all fluxes can be assumed to be positive and
  the flux cone has no lineality spaces\co{REF}. Here, we do not apply
  such a splitting because it would make it harder to then make the
  connection to thermodynamically consistent, reversible reaction
  kinetics in the underlying kinetic models.}.  I further assume that
enzyme cost depends linearly on enzyme levels (for more details about
models and their mathematical description, see the SI).

\co{In einer abb (VOR ABB 2!) klarmachen: vier elemente: stationaritaet; flussrichtungen
  (haengen zusammen mit thermodynamik); fluss benefit; flusskosten;
  werden in verschiedener weise zusammengesetzt: Classical FBA (uses
  also einzelconstraints); FBA mit min flux; FBA mit mol crowd. Wo
  liegt das problem? in kosten: kinetik und met level werden
  ignoriert; hier: nichtlineare variante; alle moeglichen
  konsequenzen} 

\myparagraph{The set of possible {\flow}s} The cost of metabolic flux
distributions can be described by a cost function in flux space. To
describe these functions, as well the set of feasible flux
distributions, we need some precise terminology.  Stationary flux
distributions are called {\flow}s, or flux modes (if their scaling
does not play a role).  \co{check this again: templates with zeros or
  not? stefan fragen?}  A {\fluxpattern} $\sign(\vv)$ is a vector
describing the actual flux directions in a flux profile $\vv$ (with
elements -1, 0, and 1). \co{sort}A \emph{flux template} $\sigmav$ is a
sign vector that defines allowed flux directions (where fluxes may
always vanish)\footnote{In this article, flux templates are understood
  to be non-strict, that is, if a flux direction is predefined, the
  flux in question is still allowed to be zero. If strict flux
  patterns are imposed (where all zero fluxes are predefined), this
  will be explicitly stated.}.  A flux distribution $\vv$ is
\emph{conformal} with $\sigmav$ if $\sign(v_{l})=\sigma_{l}$ for all
$v_{l}\ne 0$.  Flux templates may follow, for example, from
thermodynamic driving forces, which in turn depend on the chemical
potentials (and therefore on metabolite concentrations).

\co{WRITE kurz! eher konkret c, dann mu, dann
  theta, dann tope!  v conformal to -sign(theta) // beziehung zu
  segment (strict pattern) und orthant (loose pattern) // ``strict
  {\fluxpattern}'' / ``flux sign pattern'' ``TOPES''?  defines the
  allowed signs. A flux mode is conformal with a {\fluxpattern} means=
  all active fluxes (in the flux mode) have the right signs; and all
  fluxes that are required to be zero are zero; but fluxes can be off,
  even though they would be allowed by the non-zero. E.g., the loose
  thermo-constraints define a flux sign pattern. If strict matching of
  active fluxes is required, this must be marked by the word ``strict
  {\fluxpattern}'' or is ``strictly conformsal with the
  {\fluxpattern}''} \co{from ``Tope'', cite stefan et al}

\co{rename? klaeren!\\
  flux template -> flux pattern\\
  flux pattern -> strict flux pattern (if imposed)\\
flux sign profile if descriptive}

\co{Abbildung mit verschiedenen constraints und kombinationen?}

Geometrically, a {\flow} can be seen as a point in a flux space
(Figure \ref{fig:definition}).  The set of feasible {\flow}s is
defined by model constraints, including stationarity, predefined flux
directions and inactive reactions, bounds on fluxes or flux sums, or a
predefined flux benefit.  How will these constraints shape the set of
feasible {\flow}s? First, the stationarity condition defines a linear
subspace in flux space. \co{.. cone!}  Second, a given flux template
predefines zero fluxes in some of the reactions and flux directions in
others, defining a feasible segment in flux space.  \todo{The
  intersection of stationarity constraints and sign constraints yields
  a flux cone. If all flux directions are definedm the cone is inside
  of one of the orthants.}  \co{say: F-cone = directed flux cone}
\co{with EFMS as extremal rays} Instead of imposing flux directions
directly, we may require that the flux directions follow thermodynamic
constraints.  Given the equilibrium constants and external metabolite
concentrations, only some flux sign profiles remain thermodynamically
feasible (see Box 3). Each of them defines a feasible segment in flux
space. \co{orthant or n-dimensional surface of an orthant} \co{FN:
  KOMPLETT NEU SCHREIBEN ODER WEG\co{Each segment in flux space
    represents a possible {\fluxpattern}.  \co{erklaerung wo?}  Only
    some of them, however, are feasible (and can be realised by our
    kinetic model).  The set of all feasible {\flow}s consists of a
    collection of convex F-cones, each in one orthant of flux
    space. The feasible {\fluxpattern}s reflect stationarity and the
    exclusion of thermodynamic loops. With given physiological
    concentrations ranges and fixed external concentrations, defined
    in a model, even more {\fluxpattern}s can be excluded.  Each
    feasible {\fluxpattern} defines an F-cone in flux space, and an
    M-polytope, i.e.~a polytope of possible metabolite profiles in
    log-metabolite space.}}  Third, individual flux bounds define a
feasible box in flux space. Fourth, space limitations may put
constraints on the total enzyme levels in cell compartments. In mcFBA,
this density constraint leads to upper bounds on weighted sum of
fluxes. Fifth, we may score our flux profiles by a linear benefit
function, for example, the rate of biomass production. By predefining
this flux benefit\footnote{Instead of predefining the flux benefit, we
  may also enforce a positive benfit by putting a lower
  bound. However, in our models, (where an overall flux scaling leads
  to a monotonical increase of flux benefit and enzyme costs), the
  flux benefit will always reach its lower bound and so the inequality
  constraint for flux benefit can be replaced by an equality
  constraint.}, we exclude some flux profiles (those that provide no
or negative benefits) and normalise all others (e.g. to a benefit of
1). Mathematically, this restrict our {\flow}s to a feasible
hyperplane which intersects the flux cone, giving rise to the
so-called B-polytope, whose vertices correspond to EFMs\footnote{The
  flux benefit constraint defines a feasible hyperplane. Unlike the
  subspace defined by stationarity constraints, this hyperplane has an
  offset und does not cut through the origin. \co{Formally, it
    resembles a stationarity constraint with some given external
    metabolite production or consumption rates}. For flux solutions to
  exist, $\bvdir$ must be linearly independent of span$(\Nint)$; this
  allows us to rescale all {\flow}s by their flux benefits.}.

\co{benefit function yields ``B-polytope'', ; bounds (box constraints
  or bounds on weighted sums) can further reduce the B-polytope (and
  create non-EFM vertices)}

These constraints, in different combinations, lead to different
variants of FBA. By applying them to our possible flux modes, we limit
the flux space to a feasible region, consisting of a collection of
convex cones (see Box 3), or convex polytopes if flux bounds are
used. Below, the different polytopes (or cones) in this collection
will be considered one at a time. As a formal requirement, we assume
non-negative fluxes: this is just for convenience, and we can always
ensure this by reorienting the reactions. Without any other bounds, we
obtain a positively oriented flux cone (called F-cone), and if we add
a flux benefit constraint, we obtain the B-polytope.

\begin{figure*}[t!]
  \begin{center}
\parbox{10.8cm}{
    \includegraphics[width=10cm]{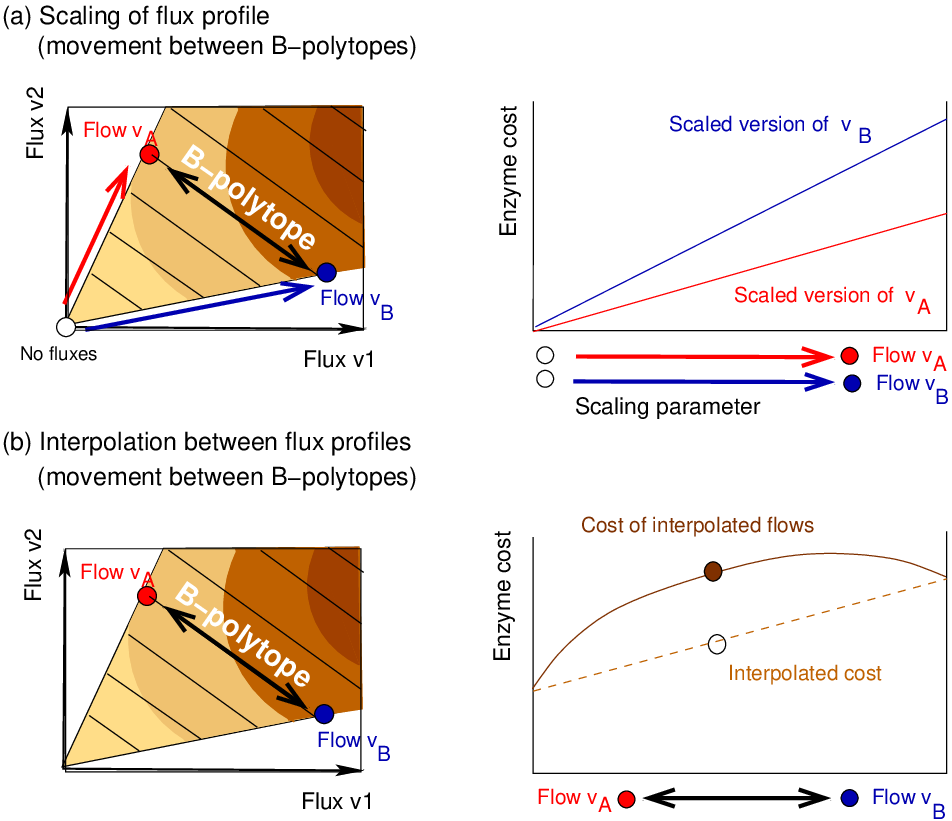}
}
\parbox{5.5cm}{
  \caption{Enzymatic flux cost function $\aenz(\vv)$ on a flux cone
    (schematic drawing).  (a) Left: a flux cone is spanned by flux
    modes $\vvA$ and $\vvB$. The cost function (shades of brown) on
    this cone is concave. Right: The enzymatic cost scales
    proportionally with the overall {\flow}: when a {\flow} $\vv$ is
    scaled by a factor $\sigma$, its enzymatic cost scales linearly,
    as $\aenz(\sigma\,\vv)=\sigma\,\aenz(\vv)$.  (b) Same as (a), but
    for movements within a B-polytope (diagonal lines).  If two
    {\flow}s $\vvA$ and $\vvB$ are kinetically distinct, i.e.~if they
    {\favour} different metabolite profiles, the cost function on the
    line between $\vvA$ and $\vvB$ is strictly concave.  To obtain a
    linear cost function instead (dashed line), the cell might run the
    {\flow}s separately, either in space (cell compartments) or in
    time (phases of a metabolic cycle) Separated in this way, each
    {\flow} can operate under its optimal biochemical conditions and
    their flux costs are additive.  \label{fig:properties}
    \co{bedeutung der form des polytopes/cones is unklar!!
      ueberlegen.  links: form wie in abbildung 1!}}
}
  \end{center}
\end{figure*}

\myparagraph{Shape of the flux cost function on a directed flux cone}
 \co{erst lieber mehr von summe
  von fluessen sprechen als von konvexkombinationen} \co{frueh von
  \compromisecost\ sprechen; bei summe und konvexkombination; daraus
  folgt strict konkav} Having defined a positive flux cone, we can
study flux cost functions on this cone. \co{erst scaling and
  interpolation of flux profiles; dann geometrisch: movement} To learn
about their shapes, we consider movements between {\flow}s (points on
the cone) and how they affect flux cost (see Figure
\ref{fig:properties}). A flux cone consists of a ``stack'' of parallel
B-polytopes, each with a different benefit.  Any movements between
flux profiles consist of two basic types of movement: a rescaling of
{\flow}s (a movement between B-polytopes) and an interpolation of
{\flow}s of equal benefit (a movement within a
B-polytope)\footnote{Interpolating between two ``basic'' {\flow}s
  $\vvA$ and $\vvB$ yields a ``combined'' {\flow}. Of course, a
  combined {\flow} can serve, again, as a basic {\flow} in another
  convex combination.}.  How do these movements affect the flux cost
$\acostenz$?  If we scale all fluxes by the same factor $\sigma$, the
enzymatic cost scales proportionally (while its flux-specific cost or
benefit-specific cost, stays the same).  If we interpolate between
{\flow}s, the cost may vary non-linearly. In general, enzymatic flux
costs are concave on the interpolation line, i.e.~the cost
$\acostenz(\vvA+\vvB)$ of a sum of {\flow}s can be higher than the sum
of costs $\acostenz(\vvA)+\acostenz(\vvB)$, but not lower. If this
superadditive interpolation holds between any two points on a
B-polytope, the flux cost functions is strictly concave within the
B-polytope.

\co{\textbf{Scaled flux cost functions, flux modes, and the
    B-polytope} JA! alles zum thema hier! sagen: allgemein fluesse
  skalierbar; das aendert den enzymbedarf (bei konstanten met conc:
  proportional!) dh zum vergleich flussprofile normieren (zB auf
  influx); hier aber meistens auf benefit, dann kostenvergleich) wird
  klar, dass skalieren die skalierte kostenfunktion unveraendert
  laesst?  wenn wir uns nur fuer das verhaeltnis (kosten/nutzen)
  interessieren, dann koennen wir alles im B-polytop machen! bei
  einheits-b-wert haben dort sogar skaplierte und nichtskalierte
  kostenfunktion die gleichen numerischen werte!  aber vorsicht, wenn
  es weitere constraints gibt - dann spielt die absolute skalierung
  eine rolle wegen der jeweiligen form des b-polytops!}
\co{einfuehren, fuer cost function on the b-polytope lieber ein
  extra-symbol! $\hat a(..)$ (``scaled''), relation to flux modes usw}
\co{soll ich das B-polytope (wenn nichts anderes gesagt wird) mit b=1
  definieren?}  \co{on b-polytope oefter ``flux mode'' statt {\flow}
  verwenden?}

\myparagraph{How flux cost scales with the overall {\flow}} Let us now
have a closer look at these properties -- proportional scaling and
concavity.  We start with proportional scaling.  With a linear enzyme
cost function $\hminus(\enzymev)$ and a proportional scaling
$\vv \sim \esymbolv$, the flux cost $\aenz(\vv)$ would scale linearly
with $\vv$.  But why can we assume that metabolite concentrations
remain fixed?  A given {\flow} $\vv$ defines an enzymatic M-cost
$\enzymemetcost(\lncv)$ on the metabolite polytope. If we scale our
{\flow} $\vv$ by $\sigma$, the function $\enzymemetcost(\lncv)$ keeps
its shape, but is scaled by the same factor $\sigma$. The {\favoured}
metabolite profile (i.e.~the minimum point of the cost function in
metabolite space), stays unchanged\footnote{What if the enzyme cost
  function $h(\esymbolv)$ itself is nonlinear, e.g.~a function
  $h(\esymbolv) = h'(\sum_l \alpha_l \, \esymbol_l)$ with a nonlinear,
  increasing function $h'$? Also in this case, the {\favoured}
  metabolite profile would be independent of flux scaling, but the
  flux cost itself would not scale linearly with $\vv$ but reflect the
  nonlinearity in $h'$.}, and the {\favoured} enzyme profile keeps its
shape, but scales proportionally with the fluxes. Altogether, the
enzymatic flux cost $\acostenz(\vv)$ scales proportionally with the
{\flow} and the scaled cost function
$\hat{\acost}^{\rm enz}(\vv) = \acostenz(\vv)/\fluxbene(\vv)$ is
constant under scaling.  Linear scaling leads to useful sum rules (see
Box 2) for enzymetic cost functions.  What about kinetic flux cost
functions $\akin(\vv)$?  If we rescale fluxes and enzyme levels and
keep the metabolite levels unchanged, the kinetic M-cost
$\hmet^{\rm kin}(\lncv) = \metcost(\lncv) + \hmet^{\rm enz}(\lncv)$,
will change linearly with an offset term because
$\metcost(\lncv)=\const$ and
$\hmet^{\rm enz}(\lncv)=\hmet(\esymbolv(\vv,\cv))$ is proportional to
$\vv$. Since this changes the optimal metabolite profile, we cannot
expect a simple scaling behaviour for $\hmet^{\rm kin}(\lncv)$.
\co{WO? enzyme demand and enzymatic cost will change proportionally
  with the overall flux.}

\co{WO? bei kinetic flux cost: quadratisches skalierungsverhalten bei skalierung von gesamtkonzentration (halbe enz und halbe met) ergibt ungefaehr (in linearem bereich) ein viertel der fluesse!}

\begin{figure*}[t!]
\mybox{\textbf{Box 2: Sum rules for enzymatic flux cost functions}\\ 

  Metabolic steady states have a simple scaling property: if all
  fluxes are proportionally scaled (e.g.~be a time rescaling or a
  scaling of all enzyme levels) and all metabolite levels stay the
  same, we obtain again a feasible steady state. In FCM, this scaling
  has consequences for optimal states.  If a {\flow} $\vv$ is scaled
  by a factor $\sigma$ and all external metabolite concentrations stay
  fixed, the optimal metabolite and enzyme profiles change in a simple
  way: the metabolite concentrations $\cv^{\rm opt}(\vv)$ stay
  constant, while the optimal enzyme levels $\esymbolv^{\rm opt}(\vv)$
  and the enzymatic cost $\aenz(\vv)$ scale proportionally with
  $\sigma$ (see Figure \ref{fig:properties}). These scaling properties
  lead to sum rules for the derivatives of these functions.  Since
  $\acostenz(\vv)$ scales proportionally to $\sigma$ with an exponent
  of 1, it is a homogeneous function of the fluxes with degree 1.
  Euler's identity for positive homogeneous functions \co{ref!} yields
  the sum rule
\begin{eqnarray}
\label{eq:SumRulePointCost}
  \sum_{l} \frac{\partial \acostenz}{\partial \ln v_{l}} =  \acostenz.
\end{eqnarray}
By defining  the flux investment
${\apointcostvl}^{\rm enz} = \frac{\partial \acostenz}{\partial \ln
  v_{l}} = \frac{\partial \acostenz}{\partial v_{l}}\,v_l$ and the total flux investment
${\asumpointcostvl}^{\rm enz} = \sum_l {\apointcostvl}^{\rm enz}$, we
obtain the equality
\begin{eqnarray}
 {\asumpointcostvl}^{\rm enz} =  \acostenz.
\end{eqnarray}
In other words: the sum of flux investments is equal to the enzymatic
flux cost $\acostenz$.  This holds in each point $\vv$ of the flux
cone.  We obtain a similar sum rule for the scaled flux cost
$\hat{\acost}^{\rm enz}(\vv)$ (for example, a flux cost per pathway
flux or a flux cost per flux benefit). Since
$\hat{\acost}^{\rm enz}(\vv)$ is invariant under a scaling of
${\sigma}$ and is therefore a homogeneous function with degree 0, the
sum rule reads \co{symbole schoener (punkt hinten), abstand weg:}
\begin{eqnarray}
  \label{eq:SumRulePointCost2}
\hat{\asumpointcostvl}^{\rm enz} = 
\sum_{l} {{\hatapointcostvl}}\ ^{\rm enz} = \sum_{l} {{\hatacostvl}}^{\rm enz} \, v_l = 0.
\end{eqnarray}
There are similar sum rules for optimal metabolite or enzyme levels:
\co{symbol $O^{u_{j}\cdot}_{v_{l}}$? ``optimistic sensitivities''? entsprechend E und C?}
\begin{eqnarray}
\label{eq:SumRulePointCost5}
  \sum_l {\esymbol\partialder}^{j,\rm opt}_{v_{l}} = \esymbol^{\rm opt}_j, \qquad
  \sum_l {c\partialder}^{i, \rm opt}_{v_{l}} = 0,
\end{eqnarray}
where \co{punkt hinten!}
${\esymbol\partialder}^{j,\rm opt}_{v_{l}}=\frac{\partial
  \esymbol^{\rm opt}_j}{\partial v_l} v_l$ and
${c\partialder}^{i, \rm opt}_{v_{l}}=\frac{\partial c^{\rm
    opt}_i}{\partial v_l} v_l$. \co{JA! Kinetic flux cost: no linear
  scaling of costs! but for opt met und enz levels same scaling as
  with the enzymatic cost!} \co{wozu ist es gut? Mathematische
  einsichten und beweise!} The sum rules for optimal metabolic states
resemble the summation theorems of Metabolic Control Theory, which can
be derived in a similar way. While flux cost functions may be hard to
compute -- we need to solve an optimality problem to evaluate them --
our sum rules makes some mathematical proofs very easy.  For details
see SI section \ref{sec:SIsumRule}.}
\end{figure*}

\co{JA! KURZER ABSCHNITT UEBER FLUX COST GRADIENTS -- dann auf box 2
  verweisem, mit FCM: erklaeren a = sum av v (av =cost weight); in
  models av = hu u/v = burden!} \co{JA! (auch in CBA fluxes; )nochmal
  durchdenken: in FBA: ``flux prices''? in a given linear flux cost
  function: ``flux cost weights''; in kinetic models direct flux
  prices (=0); if flux prices represnt (indirect) other costs, they
  are called ``indirect flux prices'' av, or equivalently ``flux
  burdens'' (analogous to ``indirect concentration value''=''load''?)}

\myparagraph{The enzymatic flux cost at a given benefit may be
  strictly concave} \co{refer to (and use) Joost hulshoffs work
  \co{FCM: ref planque et al plos comp biol 2018: protein expression
    optimisation of an EFM is a strictly convex optimisation problem}
  on convenience kinetics yielding strictly convex enzymatic cost in
  metabolite space} Now we come to the second property, the fact that
flux cost functions are concave on each B-polytope. We consider a
particular B-polytope, defined by a {\fluxpattern} $\sign(\vv)$ and a
linear benefit $\fluxbene(\vv) = \fluxbene'$, and study the flux cost
function on this polytope. Our previous result -- enzymatic flux costs
scale proportionally with the {\flow} -- may suggest that these cost
functions are linear. If they were, flux costs could be added and
interpolated between different {\flow}s (as in FBA with minimal
fluxes).  But this is generally not true.  Enzymatic flux cost
functions are known to be concave: the cost of an interpolated {\flow}
must be equal or higher than the corresponding interpolated cost
\cite{wpht:14}: this holds whenever the enzymatic M-cost
$\enzymemetcost(\lncv,\vv)$ is positive and additive between {\flow}s
(proof in section \todo{\ref{sec:SIproofElementary}}).  While linear
functions are also concave, flux cost functions are usually nonlinear
and strictly concave. This makes them superadditive: the cost of an
interpolated {\flow} is higher than the interpolated cost.  The extra
cost depends on the shape of $\enzymemetcost(\lncv,\vv)$ (and thus on
the rate laws), and arises from a compromise between the two
{\flow}s. Each of the {\flow}s requires a different optimal metabolite
profile, and when they are added, they need to ``negotiate'' a common
metabolite profile, which will be non-optimal for each single one of
them.  Due to this non-optimality, the total cost will be higher than
the sum of individual costs (an equality would require that each
{\flow} runs at its optimal metabolite profile. A similar extra cost
exists when flux profiles are linearly interpolated. An example is
shown in Figure \ref{fig:interpolation}. \co{OMIT? If two {\flow}s
  {\favour} the same metabolite concentration $c$, we can interpolate
  between them (by interpolating between the enzyme profiles and
  keeping the metabolite profile constant): the resulting cost
  function is linear. Otherwise, if {\flow}s {\favour} different
  metabolite concentrations, they need to ``negotiate''. The
  compromise metabolite profile will still be suboptimal for of the
  original {\flow}s, so the enzymes work less efficiently, and there
  is an extra \compromisecost.}  The fact that there will be a
{\compromisecost} between any two {\flow}s on the line makes the entire
flux cost function strictly concave.

\begin{figure*}[t!]
    \includegraphics[width=16.5cm]{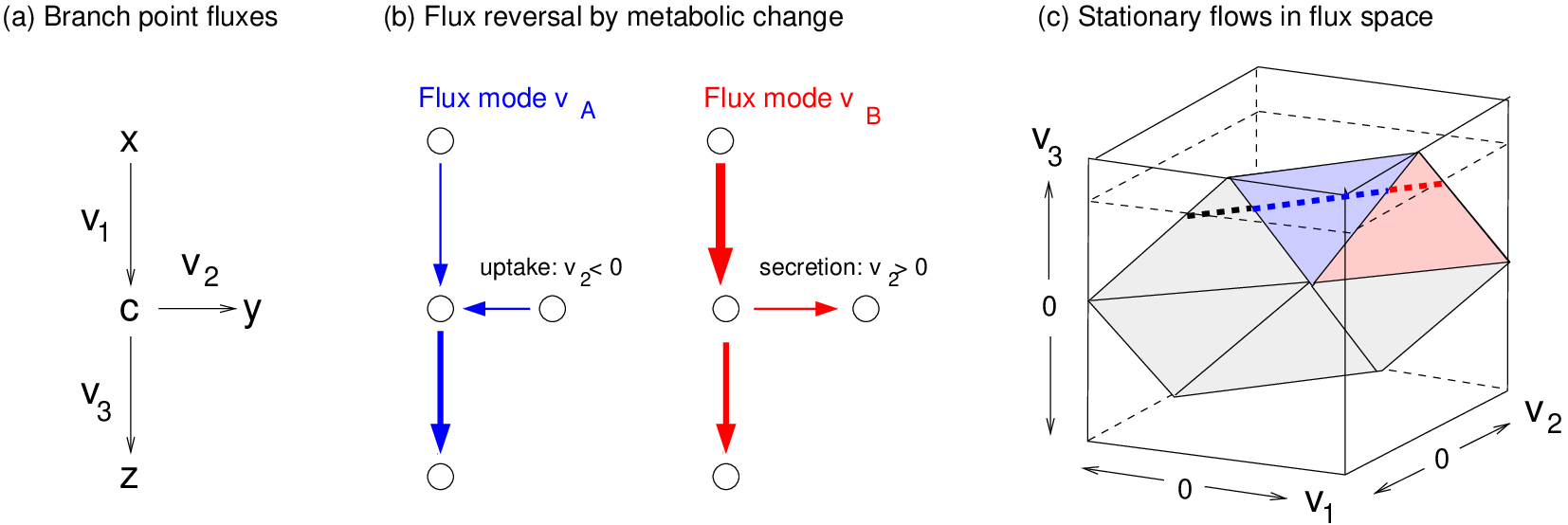}\\
 \parbox{15.5cm}{
    {(d) Jump in metabolite  levels \hspace{1cm} (e) Jump in enzyme levels}  \\[3mm]
   \includegraphics[width=5cm]{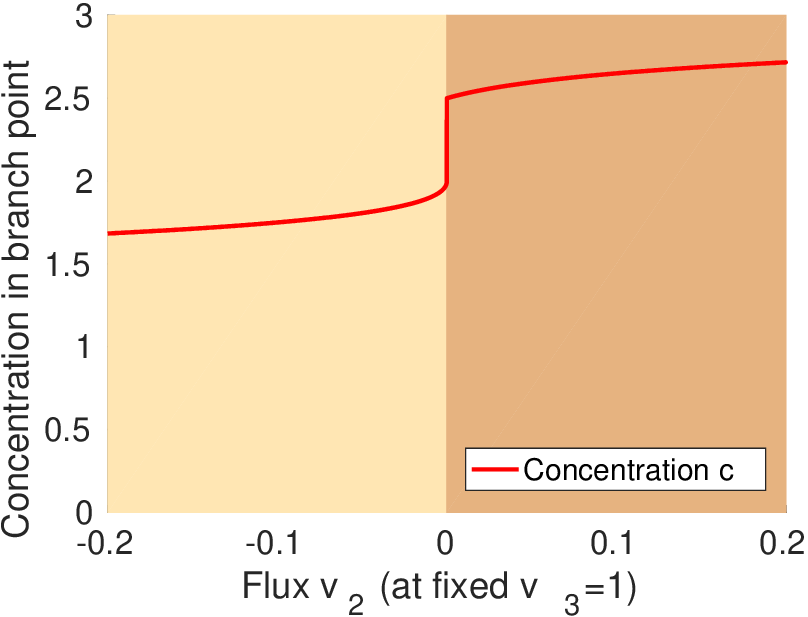}
   \includegraphics[width=5cm]{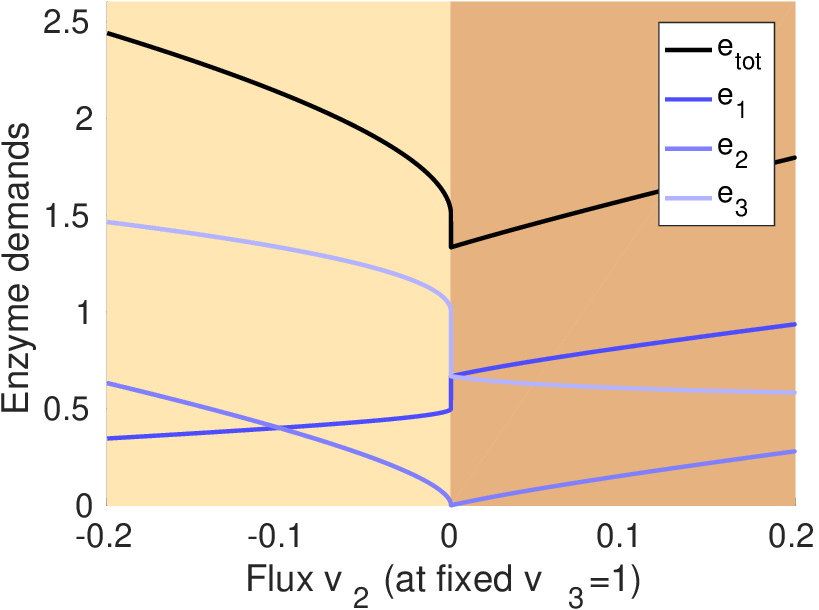}}
 \caption{\co{in c, fluxes statt flows} \co{farben in d und e blau (links) und rosa (rechts) //
     farben in legend} The shape of flux cost functions around 
   points of flux reversal. (a) Branched pathway with an internal
   metabolite C.  (b) At  given chemical potentials $\mu_{\rm X}$
   (high), $\mu_{\rm Y}$ (medium), and $\mu_{\rm Z}$ (low),  the
   fluxes $v_{1}$ and $v_{3}$ are forced to run in forward direction, while
   $v_{2}$ can change its direction (depending on the chemical potential of C). The possible
   {\fluxpattern}s can be decomposed into  (non-elementary) basic {\flow}s
   $\vvA$ and $\vvB$. the two basic profiles are scaled to units of flux benefit (flux in
   reaction 3), and   reaction rates are shown by arrow widths.  (c)
   {\Flow}s shown in flux space.  Polytope colours correspond to the
   {\fluxpattern}s shown in (b)). The thick dotted line represents the
   B-polytopes. (d) Optimal concentration of $c$, for interpolated
   {\flow}s between $\vvA$ and $\vvB$ (following the thick line in
   (c), and using the flux $v_2$ as a sweep parameter).  The optimal
   metabolite concentration shows a jump where fluxes change their direction.
   (e) The optimal (reaction-specific and total) enzyme costs as
   functions of flux $v_2$ also show a jump.\coout{aus modell
     berechnete enzymkosten auf polytop plotten?}  \coout{(d) optimal
     branch point concentration - consumption of y -production of y.}
   \coout{discuss the cool result!! evtl noch farben anpassen?  bei
     jump immer dieses beispiel
     erwaehnen!}} \label{fig:branch_point_flux_reversal}
\end{figure*}

\co{be clear about conditions for optimal states to be vertex points; considerations:\\
  ``typical case'': the FCF (on the B-polytope) is strictly concave, and there is only one global minimum (which is a vertex)
Exceptions:
1. the FCF may be strictly concave, but may have several local minima which have exactly the same cost, and are therefore are global minima (infinitely unlikely); and in fact, we don't really care about a unique global optimum here, just about local optima.

So what are the conditions under which there are optima that are NOT vertex points:
this requires that \\
1. the FCF is not strictly concave (between two points A and B), meaning that they have the same opt metabolite profile \\
2. these points must have exactly the same flux cost\\
3. these points must be  global optimum points\\
All this together is very unlikely!

Now introduce nomenclature for this.

An ECM problem is called regular if no two flux modes favour the same metabolite profile (unless these two flux modes are scaled versions of each other)
}

  \co{§OO o FCM: references for membran limitation:\\
    o Economics of membrane occupancy and respiro-fermentation Zhuang
    -> Competition for finite membrane space influences respiration and fermentation strategies.\\
    o Why do fast-growing bacteria enter overflow metabolism? Testing
    the mebrane real estate hypothesis Szenk
    -> Fermentation at higher growth rates is cell response due respiratory protein packing in the membrane at higher growth rates\\
    o Overflow metabolism in Escherichia coli results from efficient
    proteome allocation Basan -> Overflow metabolism is due to
    proteome constraint of the cell.}
\co{WO?  To decide this, we need to study the underlying ECM
  problem. The main question is whether two different flux profiles
  can favour the same metabolite profile. This, in turn, is related to
  another question: given a {\flow}, is it possible to vary the
  metabolite concentrations while leaving all individual enzyme
  demands unchanged?  \co{In fact, thsi is possible (show
    counterexample with MA kinetics and splitting / joining a
    metablite), but not very likely to happen in realistic metabolic
    models.}}

\myparagraph{Conditions for strictly concave flux cost functions}
\co{``REGULAR FCM model''} \todo{We saw that the enzymatic flux cost
  function is concave (within the B-polytope), and that it is strictly
  concave except in models with the following property: there exist
  flux profiles of different shape (not only of scaling) that favour
  (in the underlying ECM problem) the same metabolite profile.}
\co{die logik hier ist voellig unklar! den ganzen absatz editieren}
\todo{How can we know, in general, if an enzymatic flux cost function
  is strictly concave?}  To see this, we need to study the enzymatic
M-cost function, and in particular its curvature matrix.

\begin{proposition}
If the M-cost function is curved in all directions, it is
strictly convex and the resulting F-cost function in flux space is
strictly concave.
\end{proposition}
\co{Proof see 3.2 fuer ein inneres optimum\\
  2 bedingungen; die fluesse muessen genau gleichen demand haben UND gleiche opt metab profile haben! The probability for this in general models is zero!\\
Call ``normal'' models regular and others ``singular''
}

\todo{\co{CONDITIONS FOR STRICTLY CONVEX FLUX COST FUNCTIONS:}
A vanishing curvature in a direction in metabolite
space tells us that changes in this direction are
``cost-neutral''\co{UEA?}, i.e.~the cost is constant in this
direction. To relate these two properties, we now define ``kinetic
equality'': two {\flow}s are kinetically equal if they {\favour} the
same metabolite profile. Given two kinetically equal {\flow}s, there
is no compromise cost as we interpolate between the two flux profiles:
instead, their costs are additive and the cost function between them
is linear. In contrast, {\statedistinct} {\flow}s incur a positive
compromise cost, and the enzymatic flux cost is strictly concave on
the interpolation line (Proposition
\ref{prop:FluxCostIsStrictlyConcave} in SI section
\ref{sec:SIFluxCostIsStrictlyConcave}).  This means: if all {\flow}s
in a B-polytope are {\statedistinct}, the enzymatic flux cost is
strictly concave.

How can we tell whether two {\flow}s are \statedistinct?  They must
certainly have different shapes: if they differed only by scaling,
they would {\favour} the same metabolite profile and would be
{\stateequal}\co{better word? equivalent?}. But different shapes are
now enough.  For example, let two ``isoenzymes'' have exactly the same
rate laws and enzyme cost functions (i.e.~isoenzymes that are
basically identical). If two {\flow}s differ only in the usage of
these isoenzymes, they have different shapes but will still {\favour}
the same metabolite profile and are therefore {\stateequal}. This is
of course a ``pathological'' counterexample, but it \todo{shows that
 enzymatic cost functions are not generally  strictly concave.}

\co{ARGUMENT IST NOCH NICHT KLAR! lieber direkt am beweis entlang?}
Here is a general criterion for kinetically equal flux profiles: let
two {\flow}s $\vv_A$ and $\vv_B$ be conformal (i.e.~without any
opposing flux directions), and let $\lncv_A$ and $\lncv_B$ be the
corresponding optimal metabolite profiles. If a variation of $\lncv_A$
or $\lncv_B$ (at constant enzyme levels) changes a reaction rate (in
$\vv_A$ and $\vv_B$, respectively), then the same variation of $m_{A}$ or $m_{B}$ (at
constant fluxes) would change the enzyme demands. \co{aber aendern
  sich dann garantiert auch die enzymkosten?}  \co{was folgt daraus?
  argument noch expliziter} \todo{This happens, for example, if the
  enzymatic M-cost is strictly convex (Proposition
  \ref{prop:ConvexLeadsToKineticallyDistinct} in the SI) as it is the
  case in models with convenience kinetics \cite{likl:06a} or common
  modular (CM) rate laws \cite{liuk:10}. \co{(see SI section
    \ref{sec:SICMpositivelycurved}, proof by Joost Hulshof)\co{und
      bob?}}\co{new proof!}  In fact, in models with realistic rate
  laws, {\flow}s with different shapes are usually \statedistinct.}  }


\myparagraph{Flux cost functions are discontinuous at polytope
  boundaries (where fluxes change their direction)} With predefined
flux directions and a predefined benefit, our flux profiles are
confined to a B-polytope, and we already know that enzymatic flux cost
functions are concave on such polytopes.  \co{use "interior",
  "surface", and "closure" in FCM, when describing Flux polytopes}
\coout{Briefly, a {\fluxpattern} is feasible if its segment is cut by
  the subspace of stationary, benefit-realising flux modes, and if the
  corresponding metabolite polytope is not empty.  \co{The set of all
    metabolic states can be written as
    ${\mathcal S} = \{(\enzymev,\cv,\vv) | \vv \in {\mathcal V} \wedge
    (\enzymev,\cv) \in {\mathcal S}_{\vv} \}$.}}  But what happens on
the polytope boundaries, where two polytopes touch and fluxes change
their directions? An example is shown in Figure
\ref{fig:branch_point_flux_reversal} (c). Along the polytope boundary,
the flux cost is well-defined (to see this, we just have to remove the
inactive reactions and study the flux cost function in the remaining
subnetwork).  But as we cross the boundary, the flux cost function is
discontinuous\footnote{The enzymatic flux cost function is concave on
  each F-cone (and F-polytope), but  cannot be concave at the boundary between 
  F-cones (or F-polytopes).  For example, in the point $\vv=0$
  cost increases in all directions, so the function cannot be concave
  in a region containing this point.}  (see Figure
\ref{fig:branch_point_flux_reversal}): there is a jump in the enzyme
and metabolite concentrations, and therefore in the flux cost (for an
explanation, see SI Figure \ref{fig:branch_point_flux_reversal2}). The
boundary always belongs to the polytope with the lower cost\co{stimmt
  das?}. Since flux cost functions are concave within B-polytopes and
show jumps on the polytope boundaries, they favour sparse flux
profiles, in which many fluxes vanish. For instance, in Figure
\ref{fig:branch_point_flux_reversal} (e), the flux cost (black line)
decreases towards the polytope boundary from both sides, {\favour}ing
a flux $v_{2}=0$. This effect resembles an L1-regularisation and
justifies the sum of absolute fluxes as an approximative cost function
(a function with the same property, showing kinks where fluxes are
zero).

\begin{figure*}[t!]
  \mybox{ \textbf{Box 3: Screening of metabolic states to obtain states with  optimal {\flow}s}\\

\small

\textbf{Feasible {\flow}s} Metabolic states (characterised by enzyme
levels, metabolite concentrations, and fluxes) are feasible if they
satisfy all constraints defined in a model (e.g.~stationarity, rate
laws, positive concentrations, and cocentration bounds). The set
${\mathcal K}$ of feasible states  $(\vv, \cv, \esymbolv)$ can
be constructed systematically \cite{lieb:18theory}.  Each such 
state can be obtained by choosing a feasible {\fluxpattern}
(describing the flux directions), then choosing flux and metabolite
profiles consistent with this flux pattern, and finally computing the
required enzyme levels.  Given a {\fluxpattern}, feasible {\flow}s and
metabolite profiles can be found separately by linear programming. For
each combination $(\vv,\cv)$, the enzyme levels $\esymbol_{l}$ are
easily obtained from the rate laws. The difficult task is to find the
{\fluxpattern} in the first place, allowing for all constraints on
fluxes and metabolite concentrations to be respected later on!
Importantly, if we consider models with thermodynamically consistent
rate laws \cite{liuk:10}, the feasible choices of $\vv$ and $\cv$
depend only on thermodynamics, and rate laws and the enzyme levels do
not play role at this point!  We can find flux templates, fluxes and
metabolite levels exclusively based on constraints on fluxes and
metabolite levels, on the thermodynamic relationships between them.

\parbox{7cm}{\includegraphics[width=6.5cm]{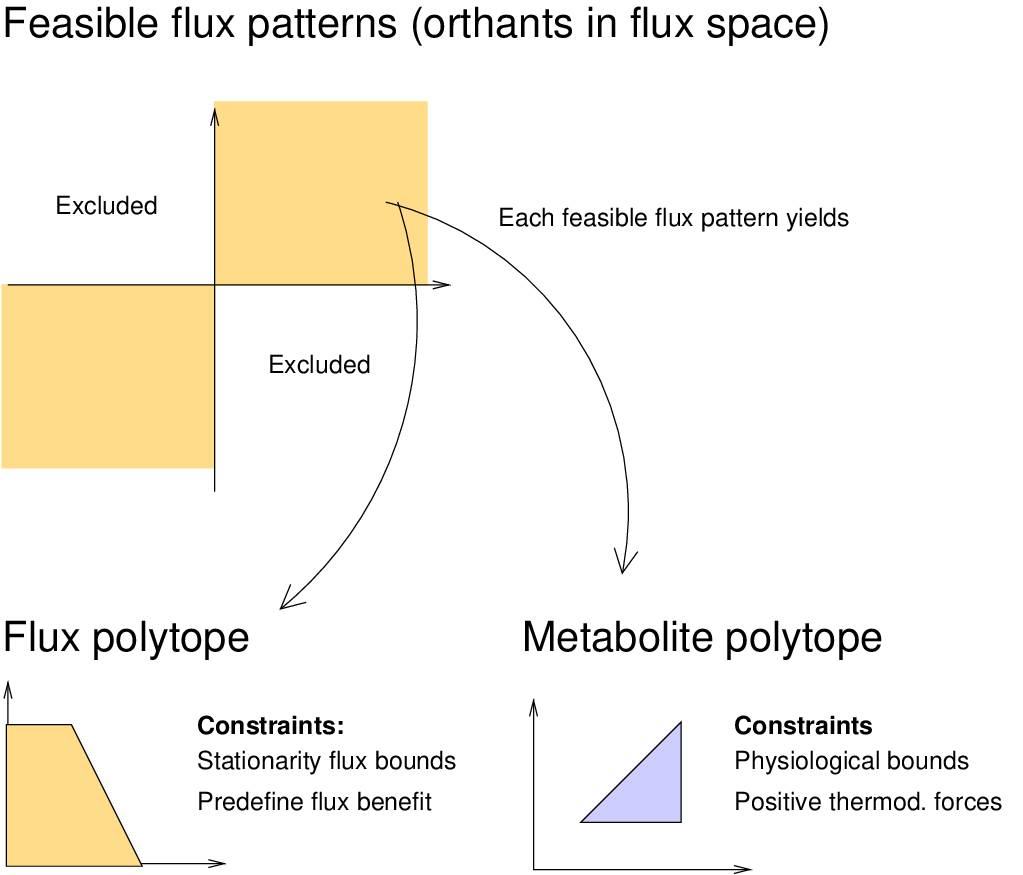}\\[7mm]}\hspace{1mm}
\parbox{8.5cm}{\textbf{Screening states of a kinetic model}
  Theoretically, metabolic states can be screened as shown on the
  left. First, we enumerate all feasible {\fluxpattern}s (i.e.~flux
  signs vectors of).  Flux patterns may be infeasible for
  thermodynamic reasons or because they do not allow for a steady
  state (i.e.~the corresponding orthants in flux space are outisde the
  hyperplane of stationary {\flow}s).  For each feasible
  {\fluxpattern}, we construct the corresponding flux and metabolite
  polytopes, and then screen the two polytopes to obtain all possible
  flux profiles $\vv$ and metabolite profiles $\cv$.  Each combination
  ($\vv,\cv$) defines a feasible state, and the corresponding enzyme
  profile $\enzymev$ is easy to compute.  We obtain all possible
  steady states $(\vv, \cv,\enzymev)$. To discard unstable states,
  each state must be checked by inspecting its Jacobian matrix.  The
  screening allows us to parametrise all metabolic states of a kinetic
  model by choices of $\vv$ and $\cv$. It shows that the collection of
  feasible flux polytopes represents all feasible steady states of the
  model, projected to flux space.\\}

\co{use m or ln c? uea einheitlich!}

\textbf{Computing the (minimal) enzymatic cost of a {\flow}} Our
screening can be used to find optimal states,
e.g.~states with a maximal flux benefit per enzyme cost. Metabolic
states $(\vv,\cv,\esymbolv)$ are scored by flux benefit
$\fluxbene(\vv)$ (a linear function of fluxes), metabolite cost
$\metcost(\lncv)$ (a convex function of logarithmic metabolite
levels), and enzyme cost $h(\enzymev)$ (a linear function of enzyme
levels).  We already saw that enzyme and metabolite costs can be
written as an effective flux cost
$\akin(\vv) = h(\enzymev(\vv)) + \metcost(\lncv(\vv))$.  To find an
optimal flux profile $\vv$ and its associated metabolite profile
$\cv$, we define a {\fluxpattern} and a flux benefit and minimise
enzyme cost across all possible states $(\vv,\cv)$. We can do this in two ways.
First,  we can screen the {\flow}s (scaled to a fixed
benefit), find the optimal metabolite profile by  ECM, and choose the {\flow} with the lowest
metabolite-optimised cost. Second, instead,  we can screen the metabolite profiles,
optimise the {\flow} for each of them (with a fixed benefit -- which
is a {\mwfFBA} problem), and choose the metabolite
profile with the lowest flux-optimised cost.\\

\parbox{7cm}{\includegraphics[width=6.5cm]{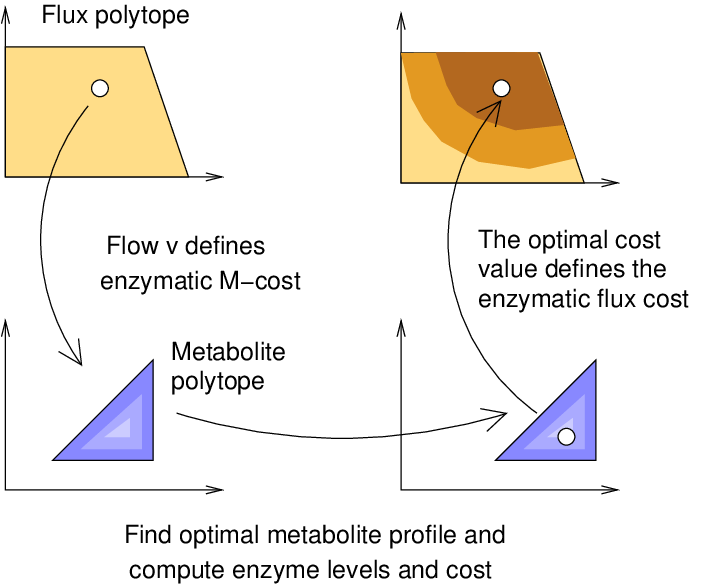}}\hspace{1mm}
\parbox{8.5cm}{ \textbf{Flux cost minimisation with a predefined flux
    pattern} To compute the enzymatic or kinetic cost of a {\flow}, we
  need to find the metabolite profile that realises these fluxes at a
  minimal enzymatic or kinetic cost.
  The M-polytope is defined by flux pattern, equilibrium constants,
  concentration ranges, and external metabolite concentrations.  Given
  our {\flow} $\vv$, each metabolite profile $\lncv$ on the M-polytope
  defines an enzyme profile $\enzymev$ (via the enzyme demand
  function) and therefore a cost.  We minimise the cost
  with respect to $\lncv$ and assign the resulting minimal cost to the
  {\flow} as a flux cost function $\acostenz(\vv)$ or a kinetic flux
  cost $\acostkin(\vv)$.\\

}

}
\end{figure*}

\section{Flux cost minimisation and flux benefit maximisation}

\subsection{Flux cost minimisation as a nonlinear FBA with minimal fluxes}

\myparagraph{\ \\Predicting metabolic states by flux cost
  minimisation} If a cell is able to produce biomass (or some other
important product) at a low enzyme demand, then protein can be
reallocated to other cell functions (e.g.~ribosomal proteins). This
allows cells to grow faster or to perform extra functions while
maintaining its growth rate. Following this logic, cells should strive
for enzyme-efficient {\flow}s, i.e.~{\flow}s that provide a given
benefit at a minimal enzyme cost. In Enzymatic Flux Cost Minimisation
\cite{wnfb:18}, such {\flow}s are computed by minimising the enzymatic
flux cost $\acost^{\rm enz}(\vv)$ at a fixed flux benefit
$\fluxbene(\vv)=\fluxbene$\co{prime?} and with given
(thermodynamically feasible) flux directions. Flux cost minimisation
is a ``layered'' procedure, with an outer optimisation in flux space
and, for each {\flow}, an inner optimisation in metabolite space.
Aside from cost functions $\aenz(\vv)$ and $\akin(\vv)$ derived from
kinetic models, we may use cost functions defined ad hoc; in any case,
flux costs must increase with the fluxes\footnote{If a flux cost
  function is also concave on each flux cone (whihc holds for the
  functions considered here), it must have a kink in the point
  $\vv=0$.}  (i.e.~$a_{v_l}\,v_l>0$ for all $v_l \ne 0$).  This
approach works for any plausible flux cost functions, and we obtain a
generalised, nonlinear version of minimal-flux FBA: we define a flux
benefit $\fluxbene(\vv)=b'$ and minimise a general flux cost
$\acost(\vv)$ on the resulting B-polytope. If the flux directions are
unknown, we may enumerate all feasible flux patterns (at least
theoretically), compute the optimal {\flow} for each of them, and
choose the best {\flow}.  \coout{Just like in FBA, we consider all
  feasible stationary {\flow}s with a given flux benefit and choose
  the one that minimises a flux cost.  The only difference is that our
  cost function is not a simple (weighted) sum of absolute fluxes, but
  a general nonlinear cost function $\acost(\vv)$.}

\myparagraph{Flux scaling and cost minimisation on the standard
  B-polytope} \co{refer back to benefit scaling on page 12} \co{MERGE
  IN \textbf{Scaling of optimal solutions} 

  In FCM, if a flux cost $a(v)$ and flux benefit $b(v)$ both scale
  proportionally (or with the same power) with an overall scaling of
  the flux profile $\vv$, the problem of minimising the ratio
  $a(v) / b(v)$ is ill-defined. In this case, the optimal flux problem
  can be arbitrarily scaled because $a(v)/b(v)$ remains constant
  despite the changing $\sigma$.  Thus, for an interior optimum, we
  cannot allow such as scaling. Similar problem arise if we minimise
  $\alpha \,a(\vv) - \beta \,b(\vv)$. But typically, $a(\vv)$ and
  $b(\vv)$ DO scale proportionally with $\vv$!  In this case, we need
  to restrict ourselves to one scaling (either fixing $a$ or $b$ at a
  predefined value, and optimising the other one)! Herem first, we
  consider fixing b and optimising a.}  §§§ FCM with enzymatic flux
cost functions has some convenient mathematical properties.  If the
flux directions are given and no flux bounds are imposed, scaling the
flux profile as a whole leads to a proportional scaling of flux cost
and flux benefit.  The cost/benefit ratio yields the scaled flux cost
$\hat{\fluxcost}_{(\fluxbene)}(\vv) = \fluxcost(\vv)/\fluxbene(\vv)$,
which is independent of a scaling of $\vv$ and corresponds to the flux
cost $\fluxcost_{(\fluxbene=1)}(\vv)$ on the standard B-polytope
(defined by $\fluxbene(\vv)=1$). In the absence of flux bounds, we can
do all our calculations on this B-polytope and compute flux costs on
other B-polytopes by simple scaling (see Figure
\ref{fig:properties}). Within our B-polytope, we can directly compare
flux profiles \co{JA! uea statt distribution} (with different active
pathways) by their costs. Comparing flux modes at a fixed benefit also
works between separate models, e.g.~models of different pathways that
produce ATP. Also in this case, we can compare {\flow}s at a fixed
flux benefit\footnote{Alternatively, we may consider one reference
  flux (e.g.~the flux in th glucose uptake reaction) and scale our
  flux distributions by this reference flux.} (or at a unit benefit,
that is, by their cost \emph{per benefit}).  \co{This comparison by
  scaled fluxes works for any quantities that scale proportionally
  with a flux scaling (or are homogeneous with degree 1).}  Aside from
comparing alternative pathways, \co{einmal klar sagen: indem wir uns
  auf ein B-polytop beziehen, faellt die moeglichkeit der linearen
  skalierung weg, und uebrig bleibt (oft) eine strukt konkave
  kostenfunktion:} there is another reason to run FCM on a B-polytope
(and not on the F-cone): on the F-cone, an enzymatic\co{FN: remember
  that we assumed linear enzyme cost functions! for convex cost
  functions of the form $h'(\sum_{l}\hul' \esymbol_{l})$ with strictly
  concave $h'$, this is different!} flux cost function will not be
stricly concave (because flux cost scales proportionally with the
{\flow}), whereas on the B-polytope it can be strictly concave.

\co{\textbf{PROPOSITION, IMPORTANT (WO?)} say: an optimum must be a polytope vertex except in the following rare case: there are two different flux modes with the same benefit (in the same B-polytope), the same favoured metabolite profiles (that is, the flux cost function is linear between them; the ECM problem is singular), and the same cost (i.e. the interpolation line is flat); in addition, for the case to matter, these flux modes must also be among the best local minima (otherwise, they would not matter); such FCM problems are called singular [or rather ``indifferent''] (others are called regular).

  Allg say, Optimal states are vertex points except for singular FCM problems; where singular is clearly defined, and have probability zero except for very symmetric, artificially constructed toy models.

also say this in abstract, intro and discussion!}

\co{formal conditions for the existence of indifferent optima:
  $\exists v_{A}\neq v_{B}$ with $b(v_{A})=b(v_{B})$ and $m^{\rm opt}(v_{A})=m^{\rm opt}(v_{B})$ (that is, flux cost is not strictly convex)
  they must further satisfy $a(v_{A})=a(v_{B})$; and in the FCM problem (global optimum!) they must actually be globally optimal;

  hence, ``red and blue enzymes'' are almost the only way to satisfy these conditions!
}

\subsection{Cost-optimal {\flow}s are polytope vertices}

\co{von daans resultat mit EFM-kombis ausgehen, hier erklaeren, dass das einfach heißt, dass die ecken so sind + hier zeigen warum (bzw unter welchen bedingungen) (nur) ecken optimal sind}

\myparagraph{\ \\Optimal metabolic {\flow}s must be flux polytope
  vertices} In FCM, we impose a flux pattern \co{without loss of
  generality: positive fluxes} and a flux benefit $b$ and search for
{\flow}s with this benefit and a minimal cost. Thus, mathematically,
we minimise a flux cost function $\acost(\vv)$ on a B-polytope.
\todo{\co{In models without any further flux bounds, the minimum cost
    is achieved on a polytope vertex. We can see this as follows. HIER
    UND AUCH SONST! concave: a concave function has no isolated
    interior minimum!  (inside cone, polytope or face): that is, a
    minimum point is either a polytope vertex on lies on an
    indifferent line or (hyper)plane (that ends in vertices) Since
    cost and benefit scale proportionally with the {\flow}, we can
    assume a unit benefit $b=1$, consider the enzyme costs for this
    benefit, and regard them as enzyme costs \emph{per flux benefit}.}
  Since the cost function is concave on the polytope, at least one
  minimum point must be a polytope vertex and at least one polytope
  vertex must be a minimum point (proof in section
  \ref{sec:SIpolytopeVertices}). If our cost function is strictly
  concave, then all minimum points must be polytope
  vertices. \co{with the assumptions made, the vertices corrrespond to
    EFMs, where typically some reaction fluxes are zero} This confirms
  what we noted before: flux profiles with a minimal enzyme cost
  {\favour}s tend to be sparse.}

\myparagraph{Elementary flux modes} \todo{We can further relate this
  to Elementary Flux Modes (EFM). Originally, EFMs were introduced to
  describe meaningful minimal routes in metabolic networks
  \cite{scdf:99, scfd:00}. It was later shown \cite{wpht:14,murs:14}
  that they also play a role as enzyme-optimal {\flow}s in two types
  of metabolic optimality problems\footnote{For the proofs, we first
    note that an optimal metabolic state defines a particular
    metabolite profile.  Given this profile, the enzyme cost is a
    linear function of the fluxes. If we optimise the flux profiles
    with respect to this cost (by solving a weighted flux-minimisation
    FBA problem), the solution must be a flux polytope
    vertex. \co{WD?:} In models without flux constraints, polytope
    vertices are elementary flux modes.} , flux maximisation at a
  limited total enzyme level; maximisation of enzyme-specific
  flux. \co{explanation: oben mention ``cone'', intersected by B-plane
    (und hier verwenden)} \co{Predefine flux directions and require
    stationarity, this will define a feasible flux cone whose edges
    are given by elementary flux modes (EFMs). If we further scale all
    {\flow}s to a predefined benefit, we obtain a B-polytope whose
    vertices are scaled versions of these EFMs.}  \co{wo?  flux
    profile, bzw nach normierung oder in b-polytop ``flux mode''}
  Hence, EFMs are not only a theoretical tool, but have biological
  relevance: there is always an EFM that provides an equal or higher
  enzyme efficiency than any non-elementary {\flow}. Here we confirmed
  this result, but for polytope vertices (instead of EFMs). We saw
  that optimal {\flow}s -- aside from exceptional cases -- are
  polytope vertices, and in models without flux constraints, these
  polytope vertices are elementary modes.  In models with flux bounds,
  these extra constraints may cut off polytope vertices (see Figure
  \ref{fig:new_vertices_3d}) and create new, non-elementary vertices
  that are candidates for optimal {\flow}s. We come back to this point
  below.  }

\subsection{Calculation of optimal flux profiles} 

\myparagraph{\ \\Locally optimal {\flow}s} A {\flow} is optimal if it
is the one (or one) with the lowest cost on the flux polytope (for a
global optimum) or within a small region around the profile (for a
local optimum). We can check local optimality by applying the KKT
conditions\co{schon erklaert?}. \co{FALSCH? ODER FALSCH ERKLAERT?
  Geometrically, if a B-polytope vertex is locally optimal, then any
  small movement towards the interior of the B-polytope must increase
  the cost (or leave it unchanged), so the cost gradient
  $\nabla \aenz(\vv)$, restricted to the subspace of stationary
  fluxes, must point towards the interior of the B-polytope (or must
  vanish\co{or be orthogonal to the B-polytope!}).} But there is also
a simpler criterion (``Manu'a criterion''): \co{ta'u-criterion aus
  biconvex herleiten?  was ist mit notwendig vs hinreichend?}
\co{call this ``mutually favoured flux + metabolite profiles''?}  A
pair $(\vv,\cv)$of a flux profile $\vv$ and a metabolite profile $\cv$
are locally optimal if (and only if) $\vv$ and $\cv$ {\favour} each
other, i.e.~$\vv$ is locally optimal given $\cv$ and $\cv$ is locally
optimal given $\vv$ (Proposition \ref{prop:segaulaCriterion} in SI,
proof in SI section \ref{sec:ProofCriterionLocallyOptimalFlow}). This
criterion can be used to test if a {\flow} $\vv$ is (locally) optimal:
we first compute its {\favoured} metabolite profile $\cv$ (by ECM);
then, we compute the {\favoured} {\flow} of $\cv$ (by {\mwfFBA}) and
check whether we reobtain our original {\flow}.  A variant of this
algorithm can be used to generate locally optimal {\flow}s: starting
from an initial, non-optimal {\flow}, we iteratively compute optimal
metabolite profiles and optimal {\flow}s until convergence (``Manu'a
algorithm'', see SI section \ref{sec:SILocallyOptimalFlows}). The
algorithm converges because the cost decreases (or remains constant)
in every step (due to biconvexity) and because there is only a finite
set of potentially optimal {\flow}s (the vertex points).
\todo{\co{WO?}In models without any flux bounds (except for a given
  flux benefit), optimal {\flow}s must be EFMs (or at least one of
  them must be an EFM). Moreover, if the flux cost function is
  strictly concave, then \emph{any} locally optimal {\flow} must be an
  EFM.  In theory, finding a global optimum is straightforward: we
  enumerate all EFMs (with any feasible flux directions) and choose
  the one with the lowest cost \cite{wnfb:18}.} \co{wo? discuss
  lineality spaces?}  \todo{In practice, the number of EFM may be
  large and enumerating them may be impossible.}

\myparagraph{Computing optimal states} How can optimal {\flow}s be
predicted in practice?  Metabolic states can be optimised by a
screening of flux profiles (Box 3). A given flux template and a
predefined numerical flux benefit defines a B-polytope; with a concave
flux cost function, we can find the optimal {\flow} by enumerating all
polytope vertices and choosing the one with the lowest
cost\footnote{The optimisation of fluxes and metabolite concentrations
  can also be done in the opposite order.  In an outer optimisation,
  we optimise metabolite profiles and for each of them, we optimise
  over fluxes \cite{lieb:18theory}.This inner optimisation is a simple
  FBA with a linear flux cost function, representing enzyme cost at
  the given metabolite profile. \co{JA! mention similarity to mueller
    approach [ref \cite{murs:14}]}}. Along with the optimal fluxes, we
obtain the optimal enzyme and metabolite profiles.  \co{To reduce the
  computational effort, we may use linear or nonlinear approximation
  of the cost function (see below). ref to wortel SI usw} \co{JA! FN:
  gibt es keine tricks um die EFM einzuschraenken?  kann man schon
  vorsortieren, welche garantiert schlecht sind?  projekt mit sabine
  schon andeuten} Enumerating and testing all polytope vertices may be
numerically expensive.  Instead of screening them all, we may directly
search for a local optimum, either by a greedy search over polytope
vertices (e.g.~using a simplex algorithm) or by the Manu'a algorithm
above.  However, to be sure to find the global optimum all polytope
vertices need to be screened.  By repeating this procedure for
different model parameters (e.g.~screening kinetic constants or
external metabolite concentrations), we can assess their effects on
optimal states. The resulting Monod curves and phase diagrams show
which metabolic pathways should be used depending on external
conditions and enzyme parameters \cite{wnfb:18}.

\myparagraph{Optimising the flux directions} Until here we assumed that
all flux directions were predefined. What if the flux directions
themselves are to be optimised? In this case, we  need to screen all possible B-polytopes, each
representing one possible pattern of flux directions.  But this can be
simplified.  \co{There are segments in flux space that do not contain a
B-polytope, either because their sign patterns are thermodynamically
infeasible or because they do not allow for stationary {\flow}s (see
the example in Figure \ref{fig:branch_point_flux_reversal}).  But
there are also adjacent B-polytopes that share some EFMs as
vertices.} To find the optimal fluxes, we can  forget about
B-polytopes and simply enumerate all thermodynamically feasible EFMs
as well as all feasible non-EFM vertices  (in models with flux
bounds), and evaluate the flux cost for all these flux profiles. Once
 the optimal flux profile is known (for the predefined external
metabolite concentrations), we also know the best {\fluxpattern} and
the globally optimal state of our kinetic model.  \co{Since our
optimisation procedure concerns flux, metabolite and enzyme profiles,
it yields an optimisation over all metabolic steady states (see Box
3).}

\co{
\begin{figure*}[t!]
  \begin{center} 
\includegraphics[width=14.5cm]{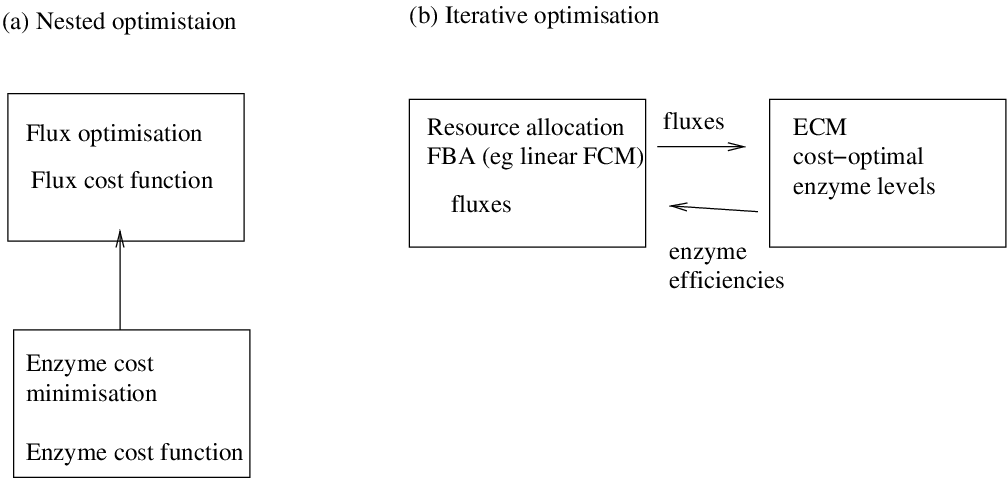}
\caption{\co{WRITE ME} \co{schoener! nested-> layered; typo in optimisation}}
\label{fig:fcmalgorithms}
\end{center}
\end{figure*}
}

\subsection{Flux  optimsiation with extra constraints can lead to non-elementary flux modes}

\co{Rewrite a lot, directly refer to Daan's work on combinations of
  EFMs} \co{REF Daan} \co{uebergang?}

\myparagraph{\ \\FCM with flux constraints can have non-EFM solutions}
In a basic version of FCM, fluxes must be stationarity and must
respect given flux directions, but no other flux bounds. The
constraints define a flux cone whose edges are elementary flux
modes. If we predefine the flux benefit, the cone is cut by a
hpyerplane and reduced to a polytope with corners given by vectors
along EFMs.  \co{FN: Why fixed benefit? remember: if flux benefit
  $b(v)$ and enzyme cost $h(e)$ are linear functions, the
  benfit-specific enzyme cost is independent of scaling, so, we can
  actually think of flux modes (instead of {\flow}s) and consider
  their \emph{benefit-specific enzymatic cost}} Hence, since FCM (with
concave flux cost functions) predicts optimal fluxes to be polytope
corners, and since polytope corners are EFMs, optimal flux modes must
be EFMs! \co{plural EFMs uea} But what if there are additional bounds
on individual fluxes or on linear combinations of fluxes (as proxies
for enzyme abundance)? As shown in Figure \ref{fig:new_vertices_3d},
such constraints can cut off parts of the polytope and create new
polytope vertices. The new non-elementary vertices are (somewhat
confusingly) called ``elementary flux vectors'' (EFV)
\cite{ksrg:17}. \co{WO? wie sollen solche fluesse genannt werden? EFV
  oder vertex {\flow}s?}  \co{cite the two papers that discuss
  elemantary flux vector: cite urbanczik und wagner \cite{urwa:05},
  elementary flux vectors, where additional constraints are mentioned?
  also new mueller paper that meike disliked?}  \co{FN: benefit
  constraint is used for scaling (yields only EFMs!). bei
  EXTRA-constraints can be mixtures of two EFMs, with 2 constraints
  mixtures of 3 EFMs and so on ; max zahl der EFMs in mischung: zahl
  ALLER constraints (scaling by benefit constraint included)} By
adding more flux constraints, we may obtain more vertices, each
representing a convex combination of EFMs (with a maximal number of
$n$ EFMs if $n$ is the number of constraints).  The convex set spanned
by these EFVs is still a polytope, but not one that is spanned by
EFMs. All its vertices are potential solutions of FCM problems.
\co{zusammenhang zu skalierung erklaeren; ohne flux constraints EFF =
  EFM; B-skalierung aendert nichts. mit flux constraints: EFV need not
  be EFMs; B-skalierung aendert form des B-polytops! (noch weiter
  unten: wie kann man basan-approach damit erklaeren?}

\co{uebergang?}

\myparagraph{Flux optimisation with constraints} \todo{We
  saw that the optimal flux profiles are B-polytope vertices. In models without other flux bounds, these
  vertices are elementary modes, but in models with flux bounds, non-elementary vertices
  can appear.} Let us see an example. In cell metabolism, ATP is typically
generated by glycolysis, whose product pyruvate is either exported (overflow
metabolism) or used to produce more ATP (via respiration). In our
model in Figure \ref{fig:definition}, the (high-yield) respiration
and (low-yield) overflow strategies are EFMs, suggesting that
one of them should be optimal  linear
combinations of them should be worse. In reality, some 
cells may ferment and respire simultaneously (e.g.~overflow metabolism in bacteria, the Crabtree effect
in yeast, and the Warburg effect in cancer cells).  Moreover, in
\emph{E.~coli} chemostat cultures the ratio between the two fluxes
(or the ratio of glucose uptake and acetate overflow) varies
gradually with the growth rate. This  cannot be explained by   a
switch between EFMs, but enatils at least a varying combination of
EFMs. \todo{So what is wrong with our model?

  In FBA or resource allocation models, non-EFMs may occur as polytope
  vertices due to flux constraints describing limited total protein
  levels, limited uptake rates, or limited membrane space
  \cite{szdg:17}. For example, non-EFM acetate overflow in
  \emph{E.~coli} has been predicted by assuming a bound on glucose
  uptake efficiency which limits respiration, whenever glucose uptake
  is high, through competition for protein resources \cite{bhoz:15}.}
\co{WO? In yeast, such a limitation is implemented by
  regulation. \co{whether or not it makes economically sense}}
\co{eigentlich: each pathway should be enzyme-efficient!  However, the
  pathways also interact: when glucose import is costly, glucose usage
  translates into transport costs, so other pathways mus also be
  glucose-efficient as well! das argument hier sehr klarmachen, an
  prominenter stelle! evtl unten in subsection on external
  concentrations} \co{wo? cite tetsuhiro's Giffen paper
  \cite{yaha:2019}} A similar possible  constraint is the competition
for bacterial membrane space (``real estate'') between glucose
transporters and the respiratory chain in bacteria \cite{zhvm:11}. In
FCM, similar constraints can be applied  to predict non-elementary
flux modes.

\todo{ \co{OK, ab jetzt: was koennen wir alles machen mit FCM und flux
    constraints?  wie koennen wir respiration/fermentation rtade-off
    beschreiben?}  \co{possible flux constraints to be considered: o
    flux constraints: put upper limit on EVRY FLUX; o constraints on
    sums: respiration + glucose transport in membrane (actually
    constraint on enzymes) o constraints on flux ratios??  (linear in
    flux space!)}  \co{mention ``membrane real-estate''
    \cite{szdg:17}} The existence of non-elementary polytope vertices
  may explain observed flux distributions like in the Crabtree
  effect. \todo{\co{WO?}  Respiro-fermentation (instead of
    fermentation alone) can be predicted by putting a bound on the
    respiration flux as shown in Figure \ref{fig:definition}
    (e.g.~representing space limitations for respiratory chain
    complexes in mitochondrial membranes).}  \coout{erstmal weg:
    However, we need to be cautious because other flux bounds may lead
    to the same vertex.}  }


\begin{figure*}[t!]
  \includegraphics[width=\textwidth]{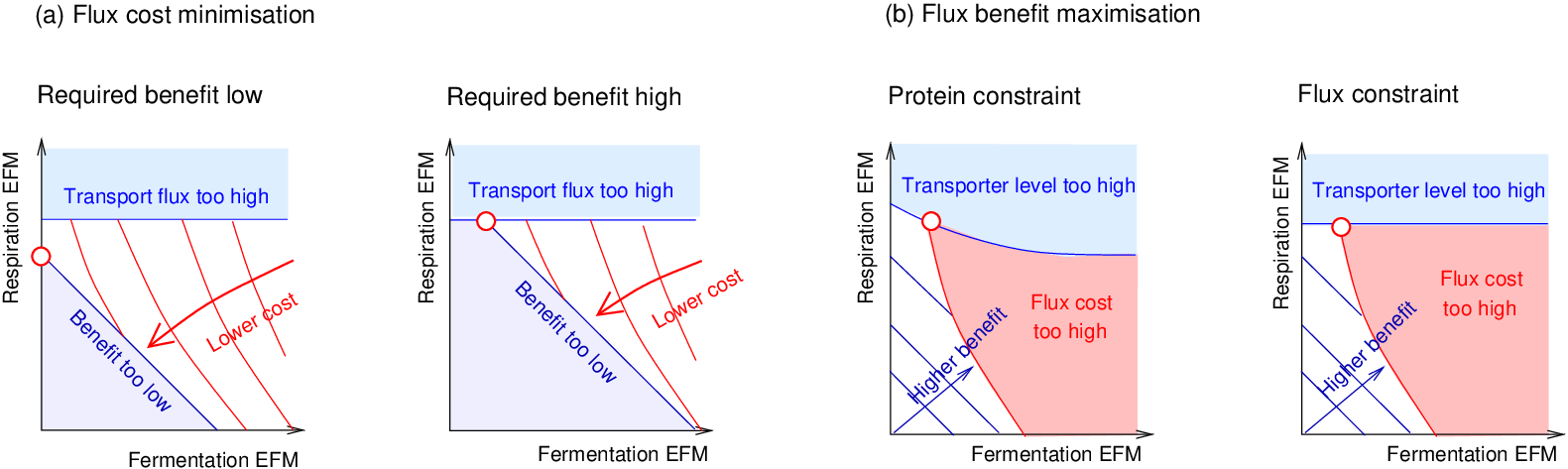}
  \begin{center}
    \caption{\co{MOVE FIGURE TO SI!!} \co{region grau (oder doch lila
        und gelborange, wie woanders?) // mehr linien - farbgradient??
        // replace the schematic drawings by a numerical example}
      Bounds on fluxes and enzyme levels can lead to non-elementary
      flux solutions. (a) Flux cost minimisation with a lower bound on
      flux benefit (schematic example).  The axes correspond to fluxes
      in EFM ``respiration'' or EFM ``fermentation'' and are scaled to
      units of flux benefit (e.g.~ATP production rate), the benefit
      constraint line is a decreasing 45 degrees diagonal. Left: The
      optimum point (red circle) represents pure respiration. Right:
      The same problem with a tighter bound on the benefit.  The new
      optimal {\flow} (red circle) is a combination of the two EFMs,
      describing respiro-fermentation (compare Figure
      \ref{fig:definition}). (b) Flux benefit maximisation (FBM).
      Left: Protein constraint (in this case, an upper bound on the
      glucose transporter). The bound is curved and concave (as a
      contour line of a concave (protein cost) function). Like in (b),
      the optimum point can be a vertex of the feasible region.
      Right: A linear flux constraint approximates the protein
      constraint: a bound on the glucose transport flux (mimicking the
      bound on the transporter level) is easy to handle
      computationally.}
  \label{fig:fluxBenefitMaximisation}
  \end{center}
\end{figure*}

\myparagraph{Is it useful to assume that ``optimal {\flow}s must be EFMs''?} \co{rather for discussion?} The
idea that optimal flows must be EFMs has led to advances in metabolic modelling.
However, how useful is it in practice? \co{nochmal kurz geom
  FBA diskutieren und EFMs gegenueberstellen?}
 \co{ITEM Geometric FBA may be practical} \co{AUCH in CBA II: Mention
    geometric FBA .. discuss epistemological pros and cons of taking
    combinations (knowledge vs actual combinations)}
\begin{enumerate}[leftmargin=5mm]
\item \todo{We saw that optimal solutions, as polytope vertices, need
    not be EFMs but can also be \emph{combinations} of EFM.} First, it
  depends on model assumptions, most notably on the absence of
  additional flux constraints. In respiro-fermentation, occurring in
  the Crabtree effect in yeast and in the Warburg effect in cancer
  cells, overflow metabolism \co{OFT statt fermentation?} takes place
  on top of (and not instead of) respiration. If a model describes
  overflow and respiration as EFMs (like the model in Figure
  \todo{1}), a non-elementary respiro-fermentation mode can still be
  explained by a bound on the respiration flux.  \co{refer to membrane
    limitation argument \cite{zhvm:11} (glucose
    transporter/respiration; which is basically of the type
    ``molecular crowding'')} \co{kurz an yield + transporter cost
    erinnern}
  \item \todo{May not hold in models with non-enzymatic reactions}
  \item Second, the claim that ``optimal flux profiles must be EFMs''
    may raise some epistemological problems. The reason is that EFMs
    are defined only within a model, with reference to a metabolic
    network structure and a choice of external metabolites. Howeverm
    the ``complete metabolic network'' of cells (which would contain
    all non-enzymatic reactions or unknown side reactions of enzymes)
    is hardly known and not even well-defined.  Also, what counts as
    ``external metabolites'' is a matter of choice: cofactors can be
    treated as strongly buffered (and therefore external), or as
    variable (and therefore ``internal'').  All these choices make the
    set of EFMs model-dependent, and so our statement that ``optimal
    {\flow}s must be EFMs'' is in fact not a statement about cells,
    but about cell models.
  \item Finally, the cost difference between an optimal elementary
    metabolic state and some other non-optimal states may be so small
    that this latter non-EFM solution may be realised by cells either
    because of other advantages, or because they are simply not costly
    enough to be suppressed.
\end{enumerate}
\co{JA! But: note that non-vertex flows require parameter fine-tuning!
  may say optimum points are FAPP EFMs?}  \co{how easily fluxes can be
  determined by regulation switching fluxes off may be easier than
  fine-tuning them.}

\subsection{Flux benefit maximisation}

\myparagraph{\ \\Different formulations of the flux cost-benefit
  problem} \co{Finally, minimising flux cost at a fixed flux benefit
  is equivalent to some other optimality problems: maximising the flux
  benefit at a given flux cost (FBM, see below); minimising the
  \co{woerter unten genauso!}  scaled flux cost
  $\hat{\fluxcost}_{(\fluxbene)}(\vv) =
  \fluxcost(\vv)/\fluxbene(\vv)$; maximising the benefit/cost
  efficiency
  $\hat{\fluxbene}_{(\fluxcost)}(\vv) =
  \fluxbene(\vv)/\fluxcost(\vv)/$ (and even, as we shall see below,
  maximising an approximation of the cellular growth rate). Please
  note that all this changes if we also consider flux constraints. We
  come back to this point below.}  \co{The compromise between flux
  cost and benefit can be described in different ways: by minimising
  cost at a given benefit (as in .. FBA with minimal fluxes),
  maximising benefit at a given cost (as in FBA with molecular
  crowding), maximising the benefit/cost ratio (leading to
  linear-fractional optimality problems) or pareto. (cite CBA opt!)}
\todo{If {\flow}s reflect a trade-off betwork (enzyme) cost and
  (production) benefit, how can we describe these trade-offs
  mathematically?  \co{One possibility is to keep one of them fixed
    and to optimise the other one.}  \co{FN: Using a linear
    combination has a big drawback: since each of the two functions
    scales linearly with the {\flow}, their linear combination would
    have an optimum at vanishing or infinite fluxes, or an indifferent
    optimum.} One possibility is to maximise the benefit/cost ratio}
\co{FN: linear-fractional programming:
  $\mbox{max}
  \stackrel{!}{=}\frac{\cv\trans\,\xv+\alpha}{\dv\trans\,\xv+\beta}\quad\mbox{s.t.}\,\Amat\,\xv\le\bv$. JA!
  in ``linearised flux cost functions''; und wo verschiedene
  problemformulierungen bepsrochen werden (kommt schon in auch CBA
  opt)} (or ``biomass/enzyme productivity'').  Like in a single
reaction, we may expect that {\flow}s should be enzyme-efficient,
showing a high biomass production per enzyme invested. As we will see,
higher metabolic efficiencies allow for a higher growth rates
\cite{sgmz:10}.  However, instead of minimising the ratio directly, we
may also minimise the cost at a fixed benefit or maximise the benefit
at a fixed cost (as in mcFBA).  \co{note correspondence (also in CBA
  theory + CBA lagrange: ``at least'': maximise! ``at most'':
  minimise!)}

\co{FN? \textbf{Objectives and constraints play similar roles and can
    be used interchangeably.}  Compromises between flux cost and flux
  benefit can be formulated in different ways: by minimising flux cost
  at a fixed (or bounded) benefit, maximising flux benefit at a fixed
  (or bounded) cost, minimising the cost/benefit ratio (which does not
  depend on the absolute scaling of {\flow}s), or by applying a
  multi-objective optimisation. In the different formulations, benefit
  and enzyme cost appear either as mathematical objectives or in a
  constraint.}  \co{enzyme amount (bounded) and cost geht
  durcheinander .. einmal gut erklaeren, was die idee ist. cost
  functions represent enzyme pools that need to be bounded in the
  model because of density constraints.} \co{NOTE THAT the two
  problems can be seen as special cases of a larger multi-objective
  optimisation of benefit and cost(s).}

\myparagraph{Flux benefit maximisation} FCM can be seen as a nonlinear
version of FBA with minimal fluxes, a version in which enzyme
efficiencies are not fixed, but are optimised along with the
fluxes. Similarly, we can imagine a nonlinear form of {\mcFBA}, in
which we maximise a flux benefit while constraining the flux cost.
This method is called Flux Benefit Maximisation (FBM).  \todo{Again,
  the flux cost can be a proxy for the amount of catalysing enzymes:
  using enzyme cost as a constraint (and not as an objective) allows
  us to implement density constraints for multiple cell compartments.}
\co{Direct optimality problem hinschreiben: max b(v) st a(v)=argminlnc
  q(ln c)}\co{again the difference to mol crowd FBA is that we use the
  real protein amount, not a simple linear proxy} \co{An example in a
  model of central metabolism would be a maximisation of ATP
  production (benefit) under a constraint on the sum of enzymes!
  (normal: total enzyme demand; other possible constraints: protein
  crowding on specific membranes (``real estate'', CITE; even between
  glucose transporter and respiration!))}  \co{To reconcile FCM with
  proteome partitional approaches} \co{klarmachen: relation between
  FCM and FBM, in the absence and presence of flux constraints)
  erwaehnen: sigma-trick, um FBM in FCM zu ueberfuehren}

\co{In some cell models, multiple enzyme constraints (e.g. density
  constraints in several compartments!) must be considered.  This
  requires that enzymes costs are used to define constraints FCM (and
  mfFBA) and not as cost terms (as in FBM or mcFBA)! [unless pareto is
  considered ..] FCM (and mfFBA) would be better suited to model
  SEVERAL BENEFITS (e.g. in a pareto setting?)}

\myparagraph{Flux constraints and enzyme constraints in flux space} In
FCM and FBM problems, flux constraints can represent enzyme
constraints. In a maximisation of flux benefit, bounds on fluxes or on
linear combinations of fluxes yield \todo{plane constraint surfaces}
that cut off parts of the polytope (see Figure
\ref{fig:fluxBenefitMaximisation} (b)). Enzyme constraints, instead,
would yield curved constraint surfaces (defined by non-concave enzyme
cost functions). \co{Hence, with bounds on the total enzyme cost or on
  weighted sums of enzyme levels (e.g.~from density constraints in
  cell compartments or on membranes) may make the remaining flux
  region non-convex. \co{Problem: different enzyme constraints would
    refer to different optimal metabolite profiles, so they are
    actually not consistent! say that this problem also underlies
    mcFBA!}  To avoid this, we may use linearised cost functions and
  constrain fluxes instead of enzyme levels.}  \co{coin a good word
  for satFBA logic} Just like constraints on flux cost, these
constraints can cut off vertices and give rise to non-elementary
vertices, leading to ``mixed strategies'' such as
respirofermentation. If we replace an enzymatic cost function
$\acostenz$ (used to define a bound in flux space) by a linear
approximation (see section \ref{sec:LinearFLuxCostFunctions}), we
obtain a linear flux constraint on the flux polytope as in
{\mcFBA}\footnote{\co{JA!  move to SI:} If flux constraints represent
  linearised enzyme constraints, and if constraints and cost terms in
  optimality problems can replace each other, why can't we simply
  replace all flux constraints by extra terms in our enzyme cost
  function?  This should yield a model without flux constraints: now
  all polytope vertices would be EFMs, apparently contradicting the
  fact hat flux constraints can lead to non-elementary
  solutions. However, there is a flaw in this argument: to mimic a
  flux constraint, i.e.~a ``steep wall'' in flux space, we would need
  a steeply increasing enzyme cost term, which implies a non-concave
  enzyme cost $\hminus(\esymbolv)$. However, in this case our proof
  for concave enzymatic flux cost functions does not apply (see
  proposition 1 in SI): with a non-concave flux cost function, the
  optimal {\flow}s may be non-vertex points, and thus
  non-EFMs. \co{JA!  similar for equality constraints (eg in mol
    crowding): a predefined sum of fluxes could be mimicked by an
    enzyme cost term $\sim (u_{\rm tot} - \sum_{l} \ul)^{2}$, which is
    convex and highly cooperative}}.  \coout{WEG: In flux benefit
  maximisation, we can consider multiple constraints, e.g.~for fluxes
  or for limited protein densities in different compartments or on
  different cell membranes (see Figure
  \ref{fig:fluxBenefitMaximisation} (b)).  Again, by linearly
  approximating these enzyme fractions, we obtain effectively linear
  flux constraints.} \co{uebergang} Thus, FBM with linearised protein
constraints effectivly resembles {\mcFBA}, \co{LIEBER: mcFBA} but with
kinetics-dependent transporter efficiencies like in satFBA. \co{schon
  erklaert?}  \co{To conclude, constraints on a sum of protein levels,
  \co{NOTE: met + enz dens constr. HUGO: kinetic cost!} for example on
  membranes, are a realistic assumption. With approximative linear
  flux-enzyme relations as described above, they can be effectively
  written as flux constraints, as in {\mcFBA}. By providing realistic
  enzyme efficiencies, FCM yields a justification for linear FBA
  methods and a way to compute realistic linear cost weights.}
\co{there is a constraint, leading to a Lagrange term
  $+ \sigma_l[ \esymbol_l(\lncv;\vv) - 1]$, again convex in metabolite
  space (where $\esymbol_l(\lncv;\vv) - 1$ also needs to be satisfied
  by the solution. Vincent fragen: WHAT DOES this mean for concave
  flux cost function?}

\myparagraph{Flux optimisation with predefined enzyme amounts}
\todo{On a fast timescale, cells cannot adjust their enzyme levels,
  but may still regulate their fluxes, for example, by enzyme
  phosphorylation.} Let us consider a variant of FBM that describes
this. \todo{The cell's strategy is to have enzyme overcapacities} that
can quickly be mobilised to adapt to changes in the
environment. \todo{The search for the right enzyme levels for such a
  preemptive expression can be formulated} as an optimality problem:
cells need to choose an enzyme profile that will enable them to
realise a number of predefined useful {\flow}s by inhibiting some of
the enzymes. Finding the cheapest enzyme profile that does this is a
convex optimality problem (see \cite{nfbd:16}, Supplementary
information). But once the enzyme profile s fixed (and the enzyme cost
spent), which fluxes (i.e. what enzyme inhibition pattern) should the
cell choose in a given environment? \todo{The problem resembles
  classical FBA (with bounds on individual fluxes, but now the bounds
  are on enzyme activities).}  \co{For example, if a cell expresses
  two enzymes that form a futile cycle, and if it can control their
  activities by inhibition, what enzyme activities would be optimal,
  depending on external metabolite concentrations?} Assuming that
enzymes can be inhibited, each enzyme activity can vary between 0 and
our predefined enzyme amount. If the aim in this is to maximise flux
benefit under these constraints, we obtain a variant of the FBM
problem \co{convex? concave?} (see SI
\ref{sec:SIFCMwithGivenEnzymeLevels}).  \co{explain that it is not
  equivalent to enz-kost min; different solution, not necessarily
  concave!}

\subsection{Linear flux cost functions} 
\label{sec:LinearFLuxCostFunctions}

\co{start by saying the linear flux cost functions are practical, and important in FBA. that we saw that realistic flux cost functions are not linear, and that we introduce linear flux cost functions here as linear approximations of realistic (non-linear) flux cost functions; that this shows what linear flux cost functions (eg in FBA) actually mean (aside from empirical ``kapp''), namely how they are related to kinetic constants and external concentrations; and what one loses in the linearisation.}

\myparagraph{\ \\Linear flux cost functions} \co{come back to enzyme
  efficiencies; mention some numbers (as rules of thumb) from EFM-FCM}
\co{catalytic rates statt enzyme efficiencies!}  In flux analysis,
enzyme efficiencies play a key role: they provide a linear conversion
formula between fluxes and enzyme levels, needed to define density
constraints (in {\mcFBA}) or linear flux cost functions (in
\todo{minimal-flux FBA}\co{UEA!}).  \co{FN: flux cost weights (flux
  prices?)  assumed to be hu/kapp; kapp is catalytic rate, between 0
  and kcat. kcat can vary widely, and the capacity utilisation
  (kapp/kcat) depends on metabolite concentrations and kinetics
  (saturation and thermodynamics, REF); for typical numbers, see
  wortel EFM paper} For a linear conversion, it is assumed that the
catalytic rates are fixed and given, as if the metabolite
concentrations were constant.  In FCM, in contrast, we acknowledge
that enzyme efficiencies can vary and are co-optimised with the
fluxes. The resulting models are more realistic, but the flux cost
functions become nonlinear, resulting from an optimality problem on
the M-polytope, and are harder to optimise than the linear cost
functions in FBA. This increases the numerical effort: in FCM, we need
to solve an optimality problem for each vertex of the B-polytope. Also
in flux sampling, e.g.~to study cell populations with
Boltzmann-distributed {\flow}s (see section \ref{sec:boltzmann}) we
need to assess many {\flow}s and compute their enzymatic costs.  In
this case, to speed up the calculations we may replace the enzymatic
flux cost by linear or quadratic approximations (see SI section
\ref{sec:linearAndNonlinearApproximations}).  To obtain a linear
approximation, we simply linearise our cost function around a
``prototype'' {\flow}. In practice, we just compute the optimal
catalytic rates $\ratelaw_{l}=v_{l}/\esymbol_{l}$ in this state and
use them for all other {\flow}s. The flux cost weights \co{UEA!}
$a_{v_{l}}= \hul/\ratelaw_{l}$, obtained from our prototype fluxes, can
be used in {\mfFBA} or with molecular crowding.  By deriving them from
kinetic models, we see what the cost weights actually mean and obtain
a justification for FBA.  \todo{By using several reference states
  instead of one, we obtain ranges \co{ref SI} or quadratic
  approximations \co{ref SI} of the flux cost.} Given the flux cost
and flux cost gradients for two {\flow}s, the cost function on the
interpolation line can be bounded by linear functions (see appendix
\ref{sec:appFluxCostFunctions}). Finally, we may choose several
{\flow}s in different regions of the flux polytope. Each of these
prototype profiles will come with a vector of enzyme efficiencies, and
by interpolating these enzyme efficiencies on the flux polytope, we
obtain a (non-convex) quadratic flux cost function.

\myparagraph{{\MfFBA} is smart} \co{discuss possible weighting of
  fluxes; relation to presumable enzyme cost} We saw that flux cost
weights for FBA can be derived from enzymatic flux cost functions
(e.g., by evaluating the optimal enzyme efficiencies for some
reference {\flow} by ECM). By doing so, we can link cost weights
directly to the parameters of underlying kinetic models, thus linking
FBA to enzyme kinetics.  Compared to a nonlinear FCM problem, FBA with
minimal fluxes is easy to solve. At the same time, it captures two
main features of FCM: since linear cost functions are concave, optimal
{\flow}s will be polytope vertices, just like in FCM! \co{also, in
  both cases, solutions need not be EFMs // non-elementary vertices ..
  unten diskutieren ..  das ist ja genau das verhalten von minimal
  flux FBA!}  \co{(except for cases in which two vertices are exactly
  equally good) - maybe call the other case ``singular'', allgmein? /
  verwechslung mit singular ECM problem? nee ist eigentlich gut!}. and
that, therefore, the flux solution jumps between discrete,
qualitatively different {\flow}s as model parameters are
changing. \co{WO? Comparison to regularisation with 1-norm (as in
  lasso regression)} \co{(compare Figure
  \ref{fig:parameterVariation})} \co{discuss kink; different cost
  weights in different segments; can be reformulated as LP problem;}
However, there is a downside: linear flux cost functions chosen
ad hoc miss some important details, e.g.~the fact that flux cost
functions are curved, the possible metabolite cost terms, and the way
flux costs wil vary with external metabolite concentrations.

\myparagraph{FCM explains how flux cost weights depend on enzyme
  parameters and external conditions} \co{MERGE IN}\co{this sounds
  similar to ``naive'' reasoning .. v = e kapp, where kapp < kcxat ..}
A main problem with linear flux cost functions is that the cost
weights are chosen ad hoc, \co{FN: In RBA, an empirical dependence on
  growth rate can be assumed; but this is only a rule of thumb, and
  only describes HOW catalytic rates are changing, and does not
  EXPLAIN why.} and that their dependence on enzyme kinetics,
metabolite concentrations, and environmental conditions remains
unclear. Methods such as {\mcFBA} \cite{bvem:07} use a flux burden
vector $\acostv=\hus/\kcat$. \todo{It \co{BUT actually hu/q(c):
    catalytic rates instead of kcat values!!  what metabolite
    concentrations?}  reflects $\kcat$ values} and enzyme cost weights
$\hul$ (e.g.~enzyme sizes), but ignore the effects of metabolite
concentrations. \co{REF TO {\mcFBA}; mFBA \cite{holz:04}} \co{Using
  ``capacity-based'' enzymatic cost scores in metabolite space
  \cite{nfbd:16}, \co{explain} we would obtain the same cost weights
  as previously proposed.}  \co{MERGE IN: ... which are otherwise only
  justified as lower bounds on the actual cost} This underestimates
the flux cost and hides its dependence on parameters such as external
metabolite concentrations.  In some FBA methods, such dependencies
have been considered for single reactions, e.g.~by assuming a
transporter kinetics $v(c) = u\,\frac{\kcat\,c}{\kM+c}$ and writing
the burden of the transport reaction as
$a_{v} = \hus \frac{u}{v(c)} = \frac{\hus}{\kcat}(1+\frac{\kM}{c})$ of
the external nutrient concentration. This flux cost function depends
on extracellular concentrations: the lower the concentration, the
higher the transporter's effective price. \co{REF satFBA} However, all
metabolite concentrations inside the cell are assumed to be constant,
which is unrealistic. This is the dilemma: a realistic expression for
flux cost weights will depend on all metabolite concentrations! An
external drop in nutrient concentration, will not only increase the
transporter demand (first-order effect); instead, also all other
enzyme leves are readjusted, resulting in a slightly higher enzyme
demand in many of the reactions.  This means that all flux cost
weights will change, and this change depends on all model details,
including possibly metabolite cost terms! If we linearise an enzymatic
(or kinetic) flux cost function, how can we predict the resulting flux
cost weights depending on all these factors?  \co{actually, on the
  enzyme efficiencies in a typical ``prototypical flux state''. Both
  factors can depend on many thing!}  FCM provides a solution: by
deriving flux cost weights from linearised enzymetic costs, it shows
how these weights depend on enzyme efficiencies and therefore on model
parameters and external metabolite concentrations.  \todo{These
  efficiencies for a ``prototype'' state. In using a linear flux cost
  function, we presume that the same efficiencies also hold for other
  flux profiles.} \co{What about metabolite costs? By directly
  linearising kinetic flux cost function in flux space, also
  metabolite costs can be naturally incorporated into flux cost
  functions! this shows that hergo's intuition is right; flux cost can
  represent other costs than enzyme.} The resulting flux cost weights
depend on all model details, including $K_{\rm M}$ values, the growth
medium and bounds on metabolite levels.  Any changes in these
parameters will change the flux cost weights.  As an example, consider
a respiring cell. If the oxygen level (a model parameter) decreases,
this alters the M-polytope and the cost function
$\enzymemetcost(\ln \cv)$ on this polytope, so also the enzymatic flux
cost function $\aenz(\vv)$ will change. A lower oxygen level leads to
a lower driving force in oxidative phosphorylation, which increases
the enzyme demand. \co{JA!  FN: This explains the ``redistribution''
  (see above in this paragraph and shows that ..)  This redistribution
  is a second-order effect. \co{REF to first-order approximation in
    wortel SI and brief explanation} refer to EFM paper and CBA opt
  (unchain rule) main effect is still local!} If we linearise
$\aenz(\vv)$ to define cost weights for FBA, the resulting cost
weights will depend on the oxygen level, and FCM describes this
dependence based on kinetic models.  Of course, this holds not only
for oxygen levels, but for any other model parameters.  In summary,
FCM assumes the same logic as satFBA (low substrate levels leads to
higher enzyme demands), but applies it to all reactions and assumes
(optimality-based) metabolite changes inside the model. The resulting
enzyme efficiencies (and flux cost weights) are more realistic (and
better justified theoretically) than the usage of $\kcat$ values or
empirical apparent $\kcat$ values.

\co{irgendwo kuez was ueber 1st order approximation of enzyme changes upon parameter changes? see wortel SI!}

\section{Protein demand and cell growth rate}

\subsection{Cell growth rate as a function in  flux space}

\myparagraph{\ \\Metabolic efficiency affects cell growth} In
metabolic models with a biomass reaction, the overall metabolic
efficiency can be described by biomass/enzyme productivity, \co{wort
  uea?}  that is, the biomass production rate divided by the total
\co{(mass or cost)? explain!} concentration of metabolic enzyme
\co{mass concentration? say this in several articles?}. Its reciprocal
value, the metabolic enzyme concentration per biomass production, is
also called biomass-specific enzyme demand\co{UEA!}. \co{diese woerter
  ueberall verwenden (auch andere artikel) 7 latex-abkuerzungen!}
Metabolic efficiency also has an effect on resource allocation: if the
growth conditions are good (e.g., good carbon
sources\footnote{\co{Here, a ``good'' carbon sources can be defined as
    one from which a given ATP or biomass output can be achieved at a
    low enzyme investment (as computed, for example, by FCM)} \co{note
    that in other contexts, a ``good carbon source'' could be one with
    a high yield!}} or high substrate levels) and allow for a given
metabolic production at a lower amount of enzyme, cells can shift of
protein investments from metabolic enzymes towards ribosomes, increase
their metabolic fluxes and translation rates simultaneously, and
therefore grow faster.  \co{hier schon scott/hwa erwaehnen, die diese
  logik klargemacht haben} The biomass-specific enzyme cost
$\hat\aenz$ can be converted into a cell growth rate $\lambda$ by a
cost-growth conversion function $\lambda(\hat\aenz)$. This function
allows us to describe cell growth rate as a function on the flux
polytope! A formula \todo{(with a monotonically decreasing function)} can be obtained from simple protein allocation
models \cite{sgmz:10}; if we use it, we conclude that FCM solutions
also maximise cell growth. Since $\lambda(\hat\fluxcost^{\rm enz})$ is
a decreasing function, maximal cell growth requires a minimal
biomass-specific enzyme demand, \co{FN: sagen, dass kinetic cost sich
  nicht fuer direkte scott-hwa methode eignet, aber hier doch
  behandelt werden kann!} which means that growth-maximising {\flow}s
can be found by FCM.\co{FN: kurz sagen, dass das anders ist als BM
  maixmisation in classical FBA (mention yield vs rate)} \co{ (naiv:
  formel umdrehen und enzymmenge festhalten; genauer:
  kosten/produktion in scott-hwa-idee einsetzen). // coin a good word
  for Scott/Hwa cost to growth rate conversion}

\co{
\begin{figure*}[t!]
  \begin{center} 
    \includegraphics[width=0.3\textwidth]{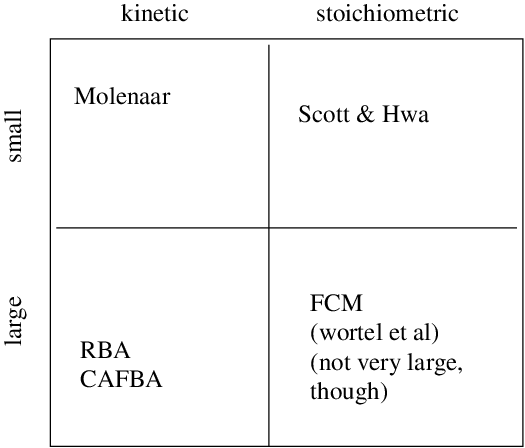}
    \caption{\co{WICHIG HIER? vielleicht in SI? JA!!!} \co{WRITE ME; 2 x 2 bild
        small/lare, kinetic/stoichiometric, with entries Molenaar,
        Scott/Hwa, RBA, ?? (maybe our EFM method!) // in legend: RBA
        is possible, but not kinetic and assumes KNOWN enzyme
        efficiencies; kinetic RBA would not be feasible.}}
    \label{fig:GrowingCellModels}
  \end{center}
\end{figure*}
}

\co{REMOVE THE FOLLOWING TWO PARAGRAPHS HERE and see where parts of
  them could fit in (discussion? other articles? FCM II und CBA opt?)}

\myparagraph{Models combining metabolism and cell growth} To make the
link between enzyme efficiencies and cell growth, let us have a look
at resource allocation models for cells.  In these models, metabolic
efficiency and choices of metabolic strategies play a main
role. \co{satz erst wenn abb gut ist!: In Figure
  \ref{fig:GrowingCellModels}, these types of models are classified
  into global (small) versus detailed (large), and kinetic versus
  stoichiometric.}  In whole-cell models, metabolism and protein
synthesis (or, generally, macromolecular processes) depend on each
other: metabolism provides precursors for protein synthesis, while
enzymes catalyse metabolic reactions. The coupled system needs to
produce all cell components (to duplicate the cell, or in other words,
to keep all compounds at constant concentrations despite their
dilution).  Finally, a density constraint (e.g.~an upper bound on the
total volume occupied by protein) makes proteins compete for space
and, indirectly, for other resources\footnote{In fact, this
  competition between metabolism and protein production resembles the
  competition between reactions or pathways described by FBM. To see
  this, we can consider a simple cell model that formally looks like a
  metabolic pathway model with three reactions (nutrient import;
  precursor production; and macromolecule production) whose fluxes
  must be balanced for a stead growth state). We assume that the
  reactions are catalysed by three types of catalysts (transporters,
  enzymes, and ribosomes), whose sum is our flux cost function. To
  maximise cell growth, we need to maximise the pathway flux at a
  bounded flux cost (total catalyst concentration). \co{A simpler
    version of this model (with two reactions, and assuming saturable
    enzymes) was considered in the seminal paper on proteome
    partitional models by Scott et al. // wieder aufgreifen in section
    on external concentrations? hier recht einfach: nur EIN
    flussmodus, also eher ein problem fuer parameterabhaengigkeit als
    fuer wahl von EFMs}}. In a simple resource allocation model, there
are just two types of proteins, metabolic enzymes and the translation
machinery, which share a limited protein budget. To maximise growth,
this budget needs to be optimally shared between the two functions,
finding a best compromise between efficient precursor and protein
production\co{explain!}. \todo{In FBA, cell growth is often associated
  with the biomass reaction rate, setting
  $\lambda=v_{\rm BM}/c_{\rm BM}$. While this is correct in principle, it ignores
  two important facts\co{and thus shifts the problem elsewhere, to
    having to know the bounds on influxes!}: first, the predicted
  biomass rate $v_{\rm BM}$ depends on the uptake fluxes (which are
  variable and generally unknown), and putting fixed bounds on the
  uptake rates will not give realistic results. This problems is
  partially solved by {\mcFBA}. Second, FBA ignores variable resource
  allocation between metabolism and protein synthesis (or any cell
  processes outside metabolism). This problem is solved by
  CAFBA. \co{AUSSCHREIBEN als 3. punkt: A third problem, the assumption of a fixed
    biomass composition in FBA, is partially solved by RBA}.  In FCM,
  like in FBA, we focus on metabolism and ignore protein
  synthesis. The resulting biomass/enzyme productivity can be related
  to cell growth in two steps \co{(CITE WORTEL)}: given a kinetic
  model and growth conditions (external compound concentrations), we
  can translate flux profiles into enzyme efficiencies, and compute an
  optimal biomass production per total metabolic enzyme. We then
  convert the biomass/enzyme productivity into a growth rate, either
  by applying an empirical formula (from growth-rate dependent
  proteomics data) or by using a simple resource allocation cell
  model. \co{HIER GUT? dann uebergang in ordnung bringen}} \co{WO?
  discuss RBA; hier bezug auf metabolisms only, other processes are
  hidden in effective fitness function (say this in abb 2 already?)}
\co{WO? introduce scott and hwa? this argument considers only
  metablism and ribosomes, but it can be extended!}  Since the 
metabolic enzyme demand depends on biomass/enzyme productivity (e.g.~biomass
production rate per total enzyme investment), the optimal resource
allocation and growth rate (in protein partitioning models or RBA) are
directly linked to the question of metabolic efficiency, i.e.~the
metabolic production per enzyme amount (which can be computed by FCM).
\co{text referring to basan approach figure, explaining how growth
  rate emerges from efficiency (give formula, refer to SI); how this
  relates to many EFMs (in figure (c))} \co{basan approach, with
  several sectors for EFMs} \co{kommt auch in FCM II: (CITE wortel!:)
  Scott-Hwa model equates the two fluxes (metabolic and protein
  synthesis); but this is equivalent to min-function
  $lambda = min lambda_i$ where
  $lambda_i = a_i * \phi_i (s.t. sum_i \phi_{i} < \phi_{{\rm tot}})$
  with proteome fractions $\phi_{i}$ and total proteome budget
  $\phi_{\rm tot}$}

\co{make it clear that here we make the direct connection between scott-hwa and kinetic models (which has not been considered by scott-hwa and would be hard to do in a single (``kinetic RBA'') model!)}

\co{\textbf{Translating metabolic efficiency into cell growth rates}}
\co{say that here, ENZYMATIC, (not kinetic) flux cost is relevant (FN
  about problems with kinetic cost in the context of scott-hwa
  approach)} \co{If we can express the cell growth rate as an
  objective function in flux space, we obtain an alternative to the
  classical FBA approach (in which biomass production is used as a
  proxy for cell growth) \co{(and runs into problems because biomass
    production, at a bounded glucose influx, does not describe growth
    rate, but yield on glucose!)}, one that actually considers an
  estimated cell growth rate.}  \co{mention that this conversion has
  already been done in wortel et al, and that the new thing here is to
  consider cell growth directly as a function in flux space.}
\co{Ziel: enzymkosten/fluss umrechnen in ``proteinkosten''; da aber
  die ``economically closed system'' assumption gemacht wird, die
  sache umdrehen, proteinmenge festsetzen und nach wachstumsrate
  aufloesen!}  \co{WO? ``if our metabolic model produces biomass as
  the end product, we can directly convert enzymatic flux cost into
  growth rate by ...}  \co{\co{WO?}  Thus, can we use kinetic
  metabolic models to compute enzyme efficiencies and convert them
  further into growth rates? In \cite{wnfb:18}, such a conversion was
  done using linear or Michaelis-Menten-like conversion functions: in
  practice, a metabolic objective (biomass production, divided by
  total enzyme) was used as a proxy for a whole-cell objective (the
  growth rate). The conversion formula was justified by a simple
  whole-cell model that describes how protein resource should be
  allocated between metabolism and protein production.}  \co{in a way,
  the enzyme demand (in metabolism) was extrapolated to a total
  protein demand (including ribosomes), which is directly related to
  growth rate // how was this done, and how is it justified?}

\myparagraph{From enzyme cost to cell growth} How can a cost-growth
formula $\lambda(\hat{\fluxcost}^{\rm enz})$ be derived? \co{WD MIT
  UNTEN?: In a simple approach, we assume that the cell has a fixed
  budget (mass concentration for metablic enzymes; with a
  biomass/enzyme productivity
  $\frac{v_{\rm BM}}{h_{\rm enz}}=k_{\rm BM}=1/\hat{a}_{\rm enz}$ and
  a biomass concentration $c_{\rm BM}$, the growth rate reads
  $\lambda=\frac{v_{\rm BM}}{c_{\rm BM}}=\frac{k_{\rm BM}}{c_{\rm
      BM}}=\frac{1}{c_{\rm BM}\,\hat{a}_{\rm enz}}$). However ,in
  realitym the enzyme budget is not fixed, but depends itself on the
  growth rate. We can model this by assuming a reallocation of protein
  resources between metabolic enzymes and ribosomes. \co{ref cafba,
    scott/hwa}} We assume a partitioning of protein resources between
enzymes and ribosomes, as used to explain bacterial growth laws
\cite{sgmz:10,wnfb:18}.  The original minimal proteome partitioning
cell model from \cite{sgmz:10} has been extended in various ways
\cite{bhoz:15,mhmm:16,murs:15}. \co{erst: list assumptions: 1. maximal
  growth rate; 2. depends on metaboic and protein production only;
  3. constant protein production efficiency; metabolic efficiency
  given by scaled flus cost function (for whole metabolic network) as
  described above.} \co{einmal klar sagen: circular dependence between
  (overall) efficiency and cell growth; a given enzyme amount allows
  for a certain biomass rate; a given growth rate requires a certain
  amount of ribosomes and thus allows for a certain enzyme amount. PUT
  This together into one formla, and solve for the growth rate}
\co{This quantity, called enzyme-specific biomass production
  $\ratelaw_{{\rm BM}}$, determines cell growth in simplified
  whole-cell models \cite{sgmz:10}.}  \co{REF and discuss caFBA + RBA}
\co{Thus, sector models can be used to translate FCM solutions (and
  enzyme demands) into growth rates} In the basic model, cell growth
is proportional to protein production, which consists of two steps,
$ \stackrel{1}{\rightarrow} \mbox{Precursors}
\stackrel{2}{\rightarrow} \mbox{Protein}$, catalysed by
metabolic enzymes (1) and ribosomes (2).  \co{FN: instead of bounds and
  optimisation, equalities may be used} We assume that protein
precursors are mass-balanced, reaction rates are proportional to the
catalysing proteins, and the  protein budget for enzymes and
ribosomes is fixed. By putting all this together, we obtain a formula that converts
\co{FN comment on possible formulation as an optimality
  problem: and a maximisation of protein production,} the
enzyme/biomass demand \co{uea} $\hat{\fluxcost}^{\rm enz}$  into
cell growth, 
\begin{eqnarray}
  \label{costGrowthConversion}
  \lambda(\hat{\fluxcost}^{\rm enz}) = \lambda^{\rm max}\, \frac{a_0}{a_0+\hat{\fluxcost}^{\rm enz}}.
\end{eqnarray}
\co{convex!}
The formula contains two parameters: a maximal growth rate
$\lambda^{\rm max}$ and an effective parameter $a_0$, denoting the
specific enzyme cost that leads to a half-maximal growth rate.  We
used the cost-growth function (\ref{costGrowthConversion}) to predict cell
growth rates in \cite{wnfb:18}. By defining the biomass/enzyme  productivity
$\ratelaw_{\rm BM}=1/\hat{\fluxcost}^{\rm enz}$, we obtain
the efficiency-growth function 
\begin{eqnarray}
  \label{efficiencyGrowthConversion}
  \lambda(\ratelaw_{\rm BM}) = \lambda^{\rm max}\, \frac{\ratelaw_{\rm BM}}{\ratelaw_{\rm BM} +k_{0}},
\end{eqnarray}
\co{concave!}
\co{FN: For the cell doubling time $T=\ln 2/\lambda$, Eq.~(\ref{efficiencyGrowthConversion}) yields the
  simple linear formula
  $T=\frac{\ln 2}{\lambda^{\rm max}}[1 + \frac{k_{0}}{\ratelaw_{\rm
      BM}}] = T_{\rm min}[1 + k_{0}\,\hat{\fluxcost}^{\rm enz}]$
  \co{REF Wortel}.}  where the parameter $k_0=1/a_{0}$ denotes the
biomass/enzyme productivity that would lead to half-maximal growth.
If $k_{\rm BM} \ll k_0$, as we found for
\emph{E.~coli}  \cite{wnfb:18}, Eq.~(\ref{efficiencyGrowthConversion}) can be
approximated by a simple proportionality
$\lambda \sim \ratelaw_{\rm BM}$.  We obtain the same linear
formula by postulating that metabolic enzymes occupy a fixed fraction
of the biomass, indpendent of the growth rate. Then we write
$\lambda = \frac{v_{\rm BM}}{c_{\rm BM}} = \frac{v_{\rm
    BM}}{\hat{\fluxcost}^{\rm enz}} \frac{\hat{\fluxcost}^{\rm
    enz}}{c_{\rm BM}}$: the first fraction is our biomass production
efficiency $\ratelaw_{\rm BM}$, and the second fraction (the protein
budget for metabolic enzymes, per biomass) is the (supposed) constant
biomass fraction.  \co{(see SI ..)} \co{Say how extra costs (eg GFP)
  can be included; maybe discuss my new idea of enforcing a RELATIVE
  proteome fraction instead of an expression level (which then
  explains interactive effects in protein expression!!)}

\myparagraph{Maximising cell growth by convex optimisation} Using the
enzyme/biomass demand $\hat{\aenz}(\vv)$ \co{symbol! lieber
  $\hat{a}^{\rm BM}_{\rm enz}(v)$? ist das gleich $k_{\rm BM}$?}
\co{ueberall an hat denken!}  and the conversion formula
(\ref{costGrowthConversion}), the growth rate can be written as a
function $\lambda(\vv)$ on the flux polytope. By maximising this
function, we obtain the {\flow} (and the corresponding metabolite and
enzyme profiles) that allows for maximal cell growth. \todo{Since the
  conversion function (\ref{costGrowthConversion}) is monotonically
  decreasing, we can simply obtain the same state by FCM.} Again, the
flux profiles that maximise growth are polytope vertices.

\co{FN : \co{To summarise: To predict the growth rate allowed by a
    given {\flow}, we define the external conditions (external
    metabolite concentrations), find the optimal enzyme allocation pattern (by
    ECM, with our flux profile), compute the enzyme investment per
    (biomass production) flux, and translate it into a maximal
    possible growth rate (possibly accounting for necessary resource
    rearrangements between metabolic enzymes and ribosomes).}  To
  predict growth-maximising strategies, we consider a metabolic model
  with biomass production as the flux benefit function, enumerate all
  F-polytope vertices, and compute for each of them the minimal enzyme
  investment per biomass rate. Using the conversion formula
  (\ref{efficiencyGrowthConversion}), we translate scaled enzymatic
  flux costs $\hat{\fluxcost}^{\rm enz}$ into growth rates
  $\lambda(\hat{\fluxcost}^{\rm enz})$, and to obtain the cell growth
  rate $\lambda(\acostenz(\vv))$ as a function on the B-polytope. If
  the function $\lambda(\hat{\fluxcost}^{\rm enz})$ is decreasing and
  convex, as in Eq.~(\ref{efficiencyGrowthConversion}), the function
  $\lambda(\vv)$ is convex as well (see SI section
  \ref{sec:SIProofGrowthRateFunction}), and if the flux cost function
  $\acostenz(\vv)$ is strictly concave, it is strictly convex.  Hence,
  growth-optimal metabolic states can be found by maximising a convex
  function on the flux polytope. Again, the optimal {\flow}s will be
  vertex points: a minimisation of relative enzyme cost and a
  maximisation of growth rate are two sides of one coin!}

\co{mention manua algo + biconvex in abstract!}

\co{\textbf{problem: fixed biomass function!} as pointed out in wortel
  et al ..}  \co{Limitation of this approach: Compared to RBA, this
  approach would still has the problem that it needs to assume a fixed
  biomass composition, in the biomass producing reaction} \co{mention
  elementary growth modes, in relation to my approach?}

\co{Vergleich mit ``wachstumsratenvorhersage'' in FBA // FBA mit flux
  minimisation // mit mol crowding .. als eventuelle NAEHERUNG DER
  wachstumsfunktion!!!}

\co{JA! note at the end of section: sagen, dass ECM klarmacht, wie
  thermodyn enzymebedarf beeinflusst; hier: verbindung zu
  zellwachstum! oliver ebenhöhs paper on thermodynamics of growth
  zitieren S(schon in CBA opt / FCM) // auch orkuns paper; }

\co{JA! SCHREIBEN SUBSTRATE-EFFICIENT INSTEAD OF FAST GROWTH
  \co{According to Eq.~(\ref{efficiencyGrowthConversion}) enzyme
    efficiency (biomass/enzyme productivity) is helpful fast
    growth. But it can also support other possible cell objectives. At
    a given growth rate, a high biomass/enzyme productivity means that
    protein resources are freed for other cell functions.}  \co{and
    also to the maximisation of other side objectives -- maximal
    production of extra substances or proteins, and even survival at
    high temperatures -- at given cell growth rate!}  \co{Relativ
    kurz: idee erwaehnen; abh von wachstumsrate= aeussere biomasse
    produktion; aber auch (in chemostat) aeussere konzentration)}
  \co{A cost-efficient metabolism saves resources, which can then be
    shifted to other cellular processes. Depending on the organism's
    ecological niche and concrete environmental conditions, various
    things can be done with these extre resources. one of them,
    frequently invoked and easy to study, is fast growth.}  \co{WO?
    make sure that fast growth is NOT presented as the main goal, just
    as an (easy-to-understand) example; enzyme effizienz erlaubt
    einfach, andere dinge zu tun, nicjt nur schnelles
    wachstusm. diskutiere ideologischen aspekt von schnellem wachstum!
    schnelles wachstum ist nur relevant insofern as (in manchen
    experimenten) einen vorteil vermitteln und sich deshalb manchmal
    durchgesetzt hat. beim modellieren praktisch: automatisch werden
    alle synthesewege gebraucht! aber das geht auch mit constraint auf
    wachstum! // also discuss that it may be ideologically preferred
    (ausfuerhlich in CBA labour)} }

\subsection{Cell populations described by probability distributions on
  the flux polytope}
\label{sec:boltzmann}

\co{boltzmann distribution has been proven to follow from population
  model by daniele de martino?}  \co{AdM lesen!}
  
\myparagraph{\ \\Statistical distribution of {\flow}s in a cell
  population} \co{gute abbildung?} In a cell population, cells will
show different flux profiles.  If we describe cells by points in flux
space, the population forms a point cloud, which may be described by a
probability distribution. If the probability density on the flux
polytope is known, we can infer the average {\flow} and the
distributions and correlations of different reaction fluxes.  To
obtain such a distribution, we can consider a simple model of cell
populations in which fast-growing cells also reproduce faster and are
more abundant in the population. We describe each cell by a point on
the B-polytope, that is, a flux profile \co{uea statt distribution?}
$\vv$ with growth rate $\lambda(\vv)$.  Inspired by statistical
physics, we define a simple probability density function: we define a
Boltzmann distribution\footnote{In the context of FBA \cite{dabg:17},
  a similar Boltzmann distribution has been defined using the negative
  biomass production rate, a linear function on the flux polytope.}
with the inverse growth rate as a ``negative energy'' function and
obtain the probability density

\begin{eqnarray}
  \label{eq:boltzmann}
  p(\vv) \sim \e^{(\lambda(\vv)-\lambda_{d})/\xi},
\end{eqnarray}
where $\lambda$ is the cell growth rate and $\xi$ describes the
strictness of selection and $\lambda_{d}$ is chosen to normalise $p$
to a total probability of 1.  The Boltzmann formula with flux cost as
an energy function can derived from a simple population model\co{refs
  to d de martino usw} (SI \ref{sec:SIPPopProbability}). In the model,
cells grow for a fixed time interval $\tau=1/\xi$ with its growth rate
$\lambda(\vv)$, then each cell switches randomly to a new flux
distribution (drawn from a uniform distribution on the flux polytope)
or dies with a death rate $\lambda_{d}$ (identical for all cells and
adjusted to maintain a constant population size).  This stochastic
process converges to an invariant distribution, the Boltzmann
distribution (\ref{eq:boltzmann}). \co{proof in SI? wo?}
\co{irrelevant?: boltzmann density is convex and monotonically
  increasing in $\lambda$} \co{FN: A similar population model has been
  based on classical FBA, with biomass production as a (negative)
  energy function (under simple flux constraints) . Compared to this
  earlier FBA approach, the present FCM approach works a bit
  diferently. It bring two main changes. First, th ecell population is
  described on a B-polytope with a predefined biomass rate $b$ (as the
  flux benefit), while the cell growth rate $\lambda$ is computed from
  the enzyme demand at this benefit. \co{Second, ..}}  Instead of the
Boltzmann distribution (\ref{eq:boltzmann}), we can also use another
possible probability distribution with the density
\begin{eqnarray}
  \label{eq:fermi}
  p(\vv) \sim \frac{1}{1-(\lambda(\vv)-\lambda_{d})/\xi}.
\end{eqnarray}
 To derive this formula, we
assume that cells can switch their flux state in any moment with a
characteristic  rate $\xi=1/\tau$, where $\tau < 1/\lambda$ is required (see appendix
\ref{sec:SIPPopProbability}).  Mathematically, the distribution (\ref{eq:fermi})
resembles the Fermi-Dirac and Bose-Einstein distributions in quantum
statistics (see appendix \ref{sec:SIProofPopProbability}).

\myparagraph{Cell population model based on flux cost functions} The
probability densities Eq.~(\ref{eq:boltzmann}) or (\ref{eq:fermi})
tell us about average fluxes, flux variability, and flux correlations
in a cell population.  They also define, implicitly, probability
distributions for the (ECM-optimised) metabolite and enzyme
levels.There is no closed formula for all these quantities\footnote{To
  obtain a closed formula for $p(\vv)$, we may replace the flux cost
  function $\acost(\vv)$ by a linear approximation. This linear energy
  function leads to a simple exponential probability distribution on
  the flux polytope. Likewise, a quadratic approximation (quadratic
  energy function) leads to a multivariate normal
  distribution. \co{mention message passing equations; ensemble cite
    jorge! JA, cite!}}, but we can use Monte Carlo sampling: by
sampling {\flow}s and recording the corresponding metabolite and
enzyme profiles, we obtain an ensemble of cell states. To characterise
this ensemble, we can use concepts from statistical mechanics, such as
the partition function, entropy, mutual information, and a quantity
analogous to the free energy, an effective population growth rate that
accounts for the dispersion of cell states. \co{A similar ``free''
  cost function, accounting for dispersion due to mutations, has been
  discussed in \cite{bnsl:11}, Supplementary Information. \co{say more
    about this!}}  \co{What can we do with these distributions?  What
  can we learn from them?  What details need to be discussed?}
\todo{Interestingly, our cell population models are linked to
  thermodynamics in two ways: on the one hand, we use tools from
  statistical mechanics for describing cell populations; on the other
  hand, within our metabolic models themselves, thermodynamics
  determines flux directions, shapes flux and metabolite profiles,
  appears in rate laws and enzyme demand functions, and thereby shapes
  flux cost function.}
\todo{When working with probability distributions, defined as in
  \cite{dabg:17} or here, an open problem is the choice of a metric or
  ``density of states''\co{ref to CBA opt}\footnote{\co{JA! MOVE TO
      MAIN TEXT} In FCM, each flux distribution is associated with
    (optimal) metabolite and enzyme profile, i.e.~with a complete
    metabolic state. However, a uniform distribution in flux space
    will not translate into a uniform distribution of profiles in
    metabolite or enzyme space. \todo{This shows that we cannot
    ``naturally'' assume a uniform in these spaces.}  Instead, we need
    to define a metric, to declare how volumes are measured in the
    different spaces. In practice this metric can be described by a
    kind of ``prior probability density'', to be included in the above
    formulae.\co{ref CBA opt}}. \co{ausfuehrlicher} \co{etwas text von CBA opt
    hierher?}} \co{“distributions + metrics on state manifold” und
  “boltzmannkram” nur kurz bezogen auf flüsse (im flussraum), aber
  konkret für populationen (sagen, inwieweit (durch kinetic flux cost)
  auch die anderen arten von variablen implizit mit einbezogen sind)}
\co{(cite orkun, and oliver's review article)} \co{what do we learn /
  conclude from this?}

\co{aus meiner email an robert matthews

Let's consider a population of growing cells. Each cell is characterised by a vector x of internal variables, which allows it to  grow at a rate $\mu(x)$. If a cell population starts with a probability density p(x, t=0) and cells grow and divide without changing their state, then after time t, the density would become p(x,t) $\sim$ exp(mu(x) t) p(x,0). Formally, this expression resembles Bayes' formula, with  p(x,t=0) as a prior, exp(mu(x) t) as a likelihood, and p(x,t) as the posterior. If the function $\mu(x)$ can be computed from models, inferring p(x,t=0) from p(x,t) would have a similar form as "reverse Bayes".
  
There is  a variant of this problem in which cells grow for a while, randomly jump to another state, grow again, and so on. In this model, a "reverse Bayes" reasoning could be applied to infer the probability of jumping to certain states from the observed distribution of states in the population;  the "jumping probability" could give a hint about the "density of cell states" behind x, i.e. the number (or rather: measure) of (viable, biologically reasonable) cell microstates behind the macrostate characterised by the vector x (a question that has not been explored).}

\iftoggle{bookversion}{\section{Conclusions}}{\section{Discussion}}

\co{discussion is too long!}  \co{some parts of the discussion are
  just summary, and a bit repetitive; move some of them to the intro
  of FCM II?}

\co{WO? Cells may avoid this extra cost by specific metabolic
  strategies: for example, by running different flux modes separately
  at different times or in different compartments.}

\myparagraph{\ \\Flux cost functions} We saw how enzymatic flux costs
$\acostenz(\vv)$ and kinetic flux costs $\acostkin(\vv)$, together
with optimal metabolite concentrations $\cv(\vv)$ and enzyme levels
$\enzymev(\vv)$, can be derived from kinetic models. While
$\aenz(\vv)$ captures enzyme cost only, $\akin(\vv)$ considers both
enzyme and metabolite costs. The costs may represent, for example,
occupied space or invested carbon or energy.  \co{Generally, higher
  catalytic rates are beneficial: the same flux benefit can be
  achieved at lower enzyme cost, or an integral benefit
  $\int b\,\md t$ can be achieved more quickly.}  Importantly, these
concentration profiles are not the only ones that can realise the
fluxes $\vv$, but they are the ones that can do that at a minimal
cost. \co{WEG? Reflecting this cost, a flux cost function may
  describe, for example, the cell volume occupied by metabolites and
  enzymes. With our model assumptions, an enzymatic space demand would
  be a convex function on the M-polytope, and by minimising it we
  obtain our kinetic flux cost $\acostkin(\vv)$: since it represents
  the lowest possible enzyme and metabolite cost
  $\metcost(\lncv)+\hminus(\enzymev)$, at flux profile $\vv$, it
  provides a lower bound on the actual cost\footnote{Other metabolite
    and enzyme profiles may have other advantages, but their enzymatic
    (or kinetic) cost would be higher.}.}  Our flux cost functions
depend on rate laws, kinetic constants, and external metabolite
levels. Although they may be hard to compute, they have some simple
general properties.  A proportional scaling of fluxes and enzyme
levels leaves metabolite levels (and therefore enzyme
efficiencies\co{JA! cat rate uea!}) unchanged. As a consequence, the
enzymatic flux cost scales proportionally with the flux profile and we
obtain simple sum rules for its derivatives. \co{summenregeln, klarere
  schluesse aus summenregeln ziehen; im zusammenhang damit punktkosten
  klar und allgemein einfuehren und diskutieren // also insgesamt: c,
  e, a, lambda auf polytop und zwischen polytopen, jeweils fuer
  enzymatic und kinetic cost optimisation!} Moreover, enzymatic flux
cost functions are concave (and often strictly concave).  When two
{\flow}s are added or interpolated, they need to ``negotiate'' a
common optimal metabolite profile; this leads to a compromise cost.
If two {\flow}s {\favour} the same metabolite profile, there is no
compromise cost between them, the flux costs are additive and the flux
cost function between them is linear. Otherwise, the flux cost
function is strictly concave, so there is a compromise cost, and any
optimal {\flow}s must be vertices of the flux polytope. This confirms
the idea that ``optimal flux distributions must be EFMs''
\cite{murs:14,wpht:14} and generalises it to models with flux
constraints.

\co{\myparagraph{Kinetic flux cost function} \co{Some of the results above
  hold only for enzymatic flux cost functions with a linear enzyme
  cost $\hminus(\esymbolv)$; beweise gehen teilweise nicht fuer
  kinetic} \co{make more separate statements about the metabolic flux
  cost function; sum rules?  scaling properties?  concavity probably
  not ..?}  \co{geht der strict concavity-beweis auch mit kinetic
  cost?  wenn ja, das sagen, wenn nein, sagen dass es nicht gilt}
Kinetic flux cost functions, comprising both enzyme and metabolite
costs, have similar properties, but there are also some
differences. \co{Some of the above results hold only for enzymatic
  flux cost functions, i.e.~in the absence of direct metabolite costs,
  and someof those hold only for concave or even linear enzyme cost
  functions $h(\enzymev)$.}  However, this has consequences: while
enzymatic flux costs scale proportionally with the {\flow}, kinetic
flux costs contain a constant offset term, which depends on the
different optimal metabolite profiles and differs between flux modes.
\co{Also the concavity proof \co{??}  holds only for enzymatic, not
  metabolic flux cost function. STIMMT DAS? // concave unclear;
  $\hminus(e)$ konvex statt linear geht nicht; dann bleibt zwar
  $\metcost(s)$ konvex, aber $\acost(v)$ ist evtl nicht mehr konkav}
However, formula (\ref{eq:FluxCostGradient}) for the cost gradient (in
SI section \ref{sec:linearAndNonlinearApproximations}) also applies to
kinetic flux cost functions.}

\co{\textbf{Criteria for strictly concave enzymatic flux cost
    functions} Here we ask whether flux cost functions to be strictly
  concave and found a number of criteria. {\Flow}s that {\favour}
  the same metabolite profile are called {\stateequal}.  \co{FALSCH?
    EDIT:} If two {\flow}s have non-unique {\favour}ed metabolite
  profiles, then at least \emph{some} of these profiles differ
  and the cost function (enzymatic or kinetic) is strictly convex on
  the metabolite polytope. The latter holds for all models with common
  modular rate laws and, more generally, for all models in which
  metabolite variations $\delta \lncv$ have kinetic effects on
  reaction rates.

  To see why certain reaction fluxes are suppressed and to
  study flux reversals, we considered the flux cost function at
  the boundary between F-cones. At this boundary, the optimal
  metabolite profile may jump between one M-polytope to another one,
  implying a jump in enzyme levels and enzyme cost.}

\myparagraph{Properties of realistic flux cost functions} The flux
cost functions defined by kinetic models also tell us about some
general features a flux cost function should have: it should increase
with the fluxes, be concave on each F-polytope, but discontinuuous
between F-polytopes. The resulting flux distributions will be sparse,
located on F-polytope boundaries rather than in its interior.  This
excludes heuristic flux cost functions like the sum of squared fluxes,
which are convex with an optimum inside a B-polytope.  Of course, such functions may be used for  other reasons
(e.g.~for obtaining a
simple unique solution based on a principle of minimal information) as
in geometric FBA, \co{REF} stating that alternative pathways fluxes
should be used simultaneously unless data are telling us
otherwise. But as proxies for actual enzyme cost, such functions are
poorly justified.  \co{Discuss relationship to Uri's ``archetypes'' in
  multi-objective approach, and discuss how EFMs, in our approach,
  could be used in a multi-objective setting; + in what respect this
  is exactly an example of Uri's approach.}

\co{\textbf{Advantages of the layered optimisation} FCM relies on a
  layered optimisation, with a convex cost function on the M-polytope
  and a concave cost function on the B-polytope.} \co{JA! praktisch:
  konvex und konkav! konvex ist gut zu loesen, eine loesung (falls
  nicht, regularisieren oder berich auswaehlen). konkav ist relativ
  gut zu loesen (ecken, koennen nur viele sein!)

  linear funktioniert
  praktisch eher wie konkav! konkave extra-kosten koennen auch durch
  constraints ausgedrueckt werden; liniearisierung ueberfuehrt
  proteinkosten/constraints in flusskosten/ constraints. mit all dem
  ergeben sich andere methoden als vereinfachungen, und die bezuge
  zwischen ihnen werden klar.}  \co{wo? vorteil in unserem ansatz:
  flux constraints are easy to handle; in stefan's + meike's approach
  difficult, but they can handle enzyme constraints more easily}

\co{\textbf{Biological results} Also, was kommt raus? nonelementary solutions
  (respiro-fermentation); flux constraints should become active at
  higher growth rates / glucose levels; so it'more likely to be
  glucose transport than respiration problems??}  \co{what if
  compromise cost is small = negligible. what does EFM result mean in
  practice? wurde schon oben gesagt .. nicht viel!}  \co{Cells can
  avoid the compromise cost by realising different {\flow}s in spatial
  or temporal compartments.}

\co{\textbf{Possible applications} mention usage for population
  dynamics models, eg very simple FBA, with enzyme cost converted to
  growth rates, accounting for simpe kinetics and thermodynamics. (wie
  in satFBA)}

\myparagraph{Compromises between flux cost and flux benefit} Flux cost
functions allow us to predict optimal fluxes and growth rates based on
metabolite and enzyme costs, kinetic laws, and physiological
constraints (e.g.~bounds on metabolite levels). By converting
metabolite and enzyme costs into flux costs, we can replace the search
for optimal states by a simple optimisation of fluxes. In different
variants of this optimisation, flux cost and flux benefit are either
treated as objectives or as constraints. \co{FN: In optimality
  problems, constraints and costs are often exchangeable (same form of
  optimality conditions); we can see this, for example, by comparing
  FCM and FBM to the corresponding mutli-objective problem!  this is
  more an argument for CBA theory - pareto as linking idea!)}  FCM (a
generalised form of {\mfFBA}) minimises flux cost at given flux
benefit, FBM (a generalised form of {\mcFBA}) maximises flux benefit
at a bounded flux cost, while linear-fractional programming optimises
the benefit/cost ratio directly.  In the absence of flux bounds (in
general, weighted sums of absolute fluxes), flux solutions are
scalable and we need to impose a scaling (e.g.~by fixing the flux
benefit) to obtain a specific solution. In contrast, with fluxes being
bounded, \co{oben ``flux bounds'' einfuehren, genau erklaeren,
  verwenden fuer ``extra constraints'', including thermodyn?? bei
  anfang 3.4!}  the absolute scaling of fluxes matters: if a flux hits
a bound, the fluxes cannot be further increased by scaling, and the
only way to (possibly) increase the flux benefit is by changing the
shape of the {\flow}. Therefore, if flux constraints are imposed
(e.g.~a bound on respiration due to limited membrane space), this does
not only affect the scaling, but also the shape of optimal flux
distributions.  Instead of assuming a single objective, we may treat
different targets (such as enzyme cost, biomass production, or biomass
yield) as separate objectives and describe optimal compromises
(e.g.~between flux-derived growth rates and yield \cite{wnfb:18}) as
Pareto-optimal points \cite{scks:07,szzh:12}. This approach can be
generally used to capture multiple cost and benefit functions. \co{REF
  schuetz + Carlson pareto \cite{carls:07}}

\co{JA! in CBA opt sagen, dass ECM schon hinweis auf optimalitaet geliefert hat!}

\co{\co{\textbf{Protein and flux constraints; non-elementary flux modes}}
As a biologically plausible standard problem, we may consider flux
benefit maximisation, e.g.~the maximisation of biomass production with
constraints on one or several protein fractions (e.g.~the total
protein mass densities in the cytosol, in other compartments, and on
cell membranes) (see SI section \ref{sec:SIFBM}). In FCM, it is easier
to put cost terms or constraints on fluxes than on enzyme
levels. Therefore, we may be inclined to replace (non-concave) protein
costs by protein constraints, then to linearise them, and to further
replace them by linear flux constraints, possibly giving rise to
non-EFM solutions.}  \co{anwendung von flux cost functions auf mol
  crowding geht; aber das waere eine lin opt mit nonlin constraints!}
\co{nochmal kurz sagen, was dann passiert: non-elementary solutions;
  ``growth rate dependent'' (actually, nutrient-level dependent)
  strategy switch can occur} \co{SORT; make it clear that according to
  my scaling logic, the maximal-growth mode (ie the winning mode) will
  always be simply the one with the minimal enzyme demand per output
  (target) flux. High yield (or ``low uptake demand''), by itself,
  does not play a role; however, a low uptake demand also means: a low
  demand for transporter; this becomes relevant at a low nutrient
  level; therefore, at a low nutrient level, it is likely that
  high-yield modes have also an enzyme cost (and thus growth rate)
  advantage. IN CBA, this is reflected in the fact that a high
  tranporter investment is EMBODIED in metabolites and later requires
  efficient usage!}  \co{nur kurz auf abschnitt oben verweisen, sagen,
  dass fECM aequivalent to wachstumsmaximierung ist} \co{REF TO EFM
  \cite{wnfb:18}, mori \cite{momm:17}, usw.}  \co{hier pareto zwischen
  verschiedenen flusskosten erwaehnen und auf ross carlsson verweisen}

\myparagraph{Flux optimisation under constraints} In flux prediction,
fluxes or enzyme levels can be constrained.  If enzyme demands are
linearly approximated, the two types of bounds play similar roles:
assuming constant enzyme efficiencies, enzyme bounds can be written as
flux bounds.  In molecular crowding FBA \co{mcFBA?} and CAFBA\co{JA!
  cafba oefters erwaehnen, auch in CBA opt und lagrange?}, this link
between fluxes and enzyme levels is used to formulate resource
allocation principles directly for fluxes.  \co{Constraints or side
  objectives on substrate demands or enzyme demands?} However, flux
constraints can also be justified in different ways.  A constraint on
glucose transport flux may reflect a bound on transporter abundance
(at a given external glucose level), but also an economical usage of
glucose (avoiding unnecessary import fluxes, and unnecessary losses by
export fluxes). \co{ref to hergo} \co{Dealing with fixed enzyme
  levels} We saw that flux optimisation under constraints can lead to
non-elementary flux distributions. \todo{But there are also other
  reasons for non-elementary fluxes.} By adapting their enzyme levels,
cells can save enzyme resources; but if enzyme levels are fixed (on a
time scale of interest), cells can still optimise their fluxes by
changing the enzyme activities (e.g.~by protein phosphorylation). Flux
optimisation in this case can be formulated as a convex problem, but
one that may lead to non-EFMs (see SI
\ref{sec:SIFCMwithGivenEnzymeLevels}).  \co{also mention this in glass
  ceiling paper!}
  
  \coout{THE EXAMPLE DOES NOT MAKE SENSE \co{\textbf{Constraints in
        different parts of the cell can have similar effects on the
        usage of flux modes} Constraints on different fluxes (or
      protein levels) may have similar effects. For example, an
      observed limit to pure respiration (see Figure
      \ref{fig:definition}) may be caused by a direct constraint on
      the respiration flux (e.g.~limited ``real estate'' for oxidative
      phosphorylation proteins on a membrane); however, it may equally
      well represent a real-estate constraint on the glucose
      transporter, since respiration involves a much higher glucose
      import per ATP production.} \co{Discuss this more, and come back
      to this. In this paper, also make the argument that high yield
      means high (benefit-specific) import rates, and therefore high
      transporter cost, possibly leading to a rate-yield trade-off at
      low nutrient levels.} \co{WD; ARGUEMNT NOT CLEARSimilarly, the
    mere fact that respiration is preferred can have various reasons;
    knowing about varying glucose and oxygen levels, one can actually
    find out!  The relation to rate-yield trade-offs will be further
    discussed below.}  \co{explicitly make the argument that FOR TWO
    EFMs only, a bound on glucose transport and on respiration can
    have equivalent effects (unless more than two modes are
    considered, or the glucose of oxygen levels a revaried and known!
    only the reponse to varying growth rates does not help, i think)}
  \co{MOEGLICH?  show that low nutrient level or pricy transporter
    upstream of a metabolite leads to yield-efficient downstream use;
    in example, show how transporter varies with external supply;
    REFER TO THIS below, in biosynthesis VS transport example}}

\co{Protein constraints and flux constraints}

  \co{Flux benefit maximisation as a basic
  problem} \co{say that optimising ONE benefit (eg biomass) under
  SEVERAL constraints (protein fractions) is the most realistic
  assumption; however, in the case of ONE constraint only, this
  reduces, effectively to FCM (which is conceptually simple).}

\co{In summary, in
  models without flux (or enzyme) bounds (on single or multiple
  reactions) FCM, FBM and cost efficiency \co{wort?}  maximisation
  (benefit/cost) are equivalent; the scaling of {\flow}s does not play
  a role, and solutions are EFMs; in models WITH such bounds, the
  different methods may yield different results (cost efficiency
  \co{wort?} maximisation can yield to small scaling!), the scaling
  plays a role, and solutions may be non-EFMs.}

\co{In fact, all these problems correspond to a common multi-objective
  optimality problem, and their optimality conditions look the same!}
\co{REF CBA theory \cite{lieb:18theory}}

\co{treat preemptive expression in a similar convex way. first attempts
  (noor 2016 + here) exist.}  \co{except for section on preemptive
  .. // also mention convexity proof in Noor 2016?}

\myparagraph{Practical usage of realistic flux cost functions: enzyme
  efficiencies and flux cost functions for FBA} The flux cost
functions in FBA are usually chosen ad hoc: it remains unclear which
actual costs they represent, whether they can capture metabolite
costs, and how they depend on real biochemical parameters.  Enzymatic
and kinetic flux costs, in contrast, are related to enzyme
efficiencies derived from kinetic models in optimal states\co{REF
  wortel}.  In molecular crowding FBA, CAFBA, and Resource Balance
Analysis \co{ref}, these enzyme efficiencies are key model parameters
with a a big impact on metabolic strategies.  If constant enzyme
efficiencies and enzyme prices $\partial h/\partial \esymbol$ are
assumed, this leads to constant flux burdens \co{$av=..$} to be used
as cost weights in linear flux cost functions. Linear flux cost
functions are widely used in flux modelling, but often without a good
justification: for example, linear relationships between fluxes and
enzyme levels are assumed, claiming that enzyme work at their maximal
capacity ($\kcat$ value) or at a constant capacity (``apparent $\kcat$
value'').  In reality, the $v/\esymbol$ ratios can vary between
metabolic states, and knowing these numbers is of utmost importance in
FBA.  We saw how enzyme efficiencies depend on kinetics, on external
conditions, and on the resulting optimal metabolite profile), and that
we can predict them by ECM.  We also learned that nonlinearities in
enzyme cost functions (which we call compromise costs) stem from the
fact that different flux profiles {\favour} different metabolite
profiles, and thus different catalytic rates.  If two {\flow}s
{\favour} the same metabolite profile, and therefore go with the same
enzyme efficiencies, flux costs can be linearly interpolated.  The
above construction of linear flux costs functions from kinetic models
justifies their usage in FBA and shows how the cost weights depend on
details of the kinetic model and how they vary between growth
conditions.

\myparagraph{Limitations of flux cost minimisation} Flux cost
minimisation relies on many simplifying assumptions. What if we
abandon some of them?  \emph{(i) Nonlinear enzyme cost functions.}
Instead of assuming a linear enzyme cost function, we may assume a
nonlinear function $\hminus(\enzymev)$. As long as this function is
convex, the ECM problem will also be convex. However, some other
results may not hold: enzymatic flux cost functions will not scale
proportionally with the flux profile and may become non-concave (the
proof in SI \ref{sec:prooflemma1} does not apply in this
case\footnote{To show that enzymatic flux cost functions are concave
  on the (positive) F-cone (SI section \ref{sec:prooflemma1}), we had
  to assume linear enzyme cost functions.  If an enzyme cost function
  is strictly convex, the enzymatic flux cost will be strictly convex
  under flux scaling, which makes it for sure non-concave.  In this
  case, our proof does not apply anymore: the equality
  Eq.~(\ref{eq:convexityonP}) is replaced by an inequality with a
  $\le$ sign, the inequality Eq.~(\ref{eq:concavityproof1}) may not
  hold anymore.}). However, the change from a linear enzyme cost
function to a convex one will not change the optimal flux solutions.
\emph{(ii) Non-enzymatic reactions.}  Non-enzymatic reactions are hard
to handle in FCM (see SI section \ref{sec:nonEnzymatic}).  Their
presence changes the shape of metabolite and flux polytopes and of the
enzymatic F-cost function. This can be explored numerically
(non-enzymatic reactions may lead to constraints on the M-polytope,
which can be handled in ECM), but the optimal solutions may be
non-EFMs.  \emph{(iii) Stabilisation of metabolic states.}  Unstable
metabolic steady states can be stabilised by feedback control,
e.g.~enzyme inhibition by pathway products. Compared to the same model
without inhibition, the enzyme level needs to be higher to yield the
same reference flux. \co{JA! say: compromise enzyme cost vs stable
  Jacobian and good control coeffs!  this aspect has NOT been
  considered so far in this paer!}  Also a simple overexpression of
enzymes, which keeps the enzymes far from saturation, can stabilise
metabolic states. In models, such extra enzyme investments would add
to the enzymatic flux cost and would possibly make it
non-concave. \co{besseres argument!: (because the extra investments
  depend on the metabolite profile and will differ between flux modes}
\co{FN JA! a similar reason holds for stabilisation (or enforcing) of
  periodic states!}  \coout{(iv)?  Isoenzymes: Isoenzymes can be
  formally described by separate chemical reactions with identical sum
  formulae.  alternative (kleiner beweis): die zelle kann einen
  reaktionsfluss durch zwei isoenzyme verwirklichen; fuer jedes davn
  gibt es eine enzymatische metabolitkostenfunktion. wenn die zelle
  jeweils (fuer jedes metabolitprofil) die billigere variante waehlt,
  ist die entstehende effektive enzymkostenfunktion (in bezug auf die
  gesamtsituation) evtl NICHT-KONVEX!! (trotzdem gibt es keine inneren
  punkte, die besser als randpunkte sind!)}

\myparagraph{Metabolic behaviour beyond maximally efficient enzyme
  usage} \co{ref to efm paper \cite{wnfb:18}? or rather say here,
  whether and how other optimality criteria could be included?}  In
this paper, we assumed that cells strive for maximal metabolic
efficiency, defined as a production flux (or other linear flux
objectives) divided by an enzyme (or enzyme plus metabolite) cost. To
implement this in models, one may minimise cost at a fixed benefit (in
FCM) or maximise benefit at a fixed cost (in FBM).  While optimising
enzyme efficiency can serve as a useful heuristics, in reality cells
do not strictly save enzyme resources or maximise catalytic rates.
\co{what does possible mean?  das ist auch die hauptfrage im glass
  ceiling paper} \co{Sometimes, apparent ``enzyme overcapacities'' are
  due to the fact that we false assume that cells could use all
  enzymes at their maximal speed (nochmal ref palsson -- kam schon
  oben?); but sometimes, cells do have overcapacities that are really
  not used.}  During a metabolic switch, \emph{Lactococcus lactis}
bacteria \cite{gepj:15} inhibit some of their enzymes (thus reducing
their efficiency) while they could have saved enzyme costs by
repressing enzymes transcriptionally \cite{gepj:15}.  How can we
explain this behaviour? Unnecessary enzymes allow cells to quickly
adapt to environmental changes. Enzymes may have also other functions
or side benefits (e.g.~serving as an amino acid storage).  By ignoring
these effects, FCM provides a \emph{theoretical minimal cost} of a
flux, and the corresponding maximal growth rate. Even if they are
overly simplified, these predictions can be usefule as a null
hypothesis for testing other optimality scenarios (e.g.~optimality
scenarios that include preemptive expression, side objectives for
proteins or metabolites, or simply non-optimality) and to compute the
difference in enzyme cost.

\section*{Acknowledgements}
I thank Elad Noor, Frank Bruggeman, Hermann-Georg Holzh\"utter, Joost
Hulshof, Stefan M\"uller, Ralf Steuer, and Meike Wortel for thinking
with me about the topic. This work was funded by the German Research
Foundation (Ll 1676/2-1 and Ll 1676/2-2).

\bibliographystyle{unsrt}
\bibliography{files/biology}

\clearpage

\begin{appendix}

\section{Metabolic {\flow}s and flux cost functions}

\co{WO?
  \includegraphics[width=14.5cm]{/home/wolfram/projekte/cba/zeichnungen/A55.jpg}
}

\subsection{Metabolic models,  metabolic {\flow}s, and flux polytopes}

\textbf{Metabolic models} We consider metabolic network models with
reversible rate laws of the form
$v_l = \esymbol_l \, \ratelaw_{l}(\cv)$ (where $\esymbol_l$
denotes enzyme levels). The catalytic rate $\ratelaw(\cv)$ of an
enzyme $l$ can be factorised into capacity, reversibility, and kinetic
terms $\ratelaw=\kcat\cdot\eta^{\rm rev}(\cv)\cdot\eta^{\rm kin}(\cv)$
\coout{superscripts runter, uea?}  \cite{nfbd:16}.  To model fitness
incentives and physiological requirements in a cell, we consider three
metabolic {\target}s: $\fluxbene(\vv)$ (linear flux benefit function),
$\metcost(\cv)$ (convex metabolite cost function), and
$h(\esymbolv)$ (linear enzyme cost function). Typically, the flux
benefit scores the production of valuable compounds or biomass, while
the cost terms describe growth disadvantages caused by higher protein
or metabolite concentrations. Under simplifying assumptions, the ratio [enzyme
cost]/[flux benefit] can be translated into a cell growth rate (see
section \ref{sec:SIProofGrowthRateFunction}).

\textbf{Metabolic {\flow}s, scaled {\flow}s, and flux modes} if a flux
distribution $\vv$ is stationary (satisfying the condition
$\Nint\,\vv=0$), it is called a \emph{metabolic {\flow}} or, if the
absolute scaling is disregarded, a \emph{flux mode}.  An elementary
flux mode is a flux mode with a minimal set of active reactions: if
one more reaction flux were restricted to zero, the remaining active
reactions would not be able to support any stationary flux
distribution.  Flux modes can be seen as \emph{scaled {\flow}s}.  For
example, with a ``target flux'' $v_r$ (e.g.~biomass production) we can
define the scaled {\flow} $\vv/v_r$ and the scaled flux cost
$\acost^{\rm sc}(\vv) = \acost(\vv)/v_r$, \co{use $\hat{a}$ instead of sc
  uea?}  i.e.~the flux cost \emph{per target flux}. \co{or, with a
  flux benefit b(v), we can ..}  If a {\flow} $\vv$ is multiplied by a
factor $\sigma$, the scaled {\flow} and the scaled cost remain
unchanged.

\co{klar definiert?  "konform": (a "konform" zu b):
   gegeben b, muss a dasselbe vorzeichen haben oder null sein.\\
   Aber wie nennt man Beziehung zwischen a und b, wenn a und b überall
   da, wo beide ungleich null sind, die gleichen Vorzeichen haben?
   Stefan: wir nennen das wieder conformal (denke dir: zu einem
   dritten vektor c). siehe unser bioinformatics paper (attached).}

\textbf{Flux patterns and conformal {\flow}s} The sign vector of a
{\flow} $\vv$ is a vector with elements 1,0, and -1, describing the
set of active reaction and flux directions. Such vectors are called
\emph{{\fluxpattern}s}.  A {\flow} is \emph{conformal} with a given
{\fluxpattern} if all active fluxes match the prescribed signs (zero
fluxes are allowed even if signs 1 or -1 are prescribed). A {\flow}
$\vv$ is said to \emph{have} (or to be ``\emph{strictly conformal}
with'') a given {\fluxpattern} if the flux signs exactly match the
prescribed signs (in this case, a prescribed sign of 0 means that the
flux must vanish).  Two {\flow}s are called \emph{conformal} if they
are conformal to a common {\fluxpattern}, i.e.~if all their shared
active reactions show the same flux directions.  \co{where
  $\tauv \sqsubseteq \sigmav$ \co{auch in FCM SI, CBA fluxes SI!}
  denotes that $\tau$ is conformal with $\sigmav$} \coout{use symbol $||$
  for ``agrees with'' (??))}  Two {\flow}s are \emph{strictly
  conformal} if they have exactly the same flux pattern. All these
definitions also hold for flux modes instead of metabolic {\flow}s.

\textbf{Feasible {\fluxpattern}s} \co{SORT until end of section}
A metabolic network allows for many
{\fluxpattern}s, but only some of them are physically and
physiologically feasible.   To be realised with
reversible kinetic rate laws, a {\flow} must be thermo-physiologically
feasible: this means, there must be a choice of metabolite concentrations
that respect the physiological concentration ranges defined in the
model and that allow  the {\flow} to  be realised under thermodynamic
constraints on the flux directions.  This requires that the {\flow} is
loopless. \co{ref}  The {\fluxpattern} of a thermo-physiologically feasible
{\flow} is called thermo-physiologically feasible as well. Briefly, a
\emph{feasible {\fluxpattern}} is a {\fluxpattern} that can be
realised by fluxes that are stationary, thermodynamically feasible,
and that realise the predefined flux benefit.

\textbf{Flux polytopes} Each {\fluxpattern} corresponds to a segment
in flux space, i.e.~to an orthant or one of its lower-dimensional
surfaces.  If a segment intersects the subspace of stationary fluxes,
the intersection set is called an F-cone (``signed flux polytope'')
\co{polytope or cone? fix and explain!}  $\sPolytope$. All {\flow}s in
the interior of an F-cone are strictly conformal.  If two {\flow}s are
conformal, they belong to a common F-cone. If a {\fluxpattern}
excludes negative fluxes, the F-cone is called a positive flux
polytope \co{cone?}  $\pPolytope = \{\vv|\Nmat \,\vv = 0; \vv\ge 0\}$.

\textbf{Reorienting the reaction directions} The flux signs in a model
depend on the reaction orientations, i.e.~which compounds are treated
as substrates or products by definition.  This is a matter of
convention, and independent of the actual flux directions.  Since
F-cones can always be converted into P-polytopes (by reorienting the
reactions\footnote{\co{in the following, we concentrate on ..}
  All statements about enyzyme or metabolite concentrations
  (in particular, the convexity proof for enzyme cost functions on the
  M-polytope and the concavity proof for flux cost functions on the
  P-polytope) remain valid after reorienting the fluxes.}), we can
assume (without loss of generality) that fluxes are non-negative.

 \textbf{Benefit-scaled polytope} If we restrict an F-cone to
 {\flow}s with a fixed benefit $\fluxbene(\vv) = b$, we obtain a
 \emph{B-polytope} (``benefit-restricted flux polytope''), which is
 convex and has one dimension less than the F-cone\footnote{One
   may also impose several benefit constraints (i.e.~a set of
   equalities $\Bmat \, \vv = \bv'$), but I do not consider such cases
   here.}.  \co{relate
   projection onto B-polytope to scaled {\flow}s} A  {\fluxpattern} is feasible if it 
 corresponds to a segment of flux space that contains a non-empty
 B-polytope, for the kinetic model, concentration ranges, and flux
 benefit function in question.

\textbf{{\Combinedflow}s} The interpolation between two {\flow}s $\vvA$
and $\vvB$, with interpolation parameter $0<\sigma<1$, yields a
{\combinedflow} $\vvC = \sigma \,\vvA + [1-\sigma]\,\vvB$; in this
combination, the {\flow}s $\vvA$ and $\vvB$ are called the
\emph{{\basicflow}s}. Combinations of three or more {\flow}s are
defined accordingly. If several {\flow}s belong to one F-cone, their
{\combinedflow}s also belong to this polytope.

\co{JA! DEF convex hull of ...; polytope vertices; relation to EFMs
  (polytope vertices of the B-polytope WITHOUT any upper or lower flux
  bounds}

\textbf{Metabolite polytope} In a given metabolic model, a {\flow}
pattern does not only define an F-cone, but also a metabolite
polytope (or ``M-polytope'') $\mPolytope$ for metabolite profiles. It
is the set of (natural) log-concentration vectors $\lncv = \ln \cv$
that respect the given bounds on individual concentrations and are
thermodynamically feasible for the given flux directions.

\subsection{Flux cost function and \compromisecost}
\label{sec:appFluxCostFunctions}

\begin{definition} 
  \textbf{Flux cost functions} We consider   a kinetic model with enzyme cost
  $h(\esymbolv)$ and metabolite cost $\metcost(\lncv)$ and define the cost functions
  \coout{in formel (und uea): superscripts runter?}
\begin{eqnarray} 
\enzymemetcost(\lncv;\vv) &=& h(\esymbolv(\lncv;\vv)) \;\,\quad \qquad \qquad \quad \mbox{(Enzymatic metabolite cost)} \nonumber \\
\hmet^{\rm kin}(\lncv;\vv) &=& h(\esymbolv(\lncv;\vv)) + \metcost(\lncv) \; \qquad  \quad \mbox{(Kinetic metabolite cost)} \nonumber \\
a^{\rm enz}(\vv)     &=& \mbox{min}_\lncv\, \enzymemetcost(\lncv;\vv) \,\qquad \qquad \mbox{(Enzymatic flux cost)} \nonumber \\
a^{\rm kin}(\vv)     &=& \mbox{min}_\lncv\, \hmet^{\rm kin}(\lncv;\vv) \; \qquad \qquad \mbox{(Kinetic flux cost)},
\end{eqnarray}
where the log-metabolite profile $\lncv = \ln \cv$ contains the
logarithmic metabolite concentrations, and the enzyme demand
$\esymbolv(\lncv;\vv)$ is the enzyme profile that realises the {\flow} $\vv$
at log-metabolite profile $\lncv$. The ``;'' sign indicates that $\vv$
is not a function argument, but a fixed function
parameter. Metabolite cost and flux cost are also called M-cost
  and F-cost.
\end{definition}

\begin{definition} 
\textbf{Concave flux cost functions} By definition, a flux cost function on a
B-polytope is concave if it satisfies, for all $\vvA$ and $\vvB$ from
the B-polytope and for all $ \sigma \in [0,1]$,
\begin{eqnarray} 
\label{eq:concavity}
 \forall \sigma \in\, ]0,1[: \quad
 \acost( [1-\sigma]\,\vvA + \sigma\, \vvB) \ge [1-\sigma]\, \acost(\vvA) + \sigma\, \acost(\vvB).
\end{eqnarray} 
Eq.~(\ref{eq:concavity}) states that the enzymatic flux cost
$\acost(\vvC)$ of the interpolated {\flow} $\vvC = [1-\sigma]\, \vvA +
\sigma\, \vvB$ is \emph{equal or larger} than the
linearly combined cost $\acost^{\rm interp}=[1-\sigma] \acost(\vvA) +
\sigma\, \acost(\vvB)$. If a flux cost function satisfies
Eq.~(\ref{eq:concavity}) with $>$ instead of $\ge$ signs, it is called
\emph{strictly concave}.
\end{definition}

\begin{figure*}[t!]
  \begin{center}
  \parbox{9.3cm}{\includegraphics[width=9cm]{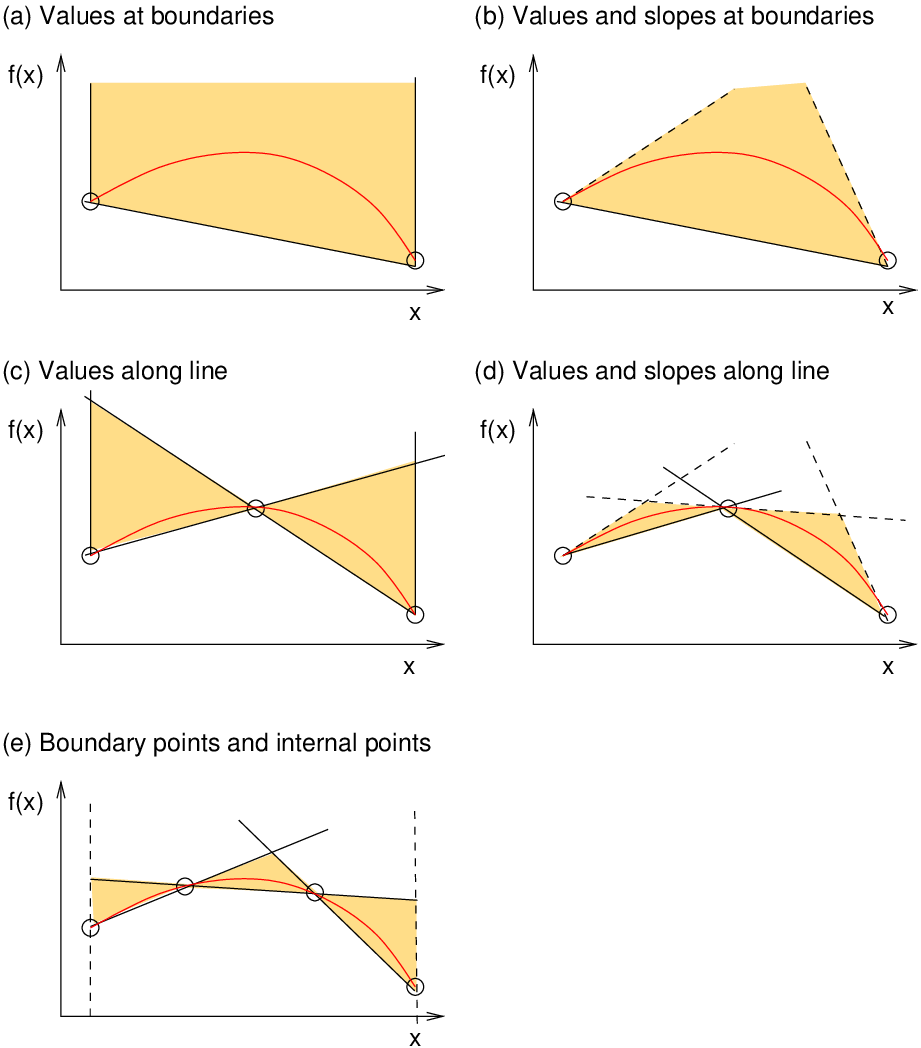}}
  \parbox{7cm}{ \caption{\co{add explanation for (e)} Concave
      functions can be approximated by linear bounds. Knowing the
      values of a function $f(x)$ in two or more points, we can find
      lower and upper bounds described by piecewise linear
      functions. (a) The values in two boundary points define a lower
      bound (the interpolation line), but no upper bound.  (b) Known
      slopes in the boundary points (dotted lines) can be used to
      further restrict the function.  (c) Several reference points on
      the line. By connecting the points, we obtain regions through
      which the function must pass. To be outside these regions,
      $f(x)$ would have to be non-concave.  (d) Again, known slopes
      can be used to determine tigher constraints.
      \label{fig:boundsOnConcave}}}
  \end{center}
\end{figure*}

\textbf{Remarks} 
\begin{enumerate}[leftmargin=5mm]
\item \textbf{Scaled flux polytopes} In the definition above, the
  benefit function defining the B-polytope need not be an actual
  biological fitness objective; it can be \emph{any} linear function
  that defines a scaling of {\flow}s.
\item \textbf{\Compromisecost} A concave flux cost function on the
  line between two {\flow}s can be split into an additive \emph{combined
    cost} plus a positive (or zero) \emph{\compromisecost}. If a flux
  cost function is strictly concave, this {\compromisecost} will be
  strictly positive.
\item \textbf{Cost of coexisting {\flow}s} A combination of {\flow}s
  $\vvA$ and $\vvB$ can be realised in two ways: as a {\combinedflow}
  (a weighted sum of the two {\flow}s), or as coexisting {\flow}s
  (e.g.~by running the {\flow}s in separate cell compartments).  Since
  the enzymatic flux cost is a concave function, the cost of a
  {\combinedflow} $\vv=\sum_\alpha \sigma_\alpha\, \vv^{(\alpha)}$
  (with coefficients $0\le \sigma_\alpha \le 1$) is at least as high
  as the combined cost
  $\acost^{\rm lin}(\vv)=\sum_\alpha \sigma_\alpha\,
  \acost(\vv^{(\alpha)})$.  In many cases, it will be strictly
  concave, which implies a positive \compromisecost.  Additive costs
  (i.e.~cost functions without compromise cost) can be achieved by
  coexisting {\flow}s, i.e.~running the {\basicflow}s separately in
  different cell compartments or at different times (with relative
  durations $\sigma$ and $1-\sigma$).
\item \textbf{Bounds on concave functions} A concave function can be
  approximated by piecewise linear lower and upper bounds (Figure
  \ref{fig:boundsOnConcave}).  In higher dimensions, this
  construction becomes difficult.
\end{enumerate}

\co{idee: boltzmann-wahl (evtl kommen bessere
  repraesentationsstrategien der kandidaten raus). Keine erst- und
  zweitstimme mehr! proporz ergibt sich aus gesetz der grossen zahlen:
  VERBinden mit punktesystem GEGEN kandidaten?}

\subsection{A sum rule for the enzymatic costs of {\combinedflow}s}
\label{sec:SIsumRule}

The sum rules for enzymatic flux cost functions,  Eqs~(\ref{eq:SumRulePointCost}) and
(\ref{eq:SumRulePointCost2}),  follow
from Euler's theorem on homogeneous functions. 

\myparagraph{Enzyme point cost and flux point cost}
Eq.~(\ref{eq:SumRulePointCost}) implies that the flux point cost
\co{aDOTvl =}
$\frac{\partial \acostenz}{\partial v_l} v_l$ is equal to the
enzyme point cost
\co{hDOTul =} $\frac{\partial \hminus}{\partial \esymbol_l}\,\esymbol_l$
(proof in SI section \ref{sec:SIproofFluxCostGradient}).  Thus, for
any {\flow} $\vv$, the enzymatic flux cost is given by the sum of
enzyme point costs of all reactions; if the enzyme cost
$\hminus(\esymbolv)$ is linear, this sum is also equal to  the absolute enzyme cost
$\hminus$.  \co{WEG? The equalities  may seem trivial, because the enzymatic flux
cost, by definition, is the sum of the enzyme costs in all
reactions. However, there is an important detail: the terms in the sum
rule do not refer to a constant metabolite profile, but account for
the optimal adjustment of metabolite concentrations as part of taking the
derivative!  \co{den letzten satz lieber etwas hoch?}}
At first glance, Eq.~(\ref{eq:SumRulePointCost}) seems to
state something obvious: that the enzyme cost
$\acostenz(\vv) = \sum_l \hminus(\esymbol_l)$ of a {\flow} is
given by the sum of all enzyme costs.  However, the sum rule is not so
obvious: the derivatives $\frac{\partial \acostenz}{\partial v_{l}}$
in the sum rule are not derivatives
$\partial \enzymemetcost_l(\vv,\lncv)/\partial v_l$ \emph{at fixed
  metabolite concentrations}, but refer to \emph{optimised} metabolite concentrations:
they imply an optimal metabolic adjustment whenever fluxes are
changing. \co{verbindung zu unchain rule erwaehnen! JA! CBA OPT} This is what distinguishes our enzymatic flux cost function
from simple additive flux costs as assumed in FBA with molecular
crowding\footnote{In taking the partial derivatives in
  Eq.~(\ref{eq:SumRulePointCost}), we consider flux variations that
  \emph{leave} the polytope of stationary {\flow}s. However, this is
  not a problem because the function $\acostenz(\vv)$ is also defined for
  non-stationary {\flow}s.}. Implicitly, these older methods assume a
proportionality between enzyme levels and stationary fluxes, as if
metabolite concentrations were constant and unaffected by changing enzyme
levels and fluxes.  Instead, our sum rule rightly assumes that fluxes
and enyzme levels are coupled through (variable) metabolite
levels. 

\myparagraph{Sum rules for coefficients in combined {\flow}s}  A
similar sum rule holds for {\flow}s that are represented by convex
combinations $\vv = \sum \sigma_\alpha \,\vv^{(\alpha)}$ of prototype
flows $\vv^{(\alpha)}$: now the cost of $\vv$ is homogenous (with
degree 1) with respect to the coefficients $\sigma_\alpha$, where the
basis may be complete or even overcomplete:
\begin{eqnarray}
\acostenz = \sum_\alpha \frac{\partial \acostenz}{\partial
  \sigma_\alpha}\,\sigma_\alpha
\end{eqnarray}
for the coefficients $\sigma_\alpha$. The derivatives are given by
$\frac{\partial \acostenz}{\partial \sigma_\alpha}= \sum_l \frac{\partial
  \acostenz}{\partial v_l}\frac{\partial v_l}{\partial \sigma_\alpha}=
\frac{\partial \acostenz}{\partial \vv} \cdot\vv^{(\alpha)}$.  Again,
there is  an analogous  sum rule for scaled flux costs.

\myparagraph{Sum rules for metabolite and enzyme levels} 
Similar sum rules hold for  optimal metabolite or enzyme
levels as functions on the flux polytope:
\begin{eqnarray}
\label{eq:SumRuleConcentrations}
  \sum_{l} \frac{\partial c_i^{\rm opt}}{\partial \ln v_{l}} 
  =  \sum_{l} \frac{\partial c_i^{\rm opt}}{\partial v_{l}}\,v_l =  0. \nonumber \\
  \sum_{l} \frac{\partial \esymbol_l^{\rm opt}}{\partial \ln v_{l}} 
  =  \sum_{l} \frac{\partial \esymbol_l^{\rm opt}}{\partial v_{l}}\,v_l =  \esymbol_l^{\rm opt}.
\end{eqnarray}
\co{also derive a sum rule for derivatives $\frac{\partial a^{enz}(v)}{\partial \kcat} \sim \sigma \Rightarrow \sum a^{enz}\frac{\partial a^{enz}(v)}{\partial \kcat} = a^{enz}$ by kcat values!  or even
  mixed sum rules for variations of kcat values in some reactions, and
  of enzyme levels in others // ein kcat hoch, alle enzyme ausgleichen?}

\subsection{Linear and nonlinear approximations of enzymatic flux cost functions} 
\label{sec:linearAndNonlinearApproximations}

\myparagraph{\ \\Linear approximation of the enzymatic flux cost} To
obtain realistic linear flux cost functions, we consider the enzymatic
flux cost and linearise it around a prototype {\flow} $\vv^{\rm pt}$.
This approximation works best near the prototype {\flow} (or, more
precisely, for {\flow}s $\vv$ that {\favour} similar metabolite
profiles as the prototype {\flow}). Therefore, we should chose a
prototype {\flow} $\vv^{\rm pt}$ that is ``typical'', resembling the
flows at which $\acostenz(\vv)$ will be evaluated. To compute the
linear cost weights, we start from the cost gradient in a point $\vv$,
\begin{eqnarray}
\label{eq:FluxCostGradient} 
\acostenz_{v_l} =  \frac{\partial \acostenz(\vv)}{\partial v_{l}} =
 \frac{\hul\,\esymbol^{\rm opt}_{l}(\vv)}{v_{l}}  =
\frac{\hul}{\ratelaw^{\rm pt}_{l}},
\end{eqnarray}
where $\esymbol^{\rm opt}_{l}(\vv)$ is the  enzyme profile
that is {\favoured} by our {\flow} $\vv$ (see SI
\ref{sec:PrototypeApproximationLinear} and
\ref{sec:SIproofFluxCostGradient}). Equation
(\ref{eq:FluxCostGradient}) holds also for kinetic flux cost functions 
 (with a direct metabolite cost).
By linearising the cost function around a prototype state $\vv^{\rm
  ref}$, we obtain the linear approximation 
\begin{eqnarray}
\label{eq:linAppr}
  \acostenz(\vv) =
   \sum_{l} \frac{\partial \acostenz}{\partial v_{l}} \,\,v_{l} 
   \approx \sum_{l} \left(\frac{\partial \acostenz}{\partial v_{l}}\right)_{\vv^{\rm pt}} \,\,v_{l}
= \sum_l \acostenz_{v_l}\,v_l
\end{eqnarray}
with the cost weights $\acostenz_{v_l}$. There is no offset term, and
so our approximated cost function is again linearly scalable.  The
formula (\ref{eq:linAppr}) holds only for non-negative fluxes. For
other {\flow}s, with negative fluxes, we may reorient the reactions and
obtain the same formula, but with different prefactors (each obtained
from a different prototype {\flow}). The signs of these prefactors must
match the flux signs. If we assume the same (absolute) prefactors for
all F-cones, we obtain the formula
$\acost^{\rm lin}(\vv)=\sum_{l} \acost_{v_l}\,|v_{l}|$, the
typical cost function in FBA with weighted flux minimisation.

\textbf{Quadratic approximation of the enzymatic flux cost} The linear
approximation (\ref{eq:linAppr}) can provide us with linear flux cost
weights for {\mwfFBA} or FBA with molecular crowding. As expected, the
linearised flux cost is exact for the prototype {\flow} and scaled
versions of it. For other {\flow}s, there is an approximation error
which depends on how much the catalytic rates ($v_l/\esymbol_l$,
for optimised enzyme levels) differ between $\vv$ and the prototype
{\flow} $\vv^{\rm pt}$.  For a more accurate, nonlinear approximation
of the flux cost $\acostenz(\vv)$, we use several prototype {\flow}s
$\vv^{(\alpha)}$.  We first approximate our {\flow} $\vv$ by a convex
combination $\vv \approx \sum_\alpha \sigma_\alpha\,\vv^{(\alpha)}$ of
the prototype {\flow}s (with positive weights \co{requirement!}
$\sigma_\alpha$ satisfying $\sum_\alpha \sigma_\alpha=1$).  Using the
same weights $\sigma_\alpha$, we define a weighted average of the
inverse catalytic rates
$1/r_l' = \sum_\alpha \sigma_\alpha /r_l^{(\alpha)}$. This yields the
approximated flux cost (proof in section
\ref{sec:PrototypeApproximation})
\begin{eqnarray}
\label{eq:quadraticApproximation}
\acost^{\rm non}(\vv) \approx \sum_l \frac{\hul}{r_l'}  \,v_l = \sum_{l\alpha} \frac{\sigma_\alpha}{r_l^{(\alpha)}}\,\hul\,v_l.
\end{eqnarray}
This cost function is nonlinear in $\vv$ because the prefactors
$1/r_l'$ are weighted averages of the prototype {\flow}s' $1/r_l$,
where the prefactors $\sigma_\alpha$ depend on $\vv$.  \co{FN? This
  method resembles neural networks based on radial basis functions}
With a single prototype {\flow}, the curvature vanishes and we
reobtain the linear approximation. With more than one prototype
{\flow}, the cost function is quadratic (see section
\ref{sec:PrototypeApproximation}).  For a good approximation
(\ref{eq:quadraticApproximation}), it may be better to use fewer
prototype {\flow}s, but to choose prototype {\flow}s that that
{\favour} similar metabolite profiles as the {\flow}s at which the
cost function will be evaluated.

\co{\subsection{Non-enzymatic reactions}

  In models with non-enzymatic reactions, the flux set may be
  non-convex, and the flux cost functions may be non-concave.}  \co{To
  see this, we may assume that non-enzymatic reactions come at a small
  (epsilon) enzyme cost?  then the entire FCM logic would still apply;
  we just need to assume that these reactions cannot be shut off; but
  this means: lower bounds on fluxes (approximately chosen ..)), and
  therefore non-elementary polytope vertices. If these bounds are
  chosen correctly, and if the costs are approximated by epsilon
  costs, the results should approximate the true solution.}
\co{Stefan also has some general statement about the optimal {\flow}s;
  but we don't.}  \co{\emph{Models with dilution} (e.g.~EVEN
  INFINITELY LOW DIFFUSION WOULD lead to zero value if the met is not
  reproduced!)}  \co{same as before AND: changing fluxes change c
  linearly (= x nonlinearly) for each single diffusion react, metab
  pol yields an upper bound on flux pol.  but there are evne stricter
  restrictions; (will they be linear in fluxes?)  the flux cost
  function itself will change in any case (may not be strictly concave
  anymore) discuss the case where efm may be dependent bcs metab in
  one is allos act of the other (and itself degraded!)}

\section{Mathematical properties of flux cost function}

\co{to see whether enzymatic flux cost functions (on the flux
  polytope) are strictky concave, we need to determine whether the
  M-cost (on the M-polytope) is strictly convex}

In this section, I study the conditions under which an enzymatic
metabolite cost (or ``briefly enzymatic M-cost'') functions have a
unique optimum. Then, I show that enzymatic flux cost functions are
concave on the F-cone and strictly concave on the interpolation
line between \statedistinct\ {\flow}s.

\subsection{Shape of the enzymatic metabolite cost function}
\label{sec:SIEnzMetCostHessian} 

\co{call the ECM problem and M-cost function ``regular'' and
  ``singular'' (also in main text). Passt gut zu regularisation}
\co{HIER UND IN MAIN TEXT: einmal klar sagen: for ... models holds: if the
  Hessian is regular in one point of the M-polytope, it is regular on
  the entire M-polytope, ie the enzymatic M-cost is strictly convex}

\textbf{\ \\Conditions for a strictly convex metabolite cost function}
An enzymatic M-cost function is called ``regular'' \co{use this
  everywhere in text} if it is strictly convex on the metabolite
polytope. Generally, in a given point of the M-polytope, the enzymatic
M-cost will be positively curved in some directions and constant in
others. These ``cost-neutral'' directions depend on network structure
and rate laws in the model. To study this further, we consider
\emph{cost curvature matrix} $\Hmat$, the Hessian matrix of the
enzymatic M-cost function. If $\Hmat$ is positive definite
(i.e.~regular) in all points of the M-polytope, the enzymatic M-cost
is strictly convex and the ECM problem has a unique solution. By
contrast, if $\Hmat$ has vanishing eigenvalues (i.e.~it is singular),
the enzyme cost is non-strictly convex and the ECM problem has a
larger set of optimal solutions. To see whether $\Hmat$ is regular, we
need to determine the cost-neutral subspace of our model.

\textbf{Shape of the enzymatic metabolite cost function} \co{teilweise
  wd mit vorigem abschnitt: sort + merge!}  Each feasible {\flow}
$\vv$ defines a polytope of feasible metabolite profiles, and each
metabolite profile $\lncv$ in this M-polytope defines an enzyme
profile, and thus metabolic state.  The enzymatic M-cost function on
the M-polytope further defines which of these states are optimal.
There are two cases: in the regular case, an M-cost function is
strictly convex, with a regular cost curvature matrix on the entire
M-polytope, leading to a single optimal metabolite profile. In the
singular case, some directions in metabolite space are cost-neutral
(i.e.~the cost curvature matrix has a non-empty nullspace), leading to
a continuous set, or subpace, of optimal metabolite profiles. What
case we are in depends on network structure and rate laws considered.
As a criterion, we search for ``cost-neutral'' metabolite variations
$\delta \lncv$ that leave enzymatic M-cost unchanged.

 \textbf{Cost-neutal variations and the nullspace of the cost
   curvature matrix} Let us have a closer look at the enzymatic M-cost
 $\enzymemetcost(\lncv)$ and its curvature matrix $\Hmat(\lncv)$. The nullspace of $\Hmat$
  is called $\mbox{Ker}(\Hmat)$.  In
 models with factorised rate laws, $\Hmat$ has the following
 properties:
\begin{enumerate}[leftmargin=5mm]
\item $\Hmat$ is positive semidefinite (proof in SI \ref{sec:ProofStrictlyConcave}).
\item Given a {\flow} $\vv$, the cost curvature matrix $\Hmat$ in a point
  $\lncv$ is a sum of matrices for the individual reactions
  $\Hmat(\lncv;\vv) = \sum_l \Hmat^{(l)}(\lncv;\vv)$. Each of these
  matrices is positive semidefinite (because the enzymatic metabolite
  cost for each reaction is convex), and so the nullspace of $\Hmat$
  is the intersection of these nullspaces of the matrices
  $\Hmat^{(l)}$.  If this intersection is empty, $\Hmat$ will be
  regular.  This is the case in all models with CM rate laws (proof in
  section \ref{sec:SICMpositivelycurved}).
\item The nullspace of $\Hmat$ is structurally determined,
  i.e.~determined by rate laws, network structure, and the
  {\fluxpattern} of $\vv$ (with strictly defined zero values). It is
  independent of the specific choices of $\lncv$ and $\vv$.
\item Let $\vvA$ and $\vvB$ be  flux modes with identical  {\flow}
  patterns, and $\lncvA$ and $\lncvB$ be the {\favoured} metabolite profiles of
  $\vvA$ and $\vvB$. Then the cost curvature matrices
  in the points $(\vvA,\lncvA)$ and $(\vvB,\lncvB)$ have identical
  nullspaces.
\item In models with reversibility-based rate laws \cite{nfbd:16}, 
  $\mbox{Ker}(\Hmat)$ is  the nullspace of the transposed
  stoichiometric matrix $\Nmat\trans$; in models with saturation-based
  rate laws, $\mbox{Ker}(\Hmat)$ is the nullspace of the
  molecularity matrix ${\Mmat^{\rm S} \choose \Mmat^{\rm P}}$; in
  models with CM rate laws, $\mbox{Ker}(\Hmat)$ is empty, i.e.,
  $\Hmat$ is regular (proof in section
  \ref{sec:SICMpositivelycurved}). \co{elad says: adding a z-score objective solves this problem for all cases, right?}
\end{enumerate}

\textbf{Conditions for singular enzymatic M-cost functions} Under what
conditions is the optimal metabolite profiles non-unique?  Two
metabolite profiles $\lncvA$ and $\lncvB$ are called cost-equal if
their difference vector $\lncvA-\lncvB$ is cost-neutral, i.e.~located
in the cost-neutral subspace (i.e.~the nullspace of $\Hmat$). As a
criterion for metabolite profiles being cost-equal, we introduce
another helpful concept, the notion of be rate-equal, \co{ueberall
  einfuehren} \co{see Proofs section; kurz sagen: wie verhaelt sich
  cost-equal zu thermodynamically equal, mass-action equal, and
  kinetically equal}

\co{DEFINITION: rate-equal metabolite profiles} Given the enzyme
concentrations $\esymbol_l=1$, a metabolite profile $\lncvA$ yields the flux
distribution $\ratev(\lncv,\onevec)$ (which may be non-stationary).
We now ask whether moving from $\lncvA$ to $\lncvB$, at fixed enzyme
levels, changes this flux distribution or not. If there is no change,
$\lncvA$ and $\lncvB$ are called rate-equal, otherwise they are
called rate-distinct. Rate-equal metabolite profiles are
cost-equal. \co{proof: if the fluxes remain unchanged at
  $\uv=\onevector$, they also remain unchanged at any $\uv$. Thus, a
  chane from $\lncvA$ to $\lncvB$ at given fluxes does not change
  $\uv$, and therefore does not change $hcost$.}

Using this notion, we obtain the following results:

\begin{enumerate}[leftmargin=5mm]
\item If a metabolite (with index $i$) has no influence on any of the
  reaction rates.  In this case, any variation vector of the form
  $\delta \lncv = (0,0, .., \delta \lnc_{i}, 0, 0 )\trans$ is a
  nullvector of $\Hmat$.
\item If the rate laws do not directly depend on metabolite concentrations,
  but on the thermodynamic forces, which are given by
  $\theta_l = \ln k^{\rm eq}_{l} - \sum n_{il} \, \ln c_i$. \co{elad:
    add that this is given by exp(sum ...)} If the rate laws depend on
  these forces only (which holds, e.g.~for the ``reversibility-based''
  rate laws in \cite{nfbd:16}), then any variation $\delta \lncv$ that
  leaves the driving forces unchanged is a nullvector of $\Hmat$.
  Such vectors $\delta \lncv$ are given by the nullvectors of
  ${\Ntot}\trans$ (i.e.~conserved moiety vectors).

\item If the rate laws do not  directly depend on metabolite concentrations,
  but only on  mass-action products. The mass-action products 
  are given by $\prod_i \,c_i^{m^S_{li}}$ (for reaction substrates) or $\prod_i
  \,c_i^{m^P_{li}}$ (for reaction products), with molecularities in the matrices $\Mmat^S$ and
  $\Mmat^P$ (e.g.~the saturation-based rate laws in
  \cite{nfbd:16}). Variations $\delta \lncv$ that do not affect the
  mass-action products are nullvectors of $\Hmat$.  Such vectors
  $\delta \lncv$ are given by the nullvectors of ${\Mmat^S \choose
    \Mmat^P}$.

\item If there exist variation vectors $\delta \lncv$ that do not
  affect any reaction rates, the enzyme cost function of a reaction
  $l$ defines, in each point of the M-polytope, a cost curvature
  matrix $\Hmat^{(l)}$ with a nullspace $\mbox{Ker}(\Hmat^{(l)})$. The
  nullspace of $\Hmat$ (for the sum of all enzyme costs) is the
  intersection of all these nullspaces. A metabolite variation that is
  cost-neutral for all reactions is therefore a nullvector of
  $\Hmat$. In contrast, if any possible variation $\delta \lncv$
  affects at least one reaction, then $\Hmat$ is regular.
\end{enumerate}

\co{the first three results refer to the rate law types defined in the ECM paper}
\co{In other words, cost functions of type ECF 2 (NAMING!) are nonlinear for any
  direction that affects a thermodynamic force; in eMC3 and eMC4,  (NAMING!)  they
  are nonlinear for any direction that affects a thermodynamic force,
  or that corresponds to the substrates or products of a reaction
  (this is still to be proven; maybe Joost has the proof already)}

To summarise, nullvectors of the cost curvature matrix can arise
\emph{structurally} and depedn only on rate laws and network
structure.  For models with factorised rate laws (and with non-constant
thermodynamic and kinetic efficiency factors), the entire nullspace of
the cost curvature matrix is determined  in this way.  This is stated by the
following lemma (proof in SI section \ref{sec:prooflemma1}).

\begin{lemma}\label{lemma1}
  \textbf{The nullspace of the enzymatic metabolite cost curvature
    matrix is given by the cost-neutral subspace} In models with
  factorised rate laws and non-constant thermodynamic and kinetic
  efficiency factors, all neutral metabolite variations $\delta \lncv$
  are nullvectors of the cost curvature matrix $\Hmat$. Conversely,
  all nullvectors of $\Hmat$ are neutral metabolite variations. This
  holds in any point $\lncv$ of the M-polytope. A nullvector of
  $\Hmat$ in a point $\lncv$ is also a nullvector of $\Hmat$ in any
  other point.
\end{lemma}

According to this lemma, the nullspace of the cost curvature matrix
for a given kinetic model (i.e.~network structure, rate laws, and
model parameters) and a predefined {\fluxpattern} is structurally
determined: it is independent of the metabolite profile $\lncv$ and
the {\flow} $\vv$. If the cost-neutral subspace is empty, then according
to Lemma \ref{lemma1} the enzymatic M-cost is regular (i.e.~strictly
convex on the M-polytope) and the ECM problem has a unique
solution. This holds for all models with common modular rate laws
because their cost-neutral subspace is empty in fact (proof in section
\ref{sec:SICMpositivelycurved}).

\subsection{Enzymatic flux cost functions are concave on the F-cone}
\label{sec:SIproofElementary}

\co{hier die sache mit der erweiterung und raendern erwaehnen: sagen,
  dass es ab jetzt um die erweiterte funktion geht; erklaeren, was mit ``inside'' gemeint ist}
\co{what about open and closed polytopes? EXTRA-abschnitt!! enzymatic M-cost only defined on partially open polytope! flux cost function defined on partially  open  F? continuation of F-cost function needed for closed polytope?}

\begin{lemma}\label{lemma2}  \textbf{Concave flux cost functions}
  (proof in section \ref{sec:prooflemma2}). We consider a cost
  function $f(\lncv;\vv)$ scoring a metabolite profile $\lncv$ (in the
  M-polytope)and a {\flow} $\vv$ (in the F-cone). Let the cost
  function be linear in the fluxes,
  $f(\lncv; \alpha\,\vvA+\beta\,\vvB) = \alpha\,f(\lncv;\vvA) +
  \beta\,f(\lncv;\vvB)$, and bounded from below (e.g.~it is
  positive). Let $g(\lncv)$ be a cost function on the M-polytope that
  is bounded from below.  Under these conditions, the flux cost
  function
  $\acost(\vv) = \min_{\lncv \in \sPolytope} f(\lncv;\vv) + g(\lncv)$
  is concave on the F-cone $\sPolytope$. This lemma also holds if $f$
  (for each given $\lncv$) is not linear, but concave in the fluxes.
\end{lemma}

From Lemma \ref{lemma2}, we  obtain

\begin{proposition} \label{prop:FluxCostIsConcave} \textbf{(Enzymatic
    and kinetic flux cost functions are concave)} Let $\hminus(\esymbolv)$
  be a linear enzyme cost function, and let $\hmet(\lncv)$ be a
  metabolite cost function, both bounded from below. We defined the
  enzymatic M-cost
  $\hmet^{\rm enz}(\lncv;\vv) = \hminus(\esymbolv(\lncv;\vv))$.  Then the
  enzymatic flux cost
  $a^{\rm enz}(\vv) = \min_\lncv \enzymemetcost(\lncv;\vv)$ and the
  kinetic flux cost
  $a^{\rm kin}(\vv) = \min_\lncv \hmet^{\rm enz}(\lncv;\vv) +
  \metcost(\lncv)$ are concave functions on the F-cone.
\end{proposition}

\textbf{Remarks} 

\begin{enumerate}[leftmargin=5mm]
\item \textbf{``Bounded from below''} means that costs cannot got to
  $-\infty$. This holds for all positive cost functions (assuming that
  fitness terms are represented on absolute, not on logarithmic scale).
\item \textbf{Linear enzyme cost function} Lemma \ref{lemma2} holds
  also for to the enzymatic M-cost $\enzymemetcost(\lncv;\vv)$, because
  it is \emph{linear} in the fluxes (since the enzyme level appears
  linearly in the rate laws) and because  the enzyme cost function
  $h(\esymbolv)$ is assumed to linear. In contrast, if $h(\esymbolv)$ were
  nonlinear and convex, the resulting  flux cost function might not be
  concave. In case  $h(\esymbolv)$ is nonlinear and concave, Lemma \ref{lemma2}
  and proposition \ref{prop:FluxCostIsConcave} will still
  hold, but  the M-cost may no further  be convex.

\item \textbf{Non-stationary fluxes} Proposition
  \ref{prop:FluxCostIsConcave} holds for any flux distributions,
  stationary or non-stationary ones.  However, the flux distributions
  $\vvA$ and $\vvB$ and all their interpolations must be
  thermo-physiologically feasible \co{die def kommt schon. aber auch
    in main text?}  (i.e.~thermodynamically feasible given the
  metabolite concentration ranges and predefined concentrations assumed
  in our model).  A thermo-physiologically feasible {\fluxpattern}
  defines an orthant in flux space.  If non-stationary flux
  distributions are allowed, then any flux distribution in this
  orthant can be a solution.  Since the flux cost is concave on this
  orthant, it is also concave on all B-polytopes within the orthant.
\end{enumerate}

\co{WO? The condition that ``the difference $\lncvA-\lncvB$ is not
  cost-neutral'' is satisfied if (i) the difference $\lncvA-\lncvB$ is
  not kinetically neutral or if (ii) the cost curvature matrix is
  regular on the entire M-polytope, i.e.~if enzymatic metabolite cost
  is a strictly convex function. // Models with CM rate laws have a
  strictly convex enzymatic M-cost function.  Therefore, any model
  with CM rate laws has a strictly concave enzymatic flux cost
  functions on the B-polytope.}

\subsection{The enzymatic flux cost function is strictly concave  between \statedistinct\ {\flow}s}
\label{sec:SIFluxCostIsStrictlyConcave} 

Depending on the model, the enzymatic flux cost may not only be
concave, but even be \emph{strictly} concave on the B-polytope: that
is, adding two {\flow}s yields an additional positive \compromisecost.
For a cost function to be strictly concave, all {\flow}s on the
B-polytope must be {\statedistinct}.

\begin{definition}
  \textbf{{\Statedistinct} metabolic {\flow}s} Consider two {\flow}s $\vvA$
  and $\vvB$ with  the same {\fluxpattern} (i.e.~with the
  same active reactions and flux directions).  The {\flow}s are called
  \emph{\statedistinct} (for the kinetic model in question) if  $\vvA$ and
  $\vvB$ have no optimal metabolite profile in common.
\end{definition}

\textbf{Remark} If the enzymatic M-cost function is regular (i.e.~if
each {\flow} {\favour}s only a single metabolite profile), then {\flow}s
with different optimal profiles are \statedistinct.  If the
enzymatic M-cost function is singular (i.e.~if {\flow}s {\favour}
several metabolite profiles), \emph{no} metabolite profile may be
{\favoured} by both of the {\flow}s.

\co{\statedistinct: either unique, different optima; or non-unique,
  different optima; in the first case, \statedistinct!  in the second
  case, the opt metab profile must change as we interpolate between
  the {\flow}s.}

\begin{lemma} \label{prop:ConvexLeadsToKineticallyDistinct} \co{beweis
    wo?}  We consider a kinetic model with a given {\fluxpattern} \co{or ``flux sign pattern''?}
  defining an M-polytope and a benefit function 
  defining a B-polytope. If the enzymatic \co{or ``enzyme'', uea?} M-cost function  is strictly convex on the interior of the 
  M-polytope, all {\flow}s in the B-polytope are
  \statedistinct.
\end{lemma}
\co{proof: M cost strictly convex: then, each flow has exactly one optim metab profile . wie weiter??}

Next, we can state that an enzymatic flux cost
function is strictly concave on the B-polytope if (and only if) all
flows in the B-polytope are \statedistinct. 

\begin{lemma} \label{prop:FluxCostIsStrictlyConcave}
  \textbf{(Conditions for strictly concave flux cost functions)} Let
  $\acostenz(\vv)$ be an enzymatic F-cost function on an F-cone. If two
  {\flow}s $\vvA$ and $\vvB$ in the interior of this polytope \co{cone?} are
  \statedistinct, then $\acostenz(\vv)$ is strictly concave on the
  line between $\vvA$ and $\vvB$ (proof in section
  \ref{sec:ProofStrictlyConcave}).
\end{lemma}

To prove this statement, we analyse
the ECM problem in a metabolic model with given rate laws and flux
directions. We assume that  external metabolite concentrations are
fixed  and that the internal metabolite concentrations need
to be optimised.

\co{Note: if the condition holds on the F-cone, then it also holds on the B-polytope}

\begin{lemma} \label{prop:FluxCostIsStrictlyConcave2} If all {\flow}s in
  the interior of a B-polytope are \statedistinct, the enzymatic
  F-cost $\acostenz(\vv)$ is strictly concave on
  the interior of this B-polytope.
\end{lemma} 

\textbf{Remarks} 
\begin{enumerate}[leftmargin=5mm]
\item The proposition \co{where is the proposition?}  also holds for
  combinations of several {\flow}s, $\vvA, \vvB, \vvC$ ... In this
  case, instead of a line we consider their convex hull, i.e.~the
  simplex spanned by the {\flow}s.
\item \emph{Kinetic} flux cost functions (i.e.~which contain enzymatic
  and direct metabolite costs) are strictly concave under the
  following conditions: the kinetic metabolite cost $\metcost(\lncv)$
  (i.e.~enzymatic plus direct metabolite cost) must be convex, and its
  cost curvature matrix must have the same nullspace everywhere in the
  M-polytope. In this case, adding the direct metabolite cost to the
  enzymatic M-cost may change the cost curvature matrices,
  but these matrices, in the entire M-polytope, will still have the
  same nullspaces. Therefore the concavity proof in SI
  \ref{sec:ProofStrictlyConcave} still applies.
\end{enumerate}

By combining Lemma \ref{prop:ConvexLeadsToKineticallyDistinct} and
Lemma \ref{prop:FluxCostIsStrictlyConcave2}, we obtain the
following proposition.  

\begin{proposition} \label{prop:ConvexLeadsToConcave} We consider a
  kinetic model with a predefined {\fluxpattern} and a benefit
  function defining an M-polytope and a B-polytope. If the enzymatic
  M-cost is strictly convex on the interior of the M-polytope, the
  enzymatic F-cost $\acostenz(\vv)$ is strictly concave on the
  interior of the B-polytope.
\end{proposition} 

\myparagraph{Examples of flux modes with vanishing {\compromisecost}s}
In a typical model, different {\flow}s are \statedistinct, i.e.~they
{\favour} different metabolite profiles. If they are not
\statedistinct, the flux cost function varies \emph{linearly} on the
line between them.

\begin{enumerate}[leftmargin=5mm]
\item \textbf{Flux profiles that differ only by scaling} The enzymatic flux
  cost scales linearly with the {\flow}. If two {\flow}s $\vvA$ and $\vvB$
  differ only scaling, the flux cost is linear, and therefore
  \emph{not} strictly concave on the interpolation line between
  them. However, such pairs of differently scaled {\flow}s cannot occur in the same
  B-polytope
\item \textbf{Flux profiles that differ only in the usage of isoenzymes} Let
  us consider all enzyme molecules that catalyse a reaction R and
  split them arbitrarily into two pools (E1, ``blue'' enzyme
  molecules) and (E2, ``red'' enzyme molecules), thus representing our
  original reaction by two separate ``isoreactions'' with identical
  sum formulae and kinetics: R1 (catalysed by blue enzyme) and R2
  (catalysed by red enzyme). Since this splitting is arbitrary, it can
  neither change the system dynamics nor the resulting flux cost. This
  means that the flux cost remains additive between the isoreactions
  R1 and R2, and is therefore \emph{not strictly concave}.
\end{enumerate}
In these two examples, the {\flow}s $\vv_A$ and $\vvB$ {\favour}
the same metabolite profile: therefore, the interpolated {\flow}s can
be optimally realised by linearly interpolating the enzyme
levels. With a linear cost function $\hminus(\esymbolv)$, this implies an
additive cost between $\vvA$ and $\vvB$.  But what if the optimal
metabolite profile of a {\flow} is not unique, and therefore two
{\flow}s may be (but do not have to be) realised by the same  profiles -- will
strict concavity still hold? \co{die naechsten beiden punkte lieber
  hoch zu non-uniqueness?}
\begin{enumerate}[leftmargin=5mm]
\item \textbf{Flux profiles that {\favour} kinetically equal
    metabolite profiles.}  In some models, there exist different
  metabolite profiles $\lncvA$ and $\lncvB$ that yield exactly the
  same reaction rates (with enzyme levels kept constant), and this may
  even hold for any metabolite profiles on the line between $\lncvA$
  and $\lncvB$.  In this case, the solution of the ECM problem will be
  non-unique. Here is an example: consider a chain of reactions A
  $\Leftrightarrow$ B + C $\Leftrightarrow$ D with reversible
  mass-action kinetics. If [B] is scaled by a factor 10 and [C] is
  scaled by 0.1, the reaction rates remain constant. And due to the
  logarithmic scale in the metabolite polytope, any convex conbination
  of the two metabolite profiles yield the same rates (and in ECM the
  same enzymatic M-cost). In fact, the enzymatic M-cost cost is
  constant on an entire subspace in log-metabolite space, as indicated by
  a nullvector of the cost curvature matrix of
  $\enzymemetcost(\lncv;\vv)$.  \co{Certain rate laws enforce such
    nullvectors, which can be determined by analysing the network
    structure.}
\item \textbf{Metabolites that do not affect the reaction rates} As an
  approximation, we may consider models in which some reaction rates
  do not depend on any metabolite concentrations. This holds, for example, if
  enzymes are assumed to be substrate-saturated and completely
  forward-driven, an approximation made by the \emph{capacity-based
    rate laws} \cite{nfbd:16}. If such rate laws are used, even small
  enzyme perturbations can completely impair a steady state.  In such
  models, it is also likely that some metabolites have no impact on
  any reaction rate.  A similar case occurs when metabolites formally
  belong to a model, but do not participate in any reactions.
\end{enumerate}

In these two examples, two {\flow}s \emph{can} {\favour} the
same metabolite profiles, but also other metabolite profiles because the optimal
metabolite profiles are not
unique. In this case, when interpolating between two {\flow}s, one could
obtain different metabolite profiles and still realise the flux
changes by linear changes in enzyme levels (i.e.~at varying metabolite
levels, but constant catalytic rates!).  Therefore, whenever two {\flow}s
\emph{can} {\favour} the same metabolite profiles, we define them
as not \statedistinct.

\subsection{The optimum points of concave flux cost functions are polytope vertices}
\label{sec:SIpolytopeVertices}

\begin{figure*}[t!]
  \begin{center}
    \includegraphics[width=15.5cm]{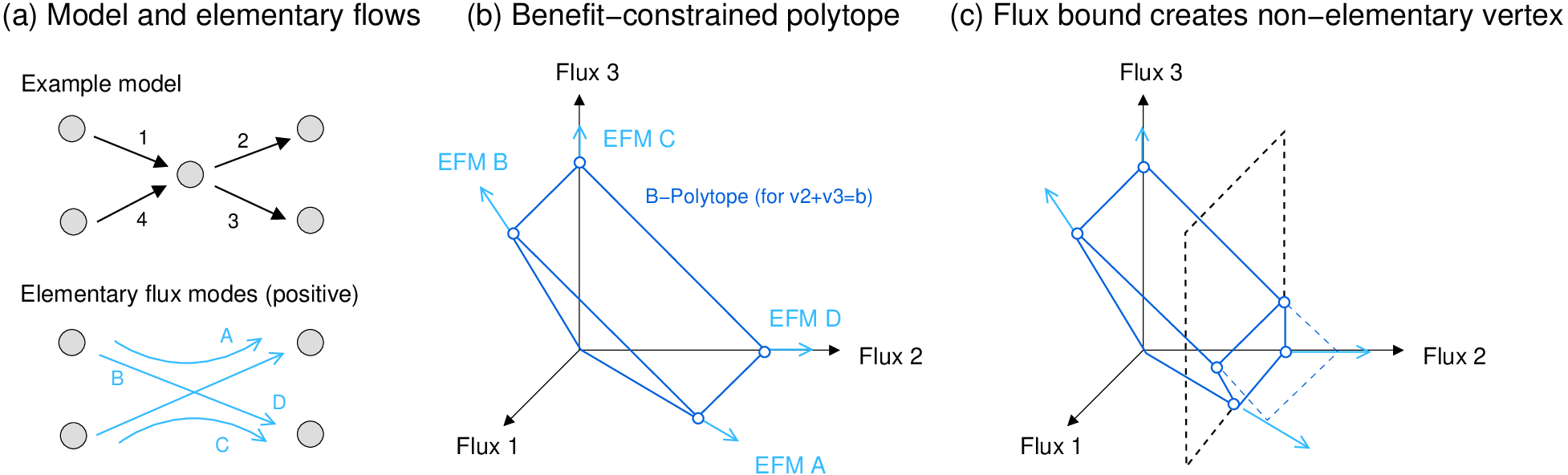}
    \caption{\co{REF in MAIN ARTICLE} Elementary and non-elementary B-polytope
      vertices. (a) Simple branch point model (top) with four
      elementary flux modes (EFMs; bottom); in the model all fluxes
      are constrained to be positive. (b) Metabolic {\flow}s as points in
      flux space (three of the four dimensions are shown, referring to
      reactions 1, 2, and 3). The cone of feasible {\flow}s is spanned by
      the four EFMs. A predefined benefit value (in the example,
      $v_2 + v_3=\const$) defines a plane that intersects the cone and
      yields the B-polytope (blue rectangle) or feasible {\flow}s. The
      vertices of this B-polytope correspond to EFMs. (c) If we add an
      upper bound on $v_2$, some of the {\flow}s are discarded; a part of
      the B-polytope (red rectangle) is cut off and two non-elementary
      vertices emerge (blue circles).}
  \label{fig:new_vertices_3d}
  \end{center}
\end{figure*}

\co{CHECK FIRST two sentences!}
If all {\flow}s in a B-polytope are \statedistinct, the optimal
(i.e.~enzyme-cost-minimising) {\flow} must be a polytope vertex if
this optimum unique. Otherwise, at least one of them must be a
polytope vertex (and they all lie on a line, plane, or
hyperplane). For the proof, we regard the F-cone as a closed set and
continue the flux cost function -- which is continuous inside the
F-cone -- to flux cone boundaries, including all vertices. After a
proof for this function, we will handle the remaining cases in which
the flux cost function is discontinuous between an F-cone and one of
its boundaries.  \co{im haupttext kurz auf dieses problem verweisen!}

\textbf{A concave flux cost function has an optimum point in a
  polytope vertex} The enzymatic flux cost function on a B-polytope  has a global 
minimum  in one of the polytope vertices. The same minimal value
can occur in non-vertex points. We can see this as follows. Any
polytope point $\vv$ can be written as a convex combination
$\sum_\alpha \sigma_\alpha \vv_{(\alpha)}$ of the vertices.  If the
flux cost function is concave on the B-polytope, the cost
$\acost(\vv)$ in a point $\vv$ must be at least as high as the
combined cost $\sum_\alpha \sigma_\alpha \acost(\vv_{(\alpha)})$ of
the vertex points, and this combined cost must be at least as high as
the lowest cost $\mbox{min}_\alpha \acost(\vv_{(\alpha)})$ among
the vertices. Thus, if an inner point $\vv$ achieves the minimum cost,
there must also be a vertex point with the same minimum cost.

\textbf{Strictly concave flux cost functions have minimum points only
  on polytope vertices} If the enzymatic flux cost function is
strictly concave on the B-polytope, all its minimum points must be
polytope vertices.

\textbf{A complication: discontinuity at polytope boundaries} \co{move
  this into proof, explain everything clearly!}  Above, we assumed
that the enzymatic flux cost is continuous on the entire (closed)
F-cone, \todo{including}\co{??} boundary points such as
vertices. However, for proving that flux cost functions must be
concave, we consider {\flow}s with identical {\fluxpattern}s 
(including zero values).  Geometrically, these proofs apply only to
the interior of an F-cone, which is an open set\footnote{To clarify
  this point, given a {\fluxpattern} $\gamma$, we can define the
  ``strict'' F-cone, in which the fluxes must strictly obey the
  {\fluxpattern} (no extra zeros allowed); it is an open set, because
  the boundaries of this polytope (where some extra fluxes become 0)
  are excluded; we can also define the ``loose'' F-cone, in which
  fluxes are allowed to become zero; in contains the boundaries and is
  a closed set.}, and may not hold for points on the polytope
boundary. In fact, we saw that flux cost functions show jumps at the
polytope boundaries, i.e.~if two neighbouring polytopes share a vertex
point, then for one of them the cost function will not be continuous
in this point.  \todo{That is, we may  have to compare a {\flow}
from the interior of an F-cone to a {\flow} on the polytope boundary,
which already shows a jump.  Unfortunately, the conclusion about
minimum points of a flux cost function concerns polytope vertices,
which are boundary points! Thus, to claim that vertex points are
optimal, we need to reformulate the problem: we consider the flux cost
function on the interior of the F-cone and define the continuous
continuation of this function, which includes all points on the
polytope boundary.}  According to our previous argument, this continued
function will have its minima in polytope vertices!  Now it is true
that the actual cost function value in a vertex point may differ from
the value of the ``continued'' function. But for optimality reasons,
the actual cost function cannot be higher than the continued function,
so the vertex point will achieve an even lower minimum value.
Therefore, our conclusions about vertex points being optimal still
hold.

\subsection{Condition for locally optimal {\flow}s}
\label{sec:SILocallyOptimalFlows}

\co{which part of the text about  locally optimal {\flow}s could be moved to cba theory?}


\begin{definition} 
\textbf{Locally optimal {\flow}.} A {\flow} is  \emph{locally optimal}
if it has a lower (or equal) flux cost than any other {\flow} in a small region
around $\vv$ within the B-polytope. 
\end{definition} 

\begin{proposition} \textbf{(``Manu'a criterion'')}
\label{prop:segaulaCriterion}
A B-polytope vertex $\vv$ is locally optimal if it has a unique
{\favoured} metabolite profile $\lncv$ and if it is the only
{\favoured} {\flow} of this profile $\lncv$ (proof in section
\ref{sec:ProofCriterionLocallyOptimalFlow}).
\end{proposition}

\textbf{Remark} Proposition \ref{prop:segaulaCriterion} suggests a
construction algorithm for locally optimal states (``Manu'a
algorithm''): (i) Start from some initial {\flow} (e.g.~obtained by
{\mwfFBA}). (ii) Use ECM to compute the {\favoured} metabolite
profile. (iii) Compute the catalytic rates, use them to define a
linear FCM problem, and solve it for the optimal {\flow}. (iv) Iterate
steps number (ii) and (iii), until convergence. \todo{\co{das auch im
    main text?}In the end}, check the resulting {\favoured}
metabolite profiles and {\flow}s for uniqueness. The flows computed by
{\mwfFBA} are likely to be sparse (in models without flux constraints,
they will be EFMs); therefore, some metabolite concentrations (e.g.~within
inactive pathways) may be undetermined. This non-uniqueness can be
resolved by regularisation during the ECM steps.

\co{MANUA ALGORITHM For small models, to run the calculation
  efficiently, we can proceed as follows.  We enumerate all
  EFMs. Then, for each EFM, we compute the optimal metabolite profile
  and the resulting catalytic rates $r_l(\lncv^{\rm
    opt})$.
  \co{ALLGEMEIN BETONEN, DASS DIE WICHTIG SIND! IN INTRO //
    DISCUSSION!}  Using these catalytic rates, we compute the cost
  matrix \co{b statt a uea}
  $a_{kl} = \metcost(\lncv^{\rm opt,k},\vv^{\rm EFM l}) = \sum_j h_j
  v^{\rm EFM, l}_j\,r_l(\lncv^{\rm opt,k})$
  between optimal metabolite profiles (rows) and all EFms (columns).
  The matrix element $b_{kl}$ describes the enzymatic cost of EFM $l$,
  evaluated with the {\favour}ed metabolite profile of EFM $k$.  To
  compute the matrix, we first define the matrix
  $R_{kj} = 1/r_j(\lncv^{\rm opt, k})$, the diagonal matrix \co{bold
    face} $H = \diag(\hv)$ containing the enzyme cost weights, and the
  EFM matrix $V$ (with EFMs as columns).  Our cost matrix is then
  given by $B = R\,H\,V$.  By construction, the diagonal elements of
  $B$ are the (unique) minimal values within their columns.  If a
  diagonal element is also the (unique) minimal value within its row,
  it corresponds to a locally optimal state.}  \co{how can one find
  the basin of attration of an EFM by using this matrix?}

\begin{figure*}[t!]
  \begin{center}
 (a) Schematic explanation\hspace{2cm} (b) Simulation results for branch point model\\[2mm]
    \parbox{6cm}{\includegraphics[width=3.5cm]{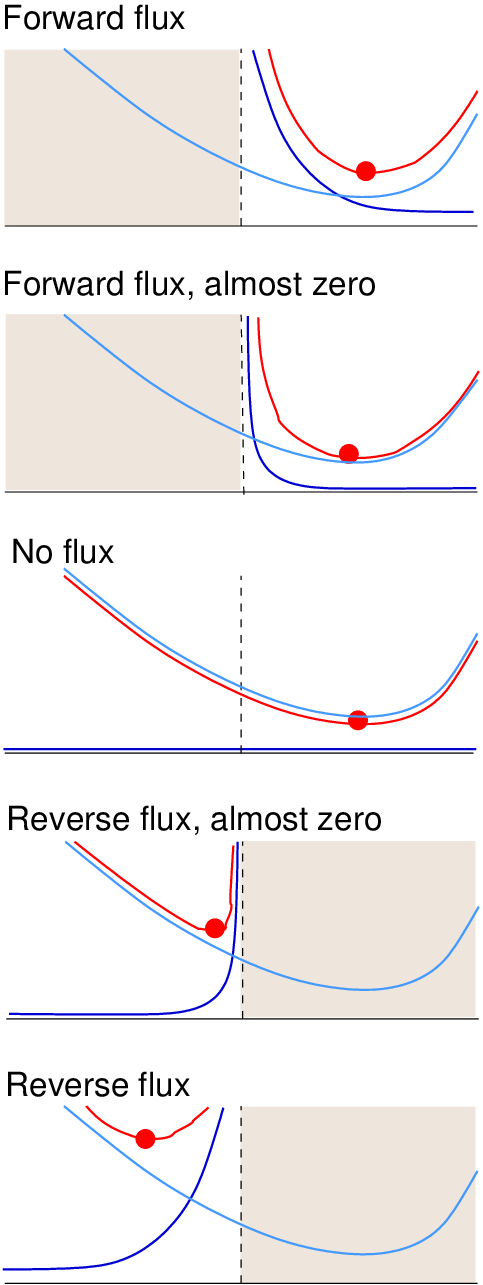}}
    \parbox{6cm}{
      \includegraphics[width=6cm]{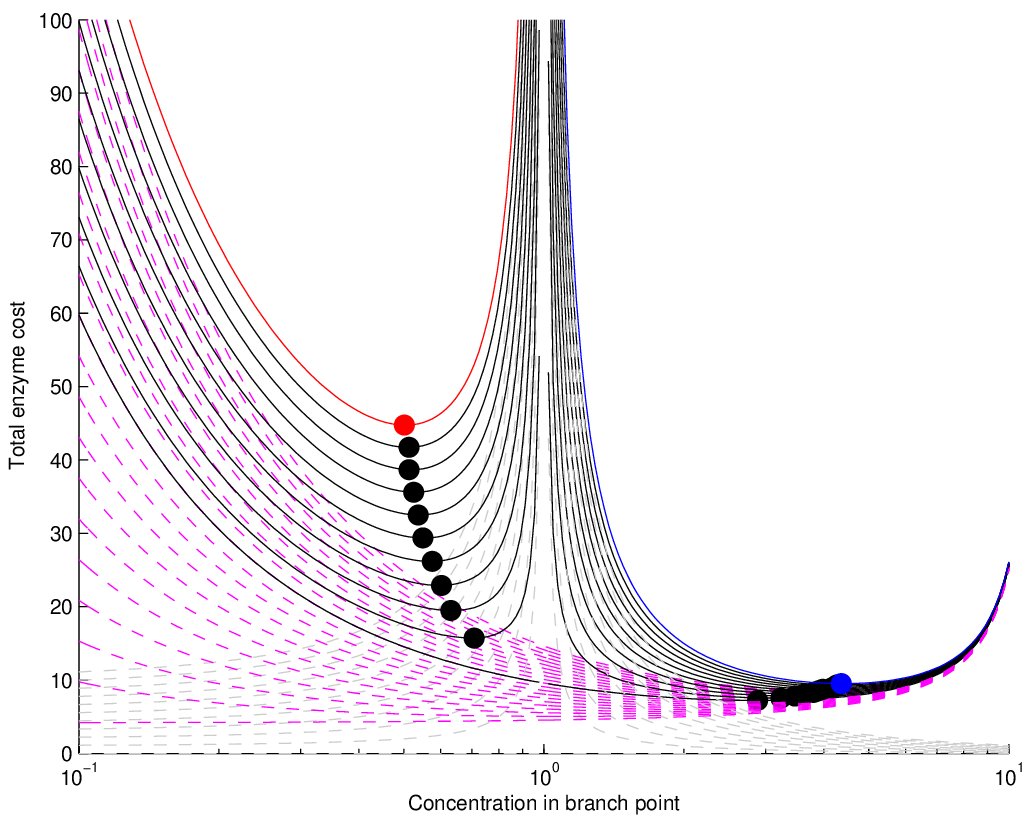}\\
      \includegraphics[width=6cm]{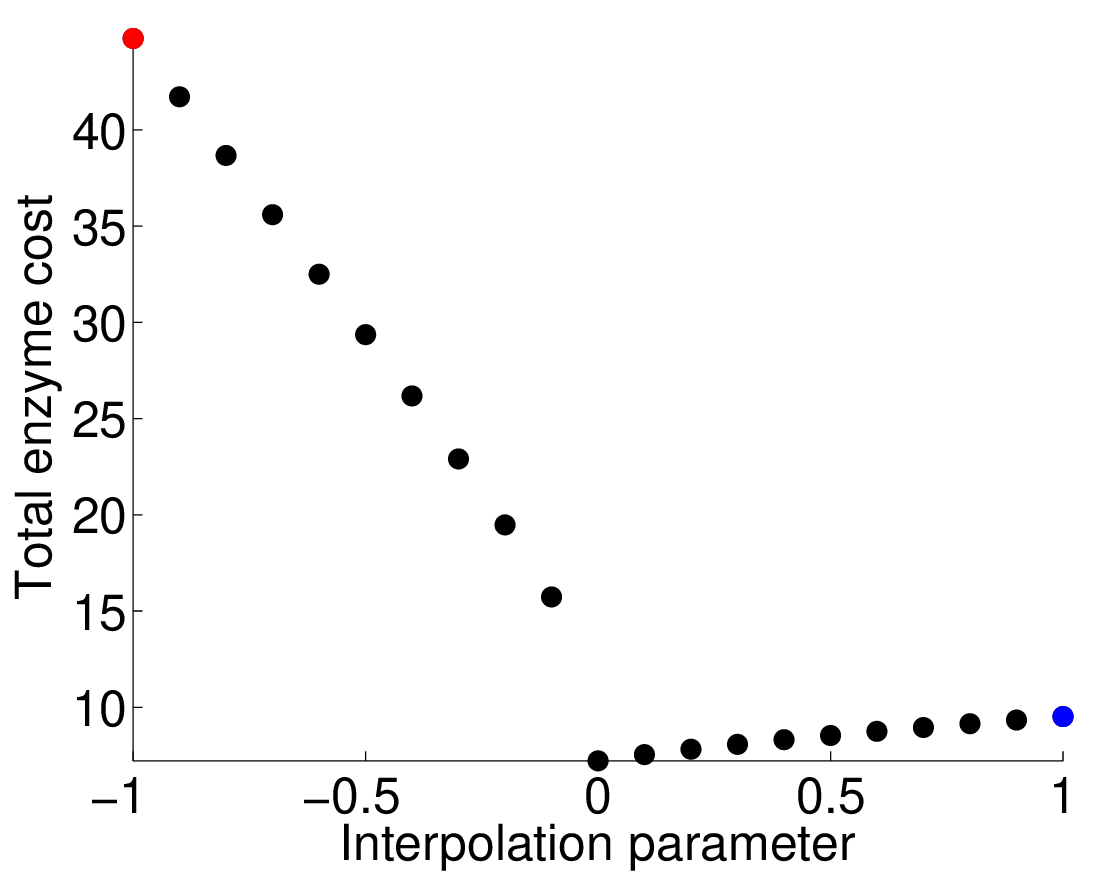}
    }
  \end{center}
  \caption{For a flux to change its direction, metabolite and enzyme
    levels must change abruptly.  (a) Schematic explanation.  Curves
    show an enzymatic M-cost function (y-axis) for one of the
    metabolites (log-concentration on x-axis) ). The dark blue curve
    shows the enzymatic cost of the reaction that changes its flux
    direction.  The light blue curve shows the sum of all enzymatic
    costs in other reactions.  Their sum (red curve) determines the
    optimal metabolite concentration (dot, x-coordinate) and the optimal
    enzyme cost (dot, y-coordinate). If the flux is positive (top
    diagram), low metabolite concentrations are thermodynamically impossible
    (shaded region). In the plots below, the flux decreases and
    changes to negative values. The cost of this reaction changes
    accordingly, while the other costs remain almost constant (a
    realistic assumption, if we consider {\flow}s close to a flux
    reversal).  Exactly at the flux reversal, our reaction has a zero
    flux and no cost, and the optimal metabolite concentration is determined
    by the other reactions. In the following state, with an
    (infinitesimally small) negative flux, the metabolite concentration must
    be \emph{below} the threshold value. This requires a jump in the
    metabolite concentration, and therefore in enzyme levels and enzyme
    cost. (b) Simulation results from the branch point model (see
    Figure \ref{fig:branch_point_flux_reversal}), showing the same effect. \co{fix colors; more
      details}}
  \label{fig:branch_point_flux_reversal2} 
\end{figure*}

\subsection{Flux cost close to a flux reversal}

\co{most likely, boundary case belongs to one of the sides} \co{a
  steady transition is unlikely (it would require that the OPTIMUM
  metabolite profile at the flux reversal (flux=0) is exactly a
  profile at which the reaction would be in equilibrium anyway,
  i.e.~exactly the thermodynmic threshold value} \co{sagen, dass glatte
  (quadratische) funktion als naeheerung doof waere (schon gesagt in
  main text, wo?}

\co{say that a flux reversal may occur in glycolysis, or in the
  excretion and uptake of ethanol in yeast}

We saw that the enzymatic flux cost function is concave on each
F-cone, i.e.~for a given flux pattern.  What shape will the cost
function have on the entire flux space? If any feasible
{\fluxpattern}s are allowed, we merge all feasible F-cones and obtain
a non-convex polytope and an F-cost on this set. What happens at the
boundaries, where F-cones touch, and where fluxes change their
directions? Will the cost function remain concave, smooth, or at least
continuous?

\co{wo? say: F-cost is piecewise concave}

\myparagraph{The flux cost function near the point thermodynamic
  equilibrium state $\vv=0$} For each flux mode, the enzymatic flux
cost increases proportionally with the {\flow}, so in the point
$\vv=0$ the flux cost increases in all directions with non-zero
slopes. If we approach this point from different directions, we obtain
different gradients, so the flux cost is not differentiable in this
point. Moreover, the kinetic flux cost function (which contains direct
metabolite costs) is not even continuous in this point; it converges
to different values when approaching $\vv=0$ from different
directions.  \co{vorher schon argument ``linear scaling supports
  linear approximation!'' hier: bestaetigung. aber was ist dann weiter
  unten mit der quadratischen approximation?  sollte die nur auf dem
  B-polytop gelten?, und deann weiter lienar skaliert werden?}  This
shows that that quadratic flux cost functions such as
$\sum_{l} v_{l}^{2}$, are not very realistic, \co{PUT references of
  people who do that} and that piecewise linear cost functions,
i.e.~linear flux cost functions on each F-cone, are a better choice.

\co{uea pareto: multi-objective}

\begin{figure*}[t!]
  \begin{center}
    \begin{tabular}{l}
      \includegraphics[width=16.5cm]{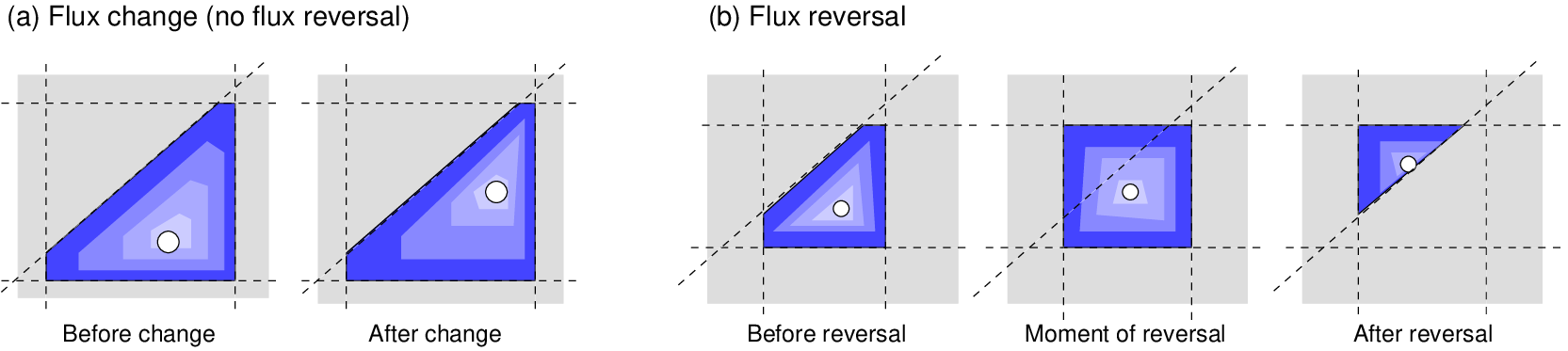}
  \end{tabular}
  \caption{\co{moment of reversal: reversal point} Metabolite profiles
    during flux reversal. The drawings show enzymatic cost functions
    on the metabolite polytope (schematic examples).  Physiological
    and thermodynamic constraints are shown by dashed lines, enzymatic
    M-cost by colours from bright (low) to dark (high).  Compare
    Figure \ref{fig:branch_point_flux_reversal2} for a one-dimensional
    example.  (a) A flux change without flux reversal leaves the
    metabolite polytope unchanged. The shape of the cost function
    changes, and the optimum point moves. (b) In a flux reversal, a
    flux changes from forward direction to a zero value and further to
    backward direction, each time defining a different M-polytope. At
    the point of flux reversal (centre), the cost barrier disappears,
    but the optimum point remains in the lower sub-polytope.  After
    the reversal, at a slightly negative flux, this optimum becomes
    inaccessible and another point, near the boundary of the new
    M-polytope, becomes optimal. At larger negative fluxes, the
    optimum moves further away from the boundary (not shown).}
  \label{fig:fluxReversal}
  \end{center}
\end{figure*}

\myparagraph{Flux reversal in a simple branch point model} For a
reaction flux to change its direction, the mass-balance ratio of
substrates and products must change. This can be achieved by a change
in external metabolite concentrations (i.e.~model parameters) or by changes in
internal metabolite concentrations (which follow from enzyme adaptations to
other parameter changes). To study the shape of the flux cost function
at the polytope boundaries, we consider again our branch point model
from Figure \ref{fig:branch_point_flux_reversal}.  If all three
reactions are active, there are $2^3=8$ possible {\flow} patterns,
corresponding to the eight orthants in flux space.  The plane of
stationary fluxes intersects six of these orthants.  The other two
orthants can be discarded, because the {\fluxpattern}s in these
orthants would lead to accumulation or depletion of the branch point
metabolite. The flux directions are directly determined by chemical
potential differences: e.g.~a positive flux $v_1$ in Figure
\ref{fig:branch_point_flux_reversal} requires that $\mu_X>\mu_c$. For
simplicity, let us assume that $\mu_x>\mu_y>\mu_z$ (this only depends
on the external metabolite concentrations).  \co{merge the next sentences} Two
possible choices of $\mu_c$ lead to stationary {\flow}s: if $\mu_c$ is
lower than $\mu_x$ and higher than $\mu_y$, all fluxes run in forward
direction; and if it is lower than $\mu_x$ and higher than $\mu_y$,
the flux $v_2$ is reversed. Thus only two of the orthants remain
feasible (blue or red), and the choice between them depends on the
chemical potential $\mu_c$. The flux cost is a function on
one of the two coloured triangles. With a predefined objective
(e.g.~flux $v_3$), we obtain the two B-polytopes (thick black line in
Figure \ref{fig:branch_point_flux_reversal2}). Within each part of the
line (blue or red), the cost function is concave.  Between the two
polytopes, the flux cost function shows a jump (see Figures
\ref{fig:branch_point_flux_reversal2} and \ref{fig:fluxReversal}).
Between two F-cones, the enzymatic flux cost function is not
continuous, except for some unlikely cases (requiring fine-tuned model
parameters). Also the metabolite concentrations, enzyme levels, and enyme cost
show jumps in their values. The jumps in the optimal metabolite and
enzyme levels, as we move smoothly between the polytopes, indicate a
zeroth-order phase transition.

\co{discuss usage of absolute value in FBA with minimal fluxes; claim
  that different prefactors should be used for different sign patterns
  (simple approximation: per reaction; better approximation: for each
  pattern!) FIGURE!}  \co{weil andere flussrichtungen andere metab
  profile und damit andere typische fluss-spec enzymekosten nach sich
  ziehen} \co{sprung in gradienten heisst: nicht konkav!  nullen
  werden bevorzugt; deshalb lieber EFMs! spricht auch fuer
  flussminimierung} \co{What are the interesting questions regarding
  flux reversals?  Can we say anything about regulation?}

\co{Hoch? leichte WD mit oben! oder in main article?} \todo{Generally, a
flux reversal requires a change of the thermodynamic driving force
(and, thus, of metabolite concentrations). At the point of reversal, the
metabolic {\flow} crosses the boundary between two F-cones and moves
to another segment of flux space. At the same time, in metabolite
space the feasible metabolite polytope flips along the boundary
associated with the reversed reaction (see Figure
\ref{fig:fluxReversal}) and the metabolite profile jumps to a new
polytope.}

\subsection{The trajectory of optimal metabolite profiles for a sweep between two metabolic {\flow}s}
\label{sec:ProofTrajectory}

\co{note that the gradient of function qA, in point cB, has a positive
  projection on the line cA -> cB; this means, it is in one
  half-plane, pointing away from cA. This holds for every point, and
  leaves different possibilities for the trajectories in different
  points.}  \co{note that the formula Fxxinv gx holds in every point
  of the trajectory! If the hessians were circular, gradient and
  trajectory tangent would be parallel. If the Hessian is
  non-circular, it relates the tangent direction to the gradient
  direction}

\co{SCHREIBEN UND BILD. BEZUG ZU URIS PARETO KLARMACHEN. Trajektorien:
  funktionen fA und fB (strikt konvex) mit optimalpunkten xA und
  xB. zeichne höhenlinie von Fb, die xA schneidet und höhenlinie von
  FA, die xB schneidet. trajektorie muss in der "linse" dazwischen
  sein. beweis: jeder punkt auf der trajektorie ist pareto-optimal;
  jeder ein punkt ausserhalb der linse waere aber dominiert von den
  punkten xA und xB.  Die gleich logik kann angewendet werden auf
  mehrere punkte (endpunkte und ein paar punkte auf der linie).  die
  trajektorie belibt gleich bei skalierung+offsetverschiebung der
  beiden funktionen (glaube ich??); das kann man dazu verwenden, alles
  ein bißchen zu normieren.  }

\begin{figure*}[t!]
  \begin{center}
    \includegraphics[width=8.5cm]{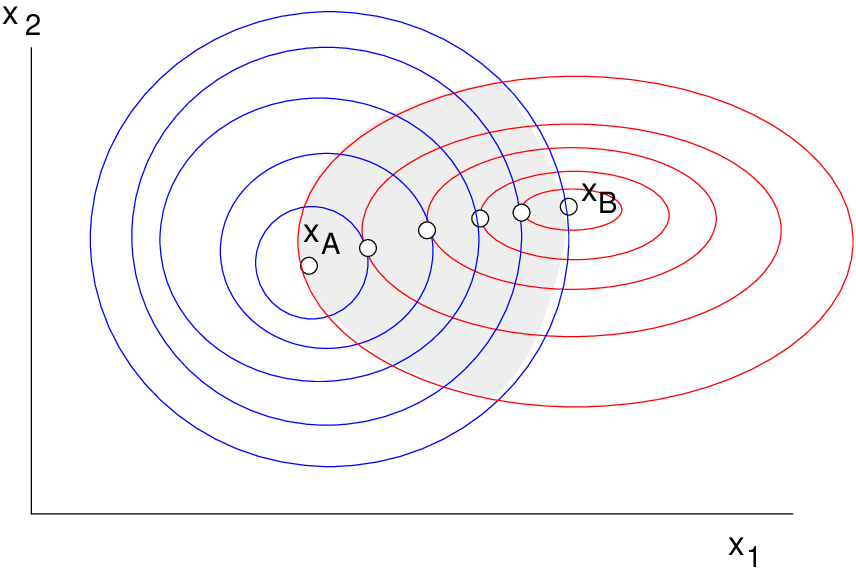}
    \caption{Trajectory of the optimal metabolite
      profiles as we interpolate between flux profiles \coout{(a)
        Enzymatic cost functions in metabolite space. Two {\flow}s
        $\vvA$ and $\vvB$ (not shown, and assumed to be conformal)
        give rise to different cost functions (shown in blue and red)
        with {\favoured} metabolite profiles $\lncv_{\rm A}$ and
        $\lncv_{\rm B}$.  During a sweep from $\vvA$ to $\vvB$, the
        cost function changes gradually and the optimal profile moves
        between $\lncv_{\rm A}$ and $\lncv_{\rm B}$. (b)} Trajectory
      of optimal metabolite profile during a flux sweep.  \co{show the
        flux sweep in inset graphics?}  In each point of the
      trajectory the gradients of the two original cost functions
      point in opposite directions.  Given the two cost functions
      (blue and red contours), this condition defines the 
      trajectory shape. \co{Uri's criterion for ``Normal'' pareto optimality
        in flux space} \co{noch extra-bild mit diesen niveaulinien:}
      The trajectory remains inside the grey region, defined by the
      contour line of function A passing through the point
      $\lncv_{{\rm B}}$ and by the contour line of function B passing
      through the point $\lncv_{{\rm A}}$.  \co{WRITE ME // Reference
        in text. note (also in text) that strict convexity is assumed;
        say what can be done if this is not the case (regularisation
        by epsilon-small strictly convex function)}
      \label{fig:trajectory1}}
  \end{center}
\end{figure*}

\co{reference this in the paper + draw the consequences // hier ref to
  main fig 4; dort ref to here} \co{check and correct this section} If
we interpolate between {\flow} $\vvA$ and {\flow} $\vvB$ in flux
space, how will the optimal metabolite profile move in metabolite
space? We assume that the two {\flow}s are in the same B-polytope
(i.e.~no reaction flux changes its sign) and that the ECM problem is
regular, i.e.~with a unique solution for each of the {\flow}s. In this
case, the metabolite profile will move along a continuous trajectory
between the two optimal  points $\lncvA = \lncv^{\rm opt}(\vvA)$
and $\lncvB = \lncv^{\rm opt}(\vvB)$.  Each trajectory point is the optimum point for some
interpolated {\flow} $\sigma\,\vvA + (1-\sigma)\,\vvB$: the 
optimality condition
$\nabla_{\lncv} \sigma \,\hmet^{\rm enz,\vvA}+ (1-\sigma) \,\hmet^{\rm
  enz,\vvB} = 0$ implies that  in each trajectory point the
gradients of the original cost functions $\hmet^{\rm enz,\vvA}$ and
$\hmet^{\rm enz,\vvB}$ must be exactly in opposite
directions. \co{put this more generally: we have to convex
  functions. a trajectory is the set of all points in which the
  gradients of the two functions are colinear.}  \co{explain relation
  to uris criterion for pareto optimal points (for more than two basic
  {\flow}s)} \co{discuss uniqueness; discuss whether the trajectory is
  one-dimensional; mention relation to kinetically disctinct {\flow}s}

 We can compute the direction of the trajectory in its
end points (e.g.~in the optimum point for $\hmet^{\rm enz,\vvA}$.),
for example, in the point $\lncv_{\rm A}$: with an infinitesimal interpolation parameter
$\md \lambda$, we obtain a displacement $\md \lncv$ of the optimal
profile. In the new point, the optimality condition reads, to second
order
\begin{eqnarray}
 (1-\md \lambda) (\underbrace{\nabla q_{\rm A}}_{0} + \Hmat_{\rm A}\,\md\lncv) + \md \lambda (\nabla q_{\rm B} + \Hmat_{\rm B}\,\md\lncv) = 0 \nonumber \\
\Rightarrow \Hmat_{\rm A}\,\md\lncv + \md \lambda \,\nabla q_{\rm B} = 0 \nonumber \\
\Rightarrow \md\lncv/\md \lambda   = - \Hmat_{\rm A}\inv\,\nabla q_{\rm B}.
\end{eqnarray}
Higher-order terms have been omitted, and the gradients
$\nabla q_{\rm A}$ and $\nabla q_{\rm B}$ and the curvature matrices
$\Hmat_{\rm A}$ and $ \Hmat_{\rm B}$ are evaluated in the point
$\lncv_{\rm A}$. The same argument can be applied to any point on the curve.

\section{Metabolic optimality problems}

\co{WEG! mehrere protein constraints gehen nicht. hoechsten BESTE
  werte fuer jedes einzelne protein; dann sind die constraints aber zu
  locker.
\co{To account for flux or protein constraints,
we need to modify FCM in three ways: (i) by introducing flux
constraints; (ii) by maximising flux benefit (at a fixed cost) rather
than minimising flux cost (at a fixed benefit); and (iii) by
considering protein constraints rather than flux constraints.}
  ALL DAS hoechstens IM SI diskutieren: \co{geht FBM mit
    mehreren kostenfunktionen ueberhaupt? (da ECM ja nur mit EINER
    kostenfunktion umgehen kann!) / falls nicht, was muss weg?}  In
  FBM, several groups of proteins can be constrained independently
  \co{WRONG!}(e.g.~the amounts of proteins can be constrained in
  different cell compartments or on different cellular
  membranes). This was not possible in FCM (unless a multi-objective
  version with several enzyme cost functions is
  considered). \co{problem: can these functions be computed
    separately?  NO!! this already implies an approximation!}  Each
  protein constraint refers to one protein cost function on the flux
  polytope. By linearising these protein costs, we obtain linear flux
  costs as proxies, and therefore linear flux constraints.  \co{ni
    some cases, the upper bounds can be replaced by fixed fluxes and
    enzyme levels).
}}

\subsection{Flux benefit maximisation}
\label{sec:SIFBM}

\begin{figure*}[t!]
\begin{center}
\includegraphics[width=13.5cm]{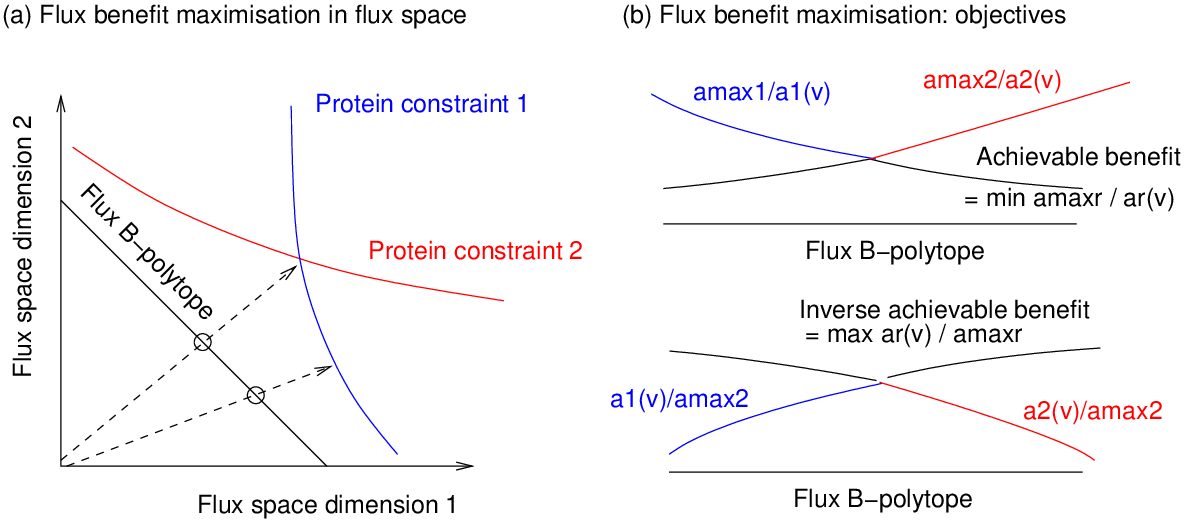}
\caption{\co{in a, shade region below curved curves; fix subscripts in
    graphics}\co{abb etwas schoener und klarer} Flux benefit
  maximisation. (a) Protein constraints and benefit optimisation in
  flux space (schematic drawing). Two protein constraints (blue and
  red line) define a feasible region (shaded).  Possible flux modes
  are defined by points on the B-polytope (diagonal lines) with a unit
  flux benefit (black line). For each of these points, the maximal
  flux benefit is given by the scaling factor that would move this
  point to the boundary defined by protein constraints. To compute
  this scaling factor, we need to solve a convex optimality problem in
  metabolite space. (b) Achievable benefit (i.e.,~scaling factor from
  (a)) as a function in flux space. The x-axis corresponds to the flux
  polytope (diagonal line in (a)). Top: red and blue lines show the
  possible scaling factor under one of the constraints. The scaling
  factor is defined by the minimum of the two functions (black
  line). The inverse achievable benefit is given by the maximum
  (black) of two concave functions (in fact, enzymatic cost
  functions), which itself is not concave.}
\label{fig:flux_benefit_maximisati}
\end{center}
\end{figure*}

In flux benefit maximisation (Figure
\ref{fig:flux_benefit_maximisati}) we maximise a linear flux benefit
$\fluxbene(\vv)$ under the constraint $\Nint\,\vv=0$ and under one or
several protein constraints
$\forall r: \acost_{r}(\vv) \le \acost_{r}^{\rm max}$ (where
\co{anderer index! uea!} $r$ denotes, e.g.~density constraints for
different cell compartments).  Without loss of generality, we can
assume that fluxes are non-negative, $\vv\ge 0$. In practice, to solve
this problem we consider scaled {\flow}s $\hat \vv$ with a nominal
benefit $b=1$. This assumption, together with stationarity and flux
sign constraints, restricts our {\flow}s to a B-polytope defined by
$\fluxbene(\hat \vv)=1$.  For each scaled {\flow} $\hat \vv$, we then
compute the maximal scaling factor $\sigma$ at which
$\forall r: \acost_{r}(\sigma \,\hat \vv) \le \acost_{r}^{\rm max}$ is
still satisfied.  This maximal factor is exactly the maximal
achievable benefit.  We can further simplify this problem: since the
enzymatic flux cost $\acostenz_{r}$ scales linearly with $\hat \vv$,
we need to maximize $\sigma$ such that
$\forall r: \sigma \, \acostenz_{r}(\hat \vv) \le \acost_{r}^{\rm
  max}$.  The solution is
$\sigma^{\rm opt}(\hat \vv) = \min_{r} \frac{\acost_{r}^{\rm
    max}}{\acostenz_{r}(\hat \vv)}$.  If we perform the same
optimisation on another B-polytope, with nominal benefit $b$, we
obtain the same formula
\begin{eqnarray}
  b^{\rm opt} = \frac{1}{b}\,\sigma^{\rm opt} = \frac{1}{b} \min_{r}  \frac{b\,\acost_{r}^{\rm max}}{\acostenz_{r}(\vv)} = \min_{r}  \frac{\acost_{r}^{\rm max}}{\acostenz_{r}(\vv)}.
\end{eqnarray}
This means: the maximal achievable benefit, as a function on a
B-polytope, is obtained by taking  the minimum  of all the functions
$ \frac{\acost_{r}^{\rm max}}{\acostenz_{r}(\vv)}$.  To find the
optimum point, it is convenient to minimise the reciprocal  benefit
\begin{eqnarray}
\label{eq:FluxBenefitMaxdum10}
 1/b^{\rm opt} =  \max_{r} \frac{\acostenz_{r}(\vv)}{\acost_{r}^{\rm max}}
\end{eqnarray}
on the flux polytope. With a single protein constraint, this leads to
exactly the same condition as our usual FCM problem (minimising a
single enzyme cost function on the flux polytope).  With several
protein constraints, the problem become more involved. First, the
protein functions cannot be determined by separate ECM
problems. Instead, they must be be determined as solutions of an ECM
problem that directly minimises
$\max_{r} \frac{q_{r}(\lncv;\vv)}{\acost_{r}^{\rm max}}$. Since the
maximum of two or more convex functions is convex, this is a convex
(typically non-smooth) problem, but the cost functions
$\acostenz_{r}(\vv)$ on the flux polytope may have complicated
shapes. Second, even if they were concave functions, the right-hand
side of (\ref{eq:FluxBenefitMaxdum10}) (to be minimised) need not be
concave, possibly giving rise to optimal {\flow}s inside the flux
polytope.  Thus, we need to employ approximations.  Approximating the
cost functions $\acostenz_{r}(\vv)$ by linear functions
$\acostenz_{r}(\vv) \approx \av^{(r)}_{\vv} \cdot \vv$, we obtain a
linear optimality problem \co{anderer variablenname als z?}
\begin{eqnarray}
\label{eq:SIFBMlinearised}
\mbox{Minimise}_{z,\vv} \quad z \quad
\mbox{subject to} && z\, \av_{r}^{\rm max} \ge \Amat_{\vv} \cdot \vv, \quad \Nint\,\vv=0, \quad \vv\ge 0. 
\end{eqnarray}
This approximation can be used within an  iterative optimisation method: we
start from some initial flux state, linarise the protein demand functions
around  this state, solve the linear problem
(\ref{eq:SIFBMlinearised}), and iterate  these steps until convergence.  The result is
 a local optimum. To find a global optimum, multiple runs with different  starting
points should be performed. \co{das scheme allgemein empfehlen, auch für FCM?
  that's more or less the Manu'a algorithm with arbitrary starting
  points!}  \co{Program this in MATLAB}

\subsection{Optimisation of enzyme activity at given protein levels}
\label{sec:SIFCMwithGivenEnzymeLevels}

When predicting optimal {\flow}s, we usually assume that cells can
choose, for each {\flow}, the best possible enzyme profile.  However,
a cell may not be able to adjust its enzyme levels, at least not very
quickly.  To be prepared for fast, unpredictable changes, cells may
employ express proteins preemptively: they produce enzyme for various
possible situations and inhibit them posttranslationally. In
\cite{nfbd:16}, we showed that this problem -- finding an enzyme
profile that can realise a given number of flux distributions, and
optimising this profile for minimal cost -- can be formulated as a
convex optimality problem.  Here we consider a different scenario: we
assume that the enzyme profile is given and that the cell needs to
find a flux mode, realisable with these enzymes, that maximises the
flux benefit (e.g.~biomass production or substance conversion), where
enzymes may be inhibited.  What is the maximal possible flux benefit,
the optimal metabolite profile, and what are the resulting enzyme
activities in this case?

\myparagraph{Flux benefit function in metabolite space} 
To answer this question, we
replace the rate law
\begin{eqnarray}
\label{eq:RateLawSplitting}
v_l = \esymbol_l\,\ratelaw_l
\end{eqnarray}
by an inequality $v_l \le \esymbol_l\,\ratelaw_l$ (assuming
that the actual reaction rates can be lower than what a non-inhibited enzyme could achieve). In our optimality problem,  the
enzyme profile \todo{$\esymbolv$} is given and we search for a {\flow} $\vv$ (and its log-metabolite profile $\lncv$) that maximise the
benefit q $\bv_{\rm v} \cdot \vv$.  To do so, we employ a trick:
we only consider {\flow}s $\vv$ with a fixed nominal benefit
$b = \bv_{\rm v} \cdot \vv$ (which brings us back to an optimality
problem on the B-polytope). We assume for each {\flow} that the benefit
can be scaled by a factor $\sigma$ and determine the maximal value
$\sigma^{\rm opt}(\vv)$ at which the enzyme constraint
Eq.~(\ref{eq:RateLawSplitting}) can still be satisfied.  A higher
value of $\sigma^{\rm opt}$ means that the nominal benefit $b$ can be
achieved at lower enzyme levels, or that a benefit time-integral
$b\,\Delta t$ can be reached in a shorter period of time. To find
the maximal scaling factor $\sigma^{\rm opt}$ for a {\flow} $\vv$, we
\begin{eqnarray}
\mbox{Maximise}~ \sigma \quad \mbox{s.t.} \quad
  \forall l: \sigma\, v_l \le \esymbol_l\,\ratelaw_l(\lncv),
\end{eqnarray}
where  fluxes $v_l$ and enzyme levels $\esymbol_l$ are given,
$\ratelaw_l(\lncv)$ denotes the catalytic rates, and the log-metabolite
levels $\lnc_i=\ln c_i$ are choice   variables. We can write this as 
\begin{eqnarray}
\mbox{Maximise}\;\sigma \quad 
\mbox{with respect to}\; \sigma\, \mbox{and}\,  \lncv\in\mPolytope  \quad
\mbox{subject to}\;
\forall l: \sigma\,  \le \frac{\esymbol_l\,\ratelaw_l(\lncv)}{v_l}.
\end{eqnarray}
To simplify this problem, we define $\sigma_{\rm
  max}(\lncv) = \min_l\, \frac{\esymbol_l\,\ratelaw_l(\lncv)}{v_l}$, the
maximal value of $\sigma$ that satisfies all the
inequality constraints, and obtain 
\begin{eqnarray}
\mbox{Maximise}\;\sigma_{\rm max}(\lncv) = \min_l\, \frac{\esymbol_l\,\ratelaw_l(\lncv)}{v_l} \quad
\mbox{with respect to}\; \lncv\in\mPolytope
\end{eqnarray}
or, equivalently,
\begin{eqnarray}
\label{eq:optimalFluxatConstantEnzyme1}
\mbox{Minimise}\;\frac{1}{\sigma_{\rm max}(\lncv)} = \max_l\, \frac{v_l/\esymbol_l}{\ratelaw_l(\lncv)} \quad
\mbox{with respect to}\; \lncv\in\mPolytope.
\end{eqnarray}
Since all ratio  terms on the right are convex in $\lncv$, their maximum is
convex, too.  the solution of this problem follows from
solving a convex optimality problem for
$1/\sigma_{\rm max}(\lncv)$ on the metabolite polytope. 
In fact, this optimality problem resembles the standard
  ECM problem
\begin{eqnarray}
\label{eq:optimalFluxatConstantEnzyme2}
\mbox{Minimise}\; \enzymemetcost(\lncv;\vv) = \sum_l\, \frac{h_l\,v_l}{\ratelaw_l(\lncv)} \quad
\mbox{with respect to}\; \lncv\in\mPolytope
\end{eqnarray}
with a sum, instead of a
maximum, over convex terms (i.e.~the enzymatic M-cost).

\myparagraph{Flux benefit function in flux space} The solution of
Eq.~(\ref{eq:optimalFluxatConstantEnzyme1}), as a function of $\vv$,
is the optimal scaling factor $\sigma^{\rm opt}(\vv)$. It is a
monotonously decreasing function\footnote{Here is a proof. Whenever a
  flux $v_l$ increases, the corresponding sum term in
  Eq.~(\ref{eq:optimalFluxatConstantEnzyme2}) increases as well; this
  can at most increase the maximum value, and so $\sigma$ can at most
  decrease. On the contrary, if a flux $v_l$ decreases, then $\sigma$
  can at most increase.}  in all fluxes $v_l$. Its reciprocal value
$a^{\rm CB ratio}(\vv)= 1/\sigma^{\rm opt}(\vv)$ is the maximally
achievable cost/benefit ratio of flux $\vv$ at fixed enzyme levels.
If we choose $\vv$ from a B-polytope with $b=1$, this is just the flux
cost itself. However, unlike the enzymatic flux cost $\acostenz(\vv)$,
this function may be non-concave in flux space. For example, if two
flux distributions $\vvA$ and $\vvB$ hit the enzyme constraints
Eq.~(\ref{eq:RateLawSplitting}) in different reactions, a convex
combination of the two {\flow}s need not hit these constraints and
could be scaled to yield higher benefits at the given enzyme
levels. \co{Its cost/benefit ratio is higher than one would expect
  from the linear combination, and so the benefit function is not
  concave. ???} As a rule of thumb, we can conclude: if the enzyme
levels are preedfined, cells should rather use the available enzyme
than strive for sparse flux distributions.

\co{Explanation: this is because of the maximum function, instead of
  the sum appearing in ECM.  Formally, write down the formula for
  1/sigma for a convex combination of {\flow}s; the combination of minimum
  und maximum functions makes it impossible to prove concavity or
  convexity!}

\subsection{Flux cost minimisation with non-enzymatic reactions}
\label{sec:nonEnzymatic}

\myparagraph{\ \\Non-enzymatic reactions} If a reaction is catalysed
by an enzyme, the reaction flux is costly, but also directly
controllable (at constant metabolite concentrations).  In cells, non-enzymatic
reactions can be important (or harmful), for example, (i) the
oxidation of molecules by free radicals; (ii) molecules leaving the
cell by passive diffusion; (iii) the dilution of substances in growing
cells (which formally resembles a linear degradation reaction).  The
presence of such reactions in models changes the set of possible
{\flow}s and the flux cost functions.  In ECM, a non-enzymatic
reaction with a given flux puts constraints on metabolite concentrations. For
example, a reaction with irreversible mass-action rate law
$v=k\,[a]\,[b]$ defines a constraint on the substrate levels $a$ and
$b$, i.e.~a linear constraint on $\ln a+\ln b$.  In general, the
constraints will be nonlinear!  The resulting equality constraints on
metabolite concentrations define a plane that intersects the M-polytope and
restricts it to a lower-dimensional subpolytope, or even to an empty
set (i.e.~stating that the {\flow} is impossible).  In FCM, this means
that some {\flow}s, even thermodynamically feasible ones, will be
discarded for kinetic reasons: parts of the flux polytope are
excluded, and the remaining flows form a non-convex set. Moreover,
different {\flow}s in an F-cone correspond to different
M-polytopes. This means that the cost functions $\metcost(\lncv;\vv)$,
obtained from different flows and defined on different sets in
metabolite space and are not comparable. Thus, the flux cost function
changes its shape, and our concavity proof does not hold anymore.

\myparagraph{Models with dilution}
In models with dilution fluxes $\lambda \,c_i$, a given
flux distribution $\vv$ determines all dilution fluxes and therefore
all metabolite concentrations, so the M-polytope shrinks to a point! \co{ instead of ECM, write directly $f(\vv)
= \fluxbene(\vv) - \metcost(\cv(\vv))-\hminus(\esymbolv(\vv,\cv(\vv)))$. a(v) zb linear; c(v)=1/lambda N v 
s.t. $\Nint \,\vv= \lambda \,\cv, \quad \cv_{{\rm min}} \le \cv \le \cv_{{\rm max}}$ und thermodyn constraints:
ein anderes polytop im flussraum!
}

\section{Cell populations described by probability distributions on the flux polytope}
 \label{sec:SIPPopProbability}

 \myparagraph{\ \\The distribution of metabolic {\flow}s} \co{evtl ist
   das standard in populationsmodellen (epidemiologie?)  dann
   zitieren!} A cell population can be described
 by a probability distribution $p_{\vv}(\vv)$ over all possible flows. To derive such a
 distribution, we assume that two {\flow}s are equally likely (i.e.~in flux space they
 show the same probability density) if they yield the
 same growth rate and that {\flow}s are more likely  if
 they yield higher growth rates.  In fact, we assume that
 probability densities depend directly on  growth rates. If a {\flow} with growth rate
 $\lambda(\vv)$ has a
 probability weight $\rho(\lambda(\vv))$, the probability
 distribution is given by
\begin{eqnarray}
\label{eq:PopProbabilityBoltzmann}
p_{\vv}(\vv) = \frac{\rho(\lambda(\vv))}{\int \rho(\lambda(\vv')) \md \vv'}.
\end{eqnarray}
To define a probability distribution, we need to  specify a formula for the
probability ratio $\frac{\rho(\lambda_{A})}{\rho(\lambda_{B})}$
between two states with growth rates $\lambda_{A}$ and
$\lambda_{B}$. Such formulae are derived in section
\ref{sec:SIProofPopProbability}.  Inspired by statistical
thermodynamics and by the usage of  Boltzmann
distributions in FBA \cite{dabg:17}, we may assume a formula
\begin{eqnarray}
\label{eq:PopProbabilityHyperbolic}
 \rho(\lambda)=\e^{(\lambda-\lambda_{d})/\xi}.
\end{eqnarray}
The formula resembles a Boltzmann distribution, where
the proliferation rate (i.e.~growth rate
$\lambda$ minus a death rate $\lambda_{d}$ equal for all cells)
represents a negative energy, while
the factor $\beta = k_{B}\,T$ is replaced by a characteristic time
$\tau = 1/\xi$.  The formula (\ref{eq:PopProbabilityHyperbolic})
follows from a simple cell population model.  We assume that each cell
starts in a randomly chosen flux state (chosen from a uniform
distribution on the flux polytope).  It grows and divides, and its
descendents remain in the same flux state until time $t=\tau$  where they
switch randomly to a new flux state (again uniformly distributed on
the flux polytope) while some of them die. The dying probability
is tuned to ensure a constant  population size. If we iterate this
procedure, the time-averaged distribution of states approaches our
Boltzmann distribution (\ref{eq:PopProbabilityHyperbolic}).  Instead
of assuming  synchronous switching, we may assume that cells can
switch their states any time, with exponentially distributed waiting
times and time constant $\tau$. This yields the alternative formula
\begin{eqnarray*}
\label{eq:probHyperbolic}
  \rho(\lambda)=\frac{1}{1- (\lambda-\lambda_{d})/\xi},
\end{eqnarray*}
which resembles the Fermi-Dirac and Bose-Einstein distributions in
quantum statistics (see \ref{sec:SIProofPopProbability}).  For consistency (avoiding
negative probability values), the characteristic switching rate $\xi=1/\tau$
must be higher than the highest proliferation rate
$\lambda-\lambda_{d}$ in the cell population.  If the switching rate
is much larger than the range of growth rates
($\xi \gg \lambda_{\rm max}-\lambda_{d}$), the distribution resembles
the Boltzmann distribution Eq.~(\ref{eq:PopProbabilityBoltzmann}).
With a linear flux cost function $\acost(\vv) = \av_{\vv} \cdot \vv$,
and assuming a linear cost-growth  formula
$\mu \sim 1/\acost$ (with constant $\gamma$), we obtain distributions
$p \sim \e^{1/(\alphav \cdot \vv)}$ or
$p \sim \frac{\alphav \cdot \vv}{\alphav \cdot \vv-1}$, respectively.

\co{clearly mention greedy search for optimal EFM. start with
  one. compute gradient of aenz, follow gradient (funkioniert auch mit
  flux constraints), iterate! check this algo for Meike's model?}

 \co{wo? some
  details on sampling: one may follow an MCMC approach (based on
  hit-and-run sampling), where a Boltzmann ensemble of {\flow}s is
  defined by $p(\vv) = \frac{1}{Z}\,e^{-k\,\acostenz(\vv)}$ with the
  partition function
  $Z = \int_{\bPolytope} \frac{1}{Z}\,e^{-k\,\acostenz(\vv)} \md \vv$
  over the flux polytope.  k may represent incomplete selection, or
  switches between {\flow}s ..  (ODER log BM-specific cost .. alle
  saehnlich ..  das ist so wie selection coefficient ..)}

\co{refer to ``Choice of a measure in flux space'' in CBA opt}

\section{Proofs}

\subsection{Proof of  Lemma \ref{lemma1}}
\label{sec:prooflemma1}

The curvature matrix $\Hmat$ of the enzymatic M-cost
$\enzymemetcost(\lncv;\vv)$ will vary depending on the metabolite
profile $\lncv$ and on the desired {\flow} $\vv$.  However, according to
Lemma \ref{lemma1}, its nullspace is structurally determined by rate
laws, network structure, and the {\fluxpattern}.  This means, the
nullspace depends on the flux and metabolite polytopes, but not on the
choices of $\vv$ and $\lncv$ within these polytopes. For the proof, we
note that factorised rate laws can be written as
\begin{eqnarray}
\label{eq:factorisedRateLaw}
\ratelaw(\lncv) = k_{\rm cat} \frac{1-\e^{-RT (\ln k_{\rm
      eq}+\nv \cdot \lncv)}}{\sum_t a_t\,\e^{\gv_t \cdot \lncv}},
\end{eqnarray}
where the vector $\lncv$ contains metabolite log-concentrations, the
vector $\nv$ contains stoichiometric coefficients, and the scalars
$a_{t}$ and vectors $\gv_t$ describe the rate law denominator (see
\cite{nfbd:16}).  We use two lemmas:
\begin{lemma}\label{lemmaHessianA2}
The function
\begin{eqnarray}
\label{eq:lemmaConstantNullSpaceH1}
f(y) = \frac{\e^{a\,y}}{1-\e^{y}}
\end{eqnarray}
with real-valued coefficient $a$  is positively curved for all $y<0$.  
To show  this, we consider the second derivative
\begin{eqnarray}
\label{eq:lemmaConstantNullSpaceH2}
\frac{\partial^2 f}{\partial y^2} 
=  \frac{a^2\,\e^{a\,y}}{1-\e^{\,y}}
 + \e^{a\,y}\,
\left( \frac{\e^{\,y}}{(1-\e^{\,y})^2} + \frac{2\,\e^{2\,\,y}}{(1-\e^{\,y})^3}\right)
 + \frac{2\,a\,\,\e^{a\,y}}{(1-\e^{\,y})^2},
\end{eqnarray}
which is  positive for all $y<0$. 
\end{lemma}

\begin{lemma}\label{lemmaHessianA2}
  \textbf{(Curvature of inverse rate law functions)} We consider a factorised rate law and study it
  inverse $1/\ratelaw(\lncv)$, the 
 as a function on the
metabolite polytope. Such functions are  positively curved in the subspace spanned
by $\nv$ and the vectors $\gv_t$ and have zero curvatures in all
directions orthogonal to that subspace. Here is the proof. After 
omitting the constant
prefactor $k_{\rm cat}$, the inverse of a rate law
Eq.~(\ref{eq:factorisedRateLaw}) can be written as
\begin{eqnarray}
\label{eq:lemmaConstantNullSpaceH3}
\frac{1}{\ratelaw(\lncv)} = \sum_t \frac{\e^{\alpha_t + \gv_t\cdot\,\lncv}}
{1-\e^{\beta + \nv'\cdot \lncv}},
\end{eqnarray}
where the vector $\nv'$ is parallel to our  vector $\nv$ and
$\beta + \nv'\cdot \lncv$ is  negative (because any point $\lncv$
in the metabolite polytope must yield positive thermodynamic forces).
If a metabolite variation $\delta \lncv$ is orthogonal on $\nv$ and on all vectors $\gv_t$,
then we obtain $1/\ratelaw(\lncv+\delta \lncv) = 1/\ratelaw(\lncv)$, and the function
Eq.~(\ref{eq:lemmaConstantNullSpaceH3}) has a vanishing curvature in
that direction. In contrast, if a variation $\delta \lncv$ lies in the space spanned
by $\nv$ and all vectors $\gv_t$, then at least one of the sum terms 
depends on $\delta \lncv$.  Then, each sum term, as a function of $\lncv$, is
identical to the function in Eq.~(\ref{eq:lemmaConstantNullSpaceH1}) except for an
affine transformation, and must therefore be positively curved with
respect to a scaling of $\delta \lncv$.
\end{lemma}

\textbf{Proof of Lemma \ref{lemma1}} The enzymatic M-cost
function is given by
$\metcost(\lncv) = \sum_l \frac{h_l\,v_l}{\ratelaw_l(\lncv)}$ and its
 curvature matrix is given by
$\Hmat(\lncv) = \sum_l h_l\,v_l \frac{\partial ^2}{\partial
  \lncv^2}\frac{1}{\ratelaw_l(\lncv)}$.
Since $\frac{\partial ^2}{\partial \lncv^2}\frac{1}{\ratelaw_l(\lncv)}$ is
convex (see \cite{nfbd:16}), all sum terms are positive semidefinite,
and so the nullspace of $\Hmat$ is given by the intersection of the
nullspaces of all matrices
$\frac{\partial ^2}{\partial \lncv^2}\frac{1}{\ratelaw_l(\lncv)}$. To prove 
that the nullspace of $\Hmat$ is independent of  $\lncv$
and $\vv$, we just need to show that the nullspace of
$\frac{\partial ^2}{\partial \lncv^2}\frac{1}{\ratelaw_l(\lncv)}$, for
single reactions $l$, is independent of  $\lncv$.  This is proven by
Lemma \ref{lemmaHessianA2}.

\subsection{Proof of Lemma \ref{lemma2}}
\label{sec:prooflemma2}

We consider a kinetic flux cost
$\fluxcost(\vv) = \min_{\lncv \in \sPolytope} f(\lncv;\vv) + g(\lncv)$,
where the cost functions $f(\lncv;\vv)$ and $g(\lncv)$ are bounded from below and where
$f(\lncv;\vv)$ is linear in $\vv$:
\begin{eqnarray}
\label{eq:convexityonP}
\forall \lncv \in \sPolytope :~ f(\lncv;
\eta\,\vvA + \mu\,\vvB) = \eta\, f(\lncv;\vvA) + \mu\,
f(\lncv;\vvB).
\end{eqnarray}
To prove that $\fluxcost(\vv)$ is concave, we need to show that 
\begin{eqnarray}
 \fluxcost(\eta\,\vvA + \mu\,\vvB) \ge
 \eta\,\fluxcost(\vvA) +  \mu\,\fluxcost(\vvB)
\end{eqnarray}
for all interpolation parameters $\eta \in [0,1]$ and $\mu=1-\eta$. We compute
\begin{eqnarray}
\label{eq:concavityproof1}
 \fluxcost(\eta\,\vvA + \mu\,\vvB) &=& 
 \min_{\lncv \in \sPolytope} f(\lncv; \eta\,\vvA + \mu\,\vvB) + g(\lncv) \nonumber\\
 &\ge& \min_{\lncv \in \sPolytope} \left[ \eta\, f(\lncv;\vvA) + \eta \,g(\lncv) + \mu\, f(\lncv;\vvB) +\mu \,g(\lncv)\right]  \nonumber\\
 &\ge& \min_{\lncv \in \sPolytope} \left[ \eta\, f(\lncv;\vvA)  + \eta \,g(\lncv) \right] + \min_{\lncv \in \sPolytope} \left[\mu\, f(\lncv;\vvB)  + \mu \,g(\lncv) \right] \nonumber\\
  &=& \eta\, \min_{\lncv \in \sPolytope} [f(\lncv;\vvA) + g(\lncv)] + \mu\, \min_{\lncv \in \sPolytope} [f(\lncv;\vvB) + g(\lncv)] \nonumber\\
  &=& \eta\,\fluxcost(\vvA) +  \mu\,\fluxcost(\vvB).
\end{eqnarray}
The inequality in the third line follows from the fact that the
minimum value of a sum of functions cannot be lower than the sum of
minimum values of these functions.  The lemma also holds if $f$ (for
each given $\lncv$) is not linear, but concave in the fluxes.

\subsection{Metabolite variations and their effect on the enzymatic metabolite cost}
\label{sec:SIHessianNullspace}

When enzyme levels are constant and metabolite concentrations are changing,
the reaction rates will usually change. However, there are exceptions.
In what cases will a metabolite variation leave all reaction rates
(or, in turn, the enzymatic cost at a given flux), unchanged? To see
this, we need some more terminology. Below we assume a given
\fluxpattern, defining feasible polytopes for flux and metabolite
profiles.

\textbf{Definition (Thermodynamically neutral metabolite variations)}
The thermodynamic force of reaction $l$, given by
$\theta_l = -\Delta_{\rm r} G_l/RT = \ln k^{\rm eq}_{l} - \sum_i
n_{il}\,\ln c_i$, is a linear function on the metabolite polytope.  A
metabolite variation $\delta \lncv$ is called \emph{thermodynamically
  neutral} if it can be added to any metabolite profile $\lncv$
without changing the thermodynamic forces $\theta_l$.  The set of
thermodynamically neutral metabolite variations $\delta \lncv$ is
called the \emph{thermodynamically neutral subspace}. It is given by
the nullspace of ${\Ntot}\trans$ (where $\Ntot$ is the stoichiometric
matrix referring to all -- internal and external -- metabolites).  If
the difference between two profiles $\lncvA$ and $\lncvB$ is
thermodynamically neutral, then $\lncvA$ and $\lncvB$ are called
\emph{thermodynamically equivalent}.

\textbf{Definition (Mass-action neutral metabolite variations)} With some  
 rate laws,  the reaction rate depends on metabolite concentrations only
through the mass-action terms $S_l = \prod_i c_i^{m^{\rm S}_{li}}$
(for substrates) and $P_l = \prod_i c_i^{m^{\rm P}_{li}}$ (for
products).  A metabolite variation $\delta \lncv$ is called
\emph{mass-action neutral} if it can be added to any metabolite
profile $\lncv$ without changing these terms in the  model.  The set of
mass-action neutral metabolite variations is called \emph{mass-action
  neutral subspace}. It is given by the nullspace of ${\Mmat^{\rm S}
  \choose \Mmat^{\rm P}}$.  
If the difference between two profiles $\lncvA$ and $\lncvB$ is
  mass-action neutral, then $\lncvA$ and $\lncvB$ are called
  \emph{mass-action equivalent}.

  \textbf{Definition (Kinetically neutral metabolite variations)} A
  metabolite variation $\delta \lncv$ is called \emph{kinetically
    neutral} if it can be added to any metabolite profile $\lncv$
  without changing any reaction rates. The set of kinetically neutral
  metabolite variations is called \emph{kinetically neutral subspace}.
  If the difference between two profiles $\lncvA$ and $\lncvB$ is
  kinetically neutral, then $\lncvA$ and $\lncvB$ are called
  \emph{kinetically equivalent}.

  \textbf{Definition (Cost-neutral metabolite variations)} A
  metabolite variation $\delta \lncv$ is called \emph{cost-neutral} if
  it can be added to any metabolite profile $\lncv$, at a fixed {\flow}
  $\vv$, without changing the enzymatic metabolite cost
  $\enzymemetcost(\lncv)$. The set of cost-neutral metabolite
  variations is called \emph{cost-neutral subspace}.  If the
  difference between two profiles $\lncvA$ and $\lncvB$ is
  cost-neutral, then $\lncvA$ and $\lncvB$ are called
  \emph{cost-equivalent}.

\co{wie verhaelt sich cost-equivalent zu thermodynamically equivalent,
  mass-action equivalent, and kinetically equivalent?}

All four criteria also apply to  single
reactions. For example, a metabolite variation $\delta \lncv$ is called
\emph{thermodynamically neutral for reaction $l$} if it can be added
to any metabolite profile $\lncv$  without
changing the thermodynamic force of reaction $l$. Using the four criteria, we can 
describe the nullspace of the cost curvature matrix $\Hmat$
for different types of rate laws:
\begin{enumerate}[leftmargin=5mm]
\item ``Reversibility-based'' rate laws depend on metabolite
  concentrations solely through thermodynamic forces  \cite{nfbd:16}. With
  such rate laws, all thermodynamically neutral metabolite
  variations are kinetically neutral and cost-neutral, and any
  nullvector of $\Nmat\trans$ is a nullvector of $\Hmat$.
\item
 ``Saturation-based'' rate laws depend on metabolite concentrations
 solely through the principal concentration terms \cite{nfbd:16}.  With
 such rate laws, all mass-action neutral metabolite
  variations are  kinetically neutral and cost-neutral, and 
any nullvector of ${\Mmat^{\rm S} \choose \Mmat^{\rm P}}$
  is a  nullvector of $\Hmat$.
\item Let $S$ be the \emph{cost-neutral subspace} of a model, i.e.~the set of
  all cost-neutral metabolite variations.  For any difference vector
  $\lncvA - \lncvB$, we obtain that  $\lncvA - \lncvB \in S
  \Rightarrow \enzymemetcost(\lncvA) = \enzymemetcost(\lncvB)$. This means: any
  kinetically neutral variation $\delta \lncv$ is  cost-neutral and
  is therefore a nullvector of $\Hmat$, i.e.~$\Hmat\,\delta
  \lncv=0$. 
\end{enumerate}
Thus, in models with reversibility-based rate laws, all thermodynamically
neutral variations are also cost-neutral and are therefore nullvectors of the cost
curvature matrix.  Likewise, with saturation-based rate laws, all
mass-action neutral variations are also cost-neutral and  nullvectors of
the cost curvature matrix.

\subsection{Proof of Proposition \ref{prop:FluxCostIsStrictlyConcave}}
\label{sec:ProofStrictlyConcave}

Proposition \ref{prop:FluxCostIsStrictlyConcave} states that enzymatic
flux cost functions $\aenz(\vv)$ are strictly concave on a straight
line between conformal, {\statedistinct} {\flow}s. For the proof, we
consider two {\flow}s $\vvA$ and $\vvB$ on an F-polytope (and assume,
without loss of generality, that all fluxes are non-negative). By
solving the ECM problem for each of these {\flow}s, we obtain the
optimal metabolite vectors $\lncvA$ and $\lncvB$, as well as the cost
curvature matrices $\Hmat_A$ and $\Hmat_B$ in the points
$(\vvA, \lncvA)$ and $(\vvB, \lncvB)$.  Since our {\flow}s $\vvA$ and
$\vvB$ are {\statedistinct}, the profiles $\lncvA$ and $\lncvB$ are
different ($\lncvA \ne \lncvB$) and cost-distinct\footnote{Without the
  latter condition, each of the two, $\lncvA$ and $\lncvB$, would be
  optimal for both of the {\flow}s.}
($\Hmat_{\rm A}\, (\lncvA-\lncvB)\ne 0$ and
$\Hmat_{\rm B}\, (\lncvA-\lncvB)\ne 0$).  In the proof, we consider a
convex combination $\vvC$ of $\vvA$ and $\vvB$ and its optimal
metabolite profile $\lncvC$.

\begin{enumerate}[leftmargin=5mm]
\item We first show that the cost function $\enzymemetcost(\lncv;\vv)$ for  a
  given {\flow} $\vv$, can be Taylor-expanded around each point $\lncv$
  inside the metabolite polytope. Here is the proof: our cost function
  is a composition of exponentials and rational functions, and thus without
  singularity points inside the metabolite polytope. It is therefore
  holomorphic in every variable $\lnc_i$, and by Hartogs' theorem it
  must therefore  be holomorphic as a function of several variables.
  Hence, it can be approximated by a convergent power series in
  each point inside the metabolite polytope.
\item To show that   $\fluxcost(\vv)$ is
  strictly concave, it is enough to  show this on a  line between  similar
  {\flow}s $\vvA \approx \vvB$. This is what we do now. Since 
  $\enzymemetcost(\lncv;\vv)$ is continuous in $\lncv$ and $\vv$, the
 optimal metabolite profiles $\lncvA$, $\lncvB$, and $\lncvC$ will be 
 similar. For  simplicity, we write
  $\enzymemetcost_A(\lncv)=\enzymemetcost(\lncv;\vvA)$ and
  $\enzymemetcost_B(\lncv)=\enzymemetcost(\lncv;\vvB)$. Now we approximate
  the cost by a quadratic expansion (justified point 1):
\begin{eqnarray}
\enzymemetcost_A(\lncv) &=& \enzymemetcost_A(\lncvA) + \half (\lncv-\lncvA)\trans \Hmat_A (\lncv-\lncvA)\nonumber  \\
\enzymemetcost_B(\lncv) &=& \enzymemetcost_B(\lncvB) + \half (\lncv-\lncvB)\trans \Hmat_B (\lncv-\lncvB).
\end{eqnarray}
For each metabolite profile $\lncv$, the two functions are
additive between $\vv_A$ and $\vv_B$, and so the enzymatic
metabolite cost for the interpolated  {\flow} $\vvC$ reads
\begin{eqnarray}
\label{eq:qformula1}
\enzymemetcost_C(\lncv) &=& [1-\eta]\, \enzymemetcost_A(\lncv) + \eta\,\enzymemetcost_B(\lncv) \nonumber \\
 &=& [1-\eta]\, [\enzymemetcost_A(\lncvA) + \half (\lncv-\lncvA)\trans \Hmat_A (\lncv-\lncvA)]
 + \eta\, [\enzymemetcost_B(\lncvB) + \half (\lncv-\lncvB)\trans \Hmat_B (\lncv-\lncvB)] \nonumber \\ 
 &=& \underbrace{[1-\eta] \enzymemetcost_A(\lncvA) + \eta\, \enzymemetcost_B(\lncvB)}_{\hmet^{\rm interp}} \nonumber \\
 && + \half[\underbrace{[1-\eta] (\lncv-\lncvA)\trans \Hmat_A (\lncv-\lncvA)
 + \eta\,  (\lncv-\lncvB)\trans \Hmat_B (\lncv-\lncvB)}_{\Delta \metcost}]
\end{eqnarray}
The first term describes the interpolated costs of $\vvA$ and $\vvB$.
The second term is our compromise cost $\Delta \metcost$. It is
non-negative, which already shows that the flux cost function must be
concave.  To show that the flux cost function is strictly concave, we
need to show that $\Delta \metcost$ is strictly  positive.

\item To find the optimal  profile $\lncv_C$ for our interpolated
  {\flow} $\vvC$, we minimise the compromise cost $\Delta \metcost$ with respect to
  $\lncv$. First, we rewrite $\Delta \metcost$ as
\begin{eqnarray}
\Delta \metcost &=& [1-\eta] \left[ \lncv\trans \Hmat_A \lncv
- 2 \lncv\trans \Hmat_A \lncvA + \lncvA)\trans \Hmat_A \lncvA \right] \nonumber \\
 && + \eta\, \left[ \lncv\trans \Hmat_B \lncv
-2  \lncv \trans \Hmat_B \lncvB
+ \lncvB \trans \Hmat_B \lncvB \right].
\end{eqnarray}
 Omitting  constant terms, we obtain 
\begin{eqnarray}
\Delta \metcost' &=& [1-\eta]\,
\left[ \lncv\trans \Hmat_A \lncv - 2 \lncv\trans \Hmat_A \lncvA  \right]
 + \eta\, \left[ \lncv\trans \Hmat_B \lncv - 2  \lncv \trans \Hmat_B \lncvB \right] \nonumber \\ 
 &=&
\lncv\trans \, 
\underbrace{\left[ [1-\eta]\, \Hmat_A  + \eta \Hmat_{B}\right]}_{\Hmat_C} \, \lncv
-2 \lncv\trans \left[ [1-\eta]\, \Hmat_A \lncvA + \eta \, \Hmat_{B} \lncvB \right].
\end{eqnarray}
By  minimising $\Delta \metcost'$ with respect to $\lncv$, we obtain the
optimal metabolite profile $\lncvC$:
\begin{eqnarray}
\lncvC &=& \Hmat_C \inv \,
 \left[ [1-\eta]\, \Hmat_A \lncvA + \eta \, \Hmat_{B} \lncvB \right].
\end{eqnarray}

\item Next, we show that $\lncvC \ne \lncvA$ and $\lncvC\ne \lncvB$
  (assuming that $0<\eta<1$).  To prove $\lncvC \ne \lncvA$ by
  contradiction, we first assume $\lncvC=\lncvA$ and obtain
\begin{eqnarray}
\lncvA &=& \lncvC  = \Hmat_C \inv \, \left[ [1-\eta]\, \Hmat_A \lncvA + \eta \, \Hmat_{B} \lncvB \right] \nonumber  \\
\Rightarrow \quad \Hmat_C\,\lncvA &=& [1-\eta]\, \Hmat_A \lncvA + \eta \, \Hmat_{B} \lncvB \nonumber  \\
\Rightarrow \quad [[1-\eta]\, \Hmat_A  + \eta \Hmat_{B}] \,\lncvA &=& [1-\eta]\, \Hmat_A \lncvA + \eta \, \Hmat_{B} \lncvB \nonumber  \\
\Rightarrow \quad  \eta \Hmat_{B} \,\lncvA &=& \eta \, \Hmat_{B} \lncvB \nonumber  \\
\Rightarrow \quad  \Hmat_{B} (\lncvA-\lncvB)  &=& 0.
\end{eqnarray}
The last line implies either  that $\lncvA=\lncvB$ or that $\lncvA-\lncvB$ is a nullvector of $\Hmat_{B}$
and  therefore cost-neutral. This  contradicts our initial 
assumptions. The proof for $\lncvC \ne \lncvB$ is analogous.
\item We now consider  the expression for $\Delta \metcost$ in
  Eq.~(\ref{eq:qformula1}), insert the optimal metabolite profile
  $\lncvC$, and obtain
\begin{eqnarray}
\Delta \metcost &=& 
[1-\eta]\, (\lncvC-\lncvA)\trans \Hmat_A (\lncvC-\lncvA)
 + \eta  (\lncvC-\lncvB)\trans \Hmat_B (\lncvC-\lncvB).
\end{eqnarray}
To finish our proof, we need to show that this expression is strictly
positive.  It cannot be negative, because the matrix products are
symmetric and the prefactors $[1-\eta]\,$ and $\eta$ are
non-negative. Thus, we need to show that it does not vanish.  We
already saw that $(\lncvC-\lncvA)$ and $(\lncvC-\lncvB)$ cannot
vanish.  Thus, in order for $\Delta \metcost$ to vanish,
$(\lncvC-\lncvA)$ would have to be in the nullspace of $\Hmat_A$, and
$\lncvC-\lncvB$ would have to be in the nullspace of $\Hmat_B$.
According to Lemma \ref{lemma1}, the two nullspaces are structurally
determined and therefore identical. Thus, for $\Delta \metcost$ to
vanish the difference $\lncvA-\lncvB$ must be in this shared
nullspace, which contradicts our assumptions.
\end{enumerate}

\subsection{The enzymatic metabolite cost for the  common modular rate law is strictly convex}
\label{sec:SICMpositivelycurved}

\co{remove acknowledgement joost, falls sein beweis nicht noetig ist}

In models with common modular (CM) rate laws the enzymatic M-cost
$\hmet^{\rm enz}(\lncv)$ has a strictly positive curvature matrix and
is therefore strictly convex. In the following proof, given by Joost
Hulshof (VU Amsterdam), this is first shown for a single  reaction and then
for an entire network. We consider a CM rate law
\begin{eqnarray}
 v = E\,\frac{k_+ \, \prod_{i \in \rm sub} (\frac{c_i}{k_i})^{n^{\rm sub}_i}(1-\e^{-\theta(x)})}
{\prod_{i \in \rm sub} (1+\frac{c_i}{k_i})^{n^{\rm sub}_i}
+ \prod_{i \in \rm prod} (1+\frac{c_i}{k_i})^{n^{\rm prod}_i}-1},
\end{eqnarray}
\co{products, index j, dafuer superscript sub und prod weglassen!}
where $n_i$ denotes  molecularities
(i.e.~positive stoichiometric coefficients) of substrates and
products. We focus on one reaction and consider all internal
metabolites (i.e.~metabolites without predefined concentrations)
appearing in the rate law.  The enzyme cost, as a function of these
metabolite concentrations only, reads
\begin{eqnarray}
\label{CMproofEnzyCost}
  h(e) = \he\,v\,\frac
{\prod_{i \in \rm sub} (1+\frac{c_i}{k_i})^{n^{\rm sub}_i}
+ \prod_{i \in \rm prod} (1+\frac{c_i}{k_i})^{n^{\rm prod}_i}-1}
{k_+\,\prod_{i \in \rm sub} (\frac{c_i}{k_i})^{n^{\rm sub}_i}(1-\e^{-\theta(x)})}.
\end{eqnarray}
The numerator is a polynomial of the form
\begin{eqnarray}
1 + \prod_{i} (\frac{c_i}{k_i})^{|n_i|} + ...,
\end{eqnarray}
where all following terms are powers of metabolite concentrations with
positive prefactors.  We now expand the denominator.  The
thermodynamic term $(1-\e^{-\Theta(\lncv)})$, as a separate factor in
(\ref{CMproofEnzyCost}), yields
$\frac{1}{1-\e^{-\theta(x)}} = 1+\e^{-\theta(x)} + \e^{-2\theta(x)} +
...$ (for positive driving forces $\theta$, this series converges).
Multiplying this by
$k_+\, \prod_{i \in \rm sub} (\frac{c_i}{k_i})^{-n^{\rm sub}_i}$, we
obtain
\begin{eqnarray}
&& k_+\,  (\prod_{i \in \rm sub} (\frac{c_i}{k_i})^{-n^{\rm sub}_i}) (1+\e^{-\theta(x)} + \e^{-2\theta(x)} + ...)  \nonumber \\
&=& k_+\, \prod_{i \in \rm sub} (\frac{c_i}{k_i})^{-n^{\rm sub}_i}
 + k_+\, \prod_{i \in \rm sub} (\frac{c_i}{k_i})^{-n^{\rm sub}_i}\, \e^{-\theta(x)} + ... \nonumber \\
&=& k_+\, \prod_{i \in \rm sub} (\frac{c_i}{k_i})^{-n^{\rm sub}_i}
 + k_- \, \prod_{i \in prod} (\frac{c_i}{k_i})^{n^{\rm prod}_i} + ...,
\end{eqnarray}
where all remaining terms are powers of metabolite concentrations with
positive cofactors.  Multiplying this with the numerator and writing
it in terms of log-concentrations, \co{aber das heisst NICHT, dass sie
  in allen richtungen gekruemmt ist! however, the numerator itself is
  positively curved in all directions! (and this holds also if
  activation and inhibition terms are added)} we obtain a series of
terms, each describing a function that is positively curved in all
log-concentrations. Thus, the cost curvature matrix for this reaction
is positive definite with respect to all metabolite log-concentrations
that appears in the rate law.  In a network model, the cost curvature
matrix is the sum of cost curvature matrices for all single
reactions. In kinetic models with CM rate laws, it will therefore be
positive definite in all metabolites that appear in the enzymatic rate
laws.

\co{In models with non-enzymatic reactions (or passive diffusion
  through membranes), there may be metabolites that do not appear in
  any enzymatic rate law, and for which convexity does not hold.}
\co{THAT MEANS: joosts beweis is nicht noetig! es geht genauso unser
  beweis mit zusatzannahmen ueber CM-denominator}

\subsection{Gradient of the enzymatic flux cost Eq.~(\ref{eq:FluxCostGradient}) and identity between flux and enzyme point cost}
\label{sec:SIproofFluxCostGradient}

Our aim is to compute the gradient $\partial \fluxcost^{\rm enz}/\partial \vv$
of an enzymatic flux cost function in a flux polytope point $\vv^{\rm ref}$ with the 
optimal metabolite profile
$\lncv^{\rm ref}= \lncv^{\rm opt}(\vv^{\rm ref})$.  A small variation
$\delta \vv$ would  change the optimal metabolite profile by
$\delta \lncv$. In a first-order approximation, the new enzymatic flux
cost reads \co{das ist ein 2. beweis fuer unchain rule! dort auf hier verweisen}
\begin{eqnarray}
 \aenz(\vv^{\rm ref}+\delta\vv) &=& \enzymemetcost(\lncv^{\rm ref}+\delta \lncv;\vv^{\rm ref}+\delta \vv) \nonumber \\
&\approx& \enzymemetcost(\lncv^{\rm ref}; \vv^{\rm ref})
+ \frac{\partial \enzymemetcost(\lncv^{\rm ref}; \vv)}{\partial \vv}|_{\vv=\vv^{\rm ref}}\,\delta \vv
+ \underbrace{\frac{\partial \enzymemetcost(\lncv;\vv^{\rm ref})}{\partial \lncv}}_0 \, \delta \lncv.
\end{eqnarray}
The prefactor of $\delta \vv$  is the desired gradient
$\partial \aenz/\partial \vv$:
\begin{eqnarray}
\frac{\partial \aenz(\vv)}{\partial v_l} = 
\frac{\partial  \enzymemetcost(\lncv^{\rm opt}(\vv); \vv)}{\partial v_l}
=
\frac{\partial}{\partial v_l} \frac{\hul\,v_l}{\ratelaw_l(\lncv^{\rm opt}(\vv))} =
\frac{\hul}{\ratelaw_l(\lncv^{\rm opt}(\vv))},
\end{eqnarray}
where $\lncv^{\rm opt}(\vv)$ is the optimal metabolite profile for the $\vv$ in question.
Thus,  the flux cost function can be expanded as
\begin{eqnarray}
  \aenz(\vv + \delta \vv) \approx 
  \enzymemetcost(\lncv^{\rm opt}(\vv); \vv + \delta \vv) 
= \aenz(\vv) +  \enzymemetcost(\lncv^{\rm opt}(\vv); \delta \vv).
\end{eqnarray}
We further obtain the flux point cost
\begin{eqnarray}
\frac{\partial \aenz}{\partial \ln \vv} &=& \frac{\partial \aenz(\vv)}{\partial v_l}\,v_l
= \frac{\hul}{\ratelaw_l}\,v_l = \hul\,\esymbol_l = \frac{\partial
  \hminus}{\partial \esymbol_l}\,\esymbol_l,
\end{eqnarray}
which is identical to the enzyme point cost $\partial \hminus/\partial \ln \todo{\ev}$.
\co{JA! This is an example of the unchain rule (ref CBA theory)} 

\subsection{Proposition \ref{prop:segaulaCriterion}: Criterion for locally optimal {\flow}s}
\label{sec:ProofCriterionLocallyOptimalFlow}

Let $\vv^{(1)}$ be a B-polytope vertex to be checked for local
optimality, and let $\vv^{(2)}, \vv^{(3)}, ..$ be all other polytope
vertices. If $\vv^{(1)}$ is locally optimal, any infinitesimal
movement $\delta \vv$ towards towards the interior of the B-polytope
must increase the cost (or leave it unchanged), i.e.,
$\fluxcost(\vv^{(1)}+\delta \vv)\ge \fluxcost(\vv^{(1)})$.  According
to the proof in \ref{sec:SIproofFluxCostGradient}, we can write the
left side as $\fluxcost(\vv^{(1)}) + \metcost(\lncv; \delta \vv)$,
where $\lncv$ is an optimal metabolite profile of $\vv^{(1)}$, and
obtain the condition $\metcost(\lncv; \delta \vv)\ge 0$. Any
infinitesimal vector $\delta \vv$ (pointing into the flux polytope)
can be written as a convex combination \co{eta is kein guter
  buchstabe, wegen effizienz-etas!}
$\delta \varepsilon \, \sum_{k\ne 1} \eta_k (\vv^{(k)} - \vv^{(1)})$
with an infinitesimal prefactor $\delta \varepsilon$ and non-negative
prefactors $\eta_{k}$. Since $\metcost$ is linear in $\vv$, we obtain
the condition
$\delta \varepsilon \, \sum_{k\ne 1} \eta_k\, \metcost(\lncv;
\vv^{(k)} - \vv^{(1)}) \ge 0$.  To verify this, we just need to show
that $\metcost(\lncv; \vv^{(k)} - \vv^{(1)})\ge>0$ for \emph{any}
other vertex $\vv^{(k)}$.  Due to linearity this is equivalent to
$\metcost(\lncv; \vv^{(k)}) \ge \metcost(\lncv; \vv^{(1)})$, which is
true by assumption (since $\vv^{(1)}$ is an optimal {\flow} given
$\lncv$).

\subsection{Alternative criterion for locally optimal {\flow}s}
\label{sec:ProofCriterionLocallyOptimalFlow2}

\co{reference this in the paper + draw the consequences} \co{anfang
  weiter hoch - hier nur beweis!}  Aside from Proposition
\ref{prop:segaulaCriterion}, there is another criterion for locally
optimal {\flow}s. For a {\flow} $\vv$ to be locally optimal, there
must be a strictly positive vector $\xv<0$ such that
\begin{eqnarray}
({\Nint}\trans, -{\Nint}\trans, \bv_{\vv}, \diag(\zetav))\, \xv =  -\nabla_{\vv} \fluxcost(\vv),
\end{eqnarray}
where $\zetav$ is the index vector for inactive reactions for our
given {\flow} $\vv$ (i.e.~$\gamma_{l}=1$ if $v_{l}=$, and $\gamma_{l}=0$
otherwise).  This criterion follows from the optimality conditions of
FCM. Here is the proof. Without loss of generality, we assume that all
active fluxes are positive.  We now consider the FCM problem
\begin{eqnarray} 
\mbox{Minimise}\,\fluxcost(\vv)\;\mbox{s.t.}\;\Nint\,\vv=0\;\mbox{and}\; b'\le \bv_{\vv} \cdot \vv\;\mbox{and}\;0\le \vv.
\end{eqnarray} 
The necessary optimality condition (Karush-Kuhn-Tucker condition) for a local optimum reads
\begin{eqnarray} 
\label{eq:ConditionKuhnTuckerdum1}
 \exists\, \alphav, \beta, \gammav: 0 = \nabla_{\vv}{\mathcal L}(\vv)= \nabla_{\vv}\fluxcost(\vv) + {\Nint}\trans\, \alphav + \bv_{\vv}\,\beta + \gammav,
\end{eqnarray} 
with Lagrange multipliers in $\alphav$ (vector of length
$n_{\rm reactions}$), $\beta$ (scalar), and $\gammav$ (vector of
length $n_{\rm reactions}$). Since we consider a minimisation 
with lower bounds on the fluxes, we know that $\gamma_{l}<0$ for all
reactions $l$ that are active in $\vv$, and  $\gamma_{l}=0$ for all
reactions $l$  inactive in $\vv$.  Similarly, since the flux
benefit is ensured by  an active lower bound, we obtain $\beta < 0$.
Hence, we can write condition (\ref{eq:ConditionKuhnTuckerdum1}) as a linear
satisfiability problem:
\begin{eqnarray} 
 \exists \alphav, \beta<0, \gammav<0:
\left(\begin{array}{lll}
{\Nint}\trans & \bv_{\vv} & \diag(\zetav) 
\end{array}\right)
\left(\begin{array}{l} \alphav\\ \beta\\ \gammav \end{array}\right)
&=& -\nabla_{\vv} \fluxcost
\end{eqnarray} 
where the known index vector $\zetav$ denotes inactive reactions in
$\vv$. By collecting all Lagrange multipliers in a vector $\xv$
\co{lieber z!} and duplicating the entried for $\alpha$ (one with a
minus sign), we can write this as
\begin{eqnarray} 
 \exists\, \xv <0:
({\Nint}\trans, -{\Nint}\trans, \bv_{\vv}, \diag(\vv=0))\, \xv =  -\nabla_{\vv} \fluxcost.
\end{eqnarray} 

\subsection{Linear approximations of the enzymatic flux cost function}
 \label{sec:PrototypeApproximationLinear}

The linearised flux cost function (\ref{eq:linAppr}) is obtained through a  linear 
 expansion
\begin{eqnarray}
\label{eq:SIlinAppr}
  \fluxcost(\vv) \approx   \fluxcost(\vv^{\rm ref})
 +  \sum_{l} \frac{\partial \fluxcost}{\partial v_{l}}|_{\vv^{\rm ref}} \,
  (v_{l}-v^{\rm ref}_{l})
=  \fluxcost(\vv^{\rm ref}) + 
  \sum_{l} \frac{\partial \fluxcost}{\partial v_{l}}|_{\vv^{\rm ref}} \,\,v_{l} 
- \underbrace{\sum_{l} \frac{\partial \fluxcost}{\partial v_{l}}|_{\vv^{\rm ref}} \,\,v^{\rm ref}_{l}}_{\fluxcost(\vv^{\rm ref}) }
=   \sum_{l} \frac{\partial \fluxcost}{\partial v_{l}}|_{\vv^{\rm ref}} \,\,v_{l}.
\end{eqnarray}
In the last step, we used  the sum rule Eq.~(\ref{eq:SumRulePointCost})
to simplify the sum expression.   The kinetic flux
cost function $\fluxcost^{\rm kin}(\vv)$ (obtained from  the kinetic
metabolite cost
$\metcost^{\rm kin}(\lncv;v) = \metcost(\lncv) + \enzymemetcost(\lncv;\vv)$)
can be linearised  similarly. Again, the gradient
$\frac{\partial \metcost(\lncv)}{\partial v_l}$ of the metabolite cost
vanishes, so the formula for $\partial \fluxcost/\partial v_l$ remains
unchanged. However, the optimal point $\lncv$ (in which the gradient needs to be
evaluated, will be different).

\subsection{Nonlinear approximation of  enzymatic flux cost, computed from prototype {\flow}s}
 \label{sec:PrototypeApproximation}

 \co{ist das nicht alles schon in b6 erklaert?}  To derive simple flux cost function, we consider
 the enzymatic flux cost function and apply  linear or nonlinear
 expansions.  For a linear approximation, we choose a prototype {\flow}
 $\vv_{\rm ref}$, define the optimal  metabolite concentrations
 $c_i^{\rm ref} = c_{i}^{\rm opt}(\vv^{\rm ref})$, and compute the catalytic rates
 $\ratelaw_l^{\rm ref} = \ratelaw_l(\cv^{\rm ref})$.  We obtain the
 flux-specific enzyme costs
\begin{eqnarray}
\left(\frac{q_l}{v_l}\right)_{\rm ref} = h_{e_l}\,\frac{e_l^{\rm ref}}{v_l^{\rm ref}} 
=  \frac{h_{e_l}}{\ratelaw_l^{\rm ref}}.
\end{eqnarray}
To approximate the flux cost function in a region around
$\vv_{\rm ref}$, we treat the flux-specific costs as constant
numbers. Based on the sum rule Eq.~(\ref{eq:SumRulePointCost}) and on
Eq.~(\ref{eq:SIlinAppr}), we obtain
\begin{eqnarray}
\fluxcost(\vv) 
      = \sum_l \frac{q_l}{v_l} \,v_l
\approx \sum_l \left(\frac{q_l}{v_l}\right)_{\rm ref} \,v_l
      = \sum_l \frac{h_{e_l}}{\ratelaw_l^{\rm ref}}  \,v_l,
\end{eqnarray}
This is a linear flux cost function, and the cost weights follow
directly from the catalytic rates $\ratelaw_l$ of our prototpye {\flow}.  For a
nonlinear approximation, we use several prototype {\flow}s
$\vv^{(\alpha)}$.  Again, we determine the optimal metabolite profile
and the resulting catalytic rates $\ratelaw_l^{(\alpha)}$ for each of the
prototype {\flow}s.  To approximated a flux cost $\acost(\vv)$, we
approximate the {\flow} $\vv$ by a convex combination
$\vv \approx \sum_\alpha \eta_\alpha\,\vv^{(\alpha)}$ of the prototype
flows.  Then we use the weights $\eta_\alpha$ for defining a weighted
mean of the inverse catalytic rates:
$1/\ratelaw_l' = \sum_\alpha \eta_\alpha /\ratelaw_l^{(\alpha)}$ and use this value
as an estimate of $1/\ratelaw_l$ for our {\flow} $\vv$. Now, we apply the
summation theorem
\begin{eqnarray}
\fluxcost(\vv)
 = \sum_l \frac{\partial \fluxcost(\vv)}{\partial v_l}  \,v_l
 =  \sum_l \frac{h_{e_l}}{\ratelaw_l(\vv)}  \,v_l.
\end{eqnarray}
Replacing the true value $\ratelaw_l(\vv)$ by our approximation $r'_l(\vv)$, we obtain
the nonlinear approximation\footnote{The approximated flux cost
  function Eq.~(\ref{eq:SINonlinApprox1}) is quadratic. To see this,
  we collect the prototype vectors $\vv_A, \vv_B, \vv_C\, ..$ in a
  matrix and set $\vv = (\vv_A\,\vv_B\, \vv_C\, ..) \etav$.  We
  can determine $\etav = (\vv_A\,\vv_B\, \vv_C\, ..)^+\,\vv$ by using
  the pseudoinverse of this matrix.  Therefore, the vector $\etav$ must be 
  linear in $\vv$, and inserting $\etav$ into the expansion $a\approx
  \sum_l \frac{h_l}{r*_l}v_l = \sum_{l\alpha}
  \frac{\eta_\alpha}{\ratelaw_l}\,h_l\,v_l$ yields a quadratic function in
  $\vv$.}
\begin{eqnarray}
\label{eq:SINonlinApprox1}
\fluxcost(\vv)  \approx \sum_l \frac{h_{e_l}}{\ratelaw_l'}  \,v_l.
\end{eqnarray}
In summary, to compute the flux cost we approximate the inverse
catalytic rate $\ratelaw_l(\vv)\inv$ \co{einfacher formulieren:} by
interpolating the inverse catalytic rates
$1/\ratelaw_l(\vv^{(\alpha)})$ of the prototype {\flow}s. This
calculation still works if some reaction fluxes in our prototype
{\flow}s vanish. However, if a catalytic rate
$\ratelaw_l(\vv^{(\alpha)})$ in a prototype {\flow} vanishes
(e.g.~because a reaction is in thermodynamic equilibrium or if its
enzyme is completely inhibited), then $\frac{1}{\ratelaw_l'}$
diverges. {\Flow}s that contain such reactions cannot be used as
prototype {\flow}s.

\subsection{The cell growth rate as a convex function on the B-polytope}
\label{sec:SIProofGrowthRateFunction}

\begin{proposition}
  We consider the 
  enzyme cost   \co{use scaled symbol for a!! uea hier in dem abschnitt; auch in main text!}
  $\aenz(\vv)$, scaled to  unit biomass production, and assume
  that our the cell growth rate $\lambda$ can be computed from this
  scaled enzyme cost by a decreasing, convex cost-growth function
  $\lambda(\aenz)$. In this case, the growth rate
  $\lambda(\aenz(\vv))$ is a convex function on the
  B-polytope. Moreover, if the scaled enzymatic flux cost
  $\aenz(\vv)$ is strictly concave, the growth rate
  $\lambda(\aenz(\vv))$ is strictly convex.
\end{proposition}

\textbf{Proof} We consider a model with strictly concave  flux cost
$\aenz(\vv)$ on the unit B-polytope (defined by unit biomass
production) as well as  a monotonically decreasing, convex cost-growth
function $\lambda(a)$.  We further consider two {\flow}s $\vvA$ and
$\vvB$ in the B-polytope. Since $\aenz(\vv)$ is strictly concave,
it must satisfy
\begin{eqnarray}
a(\eta\, \vvA + [1-\eta]\,\vvB) > \eta\, \aenz(\vvA) + [1-\eta]\, \aenz(\vvB)
\end{eqnarray}
for all $0 < \eta < 1$. Since $\lambda(a)$ is monotonically decreasing, we obtain
\begin{eqnarray}
\lambda(a(\eta\, \vvA + [1-\eta]\,\vvB)) < \lambda(\eta\, \aenz(\vvA) + [1-\eta]\, \aenz(\vvB))
\end{eqnarray}
and  since $\lambda(a)$  is convex,
\begin{eqnarray}
\lambda(a(\eta\, \vvA + [1-\eta]\,\vvB)) < \eta\, \lambda(\aenz(\vvA)) + [1-\eta]\, \lambda(\aenz(\vvB)).
\end{eqnarray}
 This means that  $\lambda(\aenz(\vv))$ is strictly convex. Likewise, if $\aenz(\vv)$
is (non-strictly) concave, then $\lambda(\aenz(\vv))$ will be
(non-strictly) convex (same proof, with an inequality $\ge$ instead of $>$).

\myparagraph{Growth formulae, Eq.~(\ref{efficiencyGrowthConversion})}
Typical cost-growth functions \co{weiter uea hat a, wegen scaled!}
$\lambda(\aenz)$ proposed in the literature are in fact decreasing and
convex \cite{sgmz:10,wnfb:18}.  To derive such a function, we consider
a cell with biomass concentration $c_{\rm BM}$ (e.g.~in carbon moles
per cell volume) and biomass production rate $v_{\rm BM}$.  To obtain
a simple growth formula, we assume a fixed total concentration
$c_{\rm ME}$ of metabolic enzymes in the cell. The cell will grow at a
rate
$\lambda = v_{\rm BM}/c_{\rm BM} = \frac{v_{\rm BM}}{c_{\rm ME}}
\frac{c_{\rm ME}}{c_{\rm BM}} = \ratelaw_{\rm BM} \; \rho_{\rm ME}$,
with enzyme-specific biomass production rate
$\ratelaw_{\rm BM} = 1/\aenz$ (where the enzymatic cost $a$ refers to
fluxes with unit biomass production) and enzyme fraction (metabolic
enzyme concentration / biomass concentration) $\rho_{\rm ME}$. This
growth rate is proportional to $\ratelaw_{\rm BM}$ (and therefore
inversely proportional to the enzymatic flux cost $a$).  To obtain a
more realistic, nonlinear growth formula, we assume that the biomass
fraction of metabolic enzyme is growth-rate dependent, possibly due to
protein allocation between metabolism and ribosomes. Following Scott
et al.~ \cite{sgmz:10}, we obtain a saturable (Michaelis-Menten-like)
formula
$\lambda = \lambda^{\rm max}\, \frac{\ratelaw_{\rm BM}}{\ratelaw_{\rm
    BM}+k_{\rm growth}}$ with a maximal growth rate
$\lambda^{\rm max}$ and a scale parameter
$k_{\rm growth}$\co{umbenennen?}. The resulting cost-growth function
has the form
$\lambda(a) = \lambda^{\rm max}\, \frac{k'}{k_{\rm growth}'+a}$ (see
Eq.~(\ref{efficiencyGrowthConversion}) ).

\subsection{Cell populations described by probability distributions on the flux polytope}
 \label{sec:SIProofPopProbability}

 \myparagraph{\ \\Probability distributions Eqs
   (\ref{eq:PopProbabilityBoltzmann}) and
   (\ref{eq:PopProbabilityHyperbolic}) from a cell population model
   with replication and mixing} The probability ratio between two
 {\flow}s $\vv_{A}$ and $\vv_{B}$ (with growth rates $\lambda_{A}$ and
 $\lambda_{B}$, respectively) can be derived from the forward and
 reverse (conditional) transition rates between the two states.  To
 define the transition rates, we assume that the cell population
 alternates between a reproduction phase (duration $\tau$) and a
 random switching of states, where cells can switch to any possible
 state with equal probability.  To keep the population size constant,
 we assume a death rate $\lambda_{d}$, identical for all cells and
 chosen to balance the growth of the cell population. The difference
 $\lambda-\lambda_{d}$, for a cell state $\vv$, is called
 proliferation rate.  We start with one cell in state $A$ and define
 the transition rate $w_{A \rightarrow B}$ as the average number of
 descendents of this cell ending up in state $B$, divided by the time
 $\tau$.  The number of cells after the reproduction phase is given by
 $\e^{(\lambda_{A}-\lambda_{d}) \tau}$, and with a finite number
 $\Omega$ of possible cell states, the transition rate reads
 $w_{A \rightarrow B} = \frac{1}{\Omega}\e^{(\lambda_{A}-\lambda_{d})
   \tau}$. Thus, the ratio for both directions reads
 \begin{eqnarray*}
   \frac{w_{A \rightarrow B}}{w_{B \rightarrow A}}  = \frac{\e^{(\lambda_{A}-\lambda_{d}) \tau}}{\e^{(\lambda_{B}-\lambda_{d}) \tau}}.
 \end{eqnarray*}
 Since $\Omega$ cancels out in the formula, the formula also holds for
 a continuous state space (with an infinite number of states).  To
 show that this leads to a Boltzmann distribution, we employ the
 principle of detailed balance from statistical mechanics, the
 postulate that the absolute transition rates in the two directions
 between any two states A and B, must cancel out:
 $w_{A \rightarrow B}\,\rho_{A} - w_{B \rightarrow A}\, \rho_{B} = 0$,
 implying that
 $\frac{\rho_{A}}{\rho_{B}} = \frac{w_{B \rightarrow A}}{w_{A
     \rightarrow B}}$. For our cell population model, we can make the
 same argument.  If we describe cell death by transitions to a
 hypothetical dead state, then for all remaining states the
 occupation probabilities must remain constant in time, and we can
 postulate detailed balance between any two states.

 To derive our second distribution, Eq.~(\ref{eq:probHyperbolic}), we
 make a similar argument, but we assume that cells can switch their
 state at any moment with exponentially distributed waiting times and
 a characteristic waiting time $\tau$. We consider a single cell,
 starting in state $A$ at time $t=0$, and count the average number of
 descendents (i.e.~the average total number of cells that develop before leaving state
 $A$). This number is given by
 \begin{eqnarray*}
\langle n \rangle &=& \frac{\int_{0}^{\infty} \e^{t/\tau}\,\e^{(\lambda-\lambda_{d}) \,t} \md t}{\int_{0}^{\infty} \e^{t/\tau} \md t}
= \frac{\frac{1}{(\lambda-\lambda_{d})-1/\tau} \e^{((\lambda-\lambda_{d})-1/\tau) t}|_{0}^{\infty}}{\frac{1}{-1/\tau} \e^{(-1/\tau) t}|_{0}^{\infty}}
= \frac{\frac{-1}{(\lambda-\lambda_{d})-1/\tau}}{-(-\tau)} = \frac{1}{1-(\lambda-\lambda_{d})\,\tau}.
 \end{eqnarray*}
 The formula tells us that we need to assume
 $\tau<1/(\lambda-\lambda_{d})$, because otherwise the number of
 descendants would not be finite and positive. Now the ratio of
 transition rates reads
 \begin{eqnarray*}
   \frac{w_{A \rightarrow B}}{w_{B \rightarrow A}} =  \frac{1-(\lambda_{B}-\lambda_{d})\,\tau}{1-(\lambda_{A}-\lambda_{d})\,\tau}.
 \end{eqnarray*}
 Again, we obtain the desired
 probability distribution by by postulating detailed balance.

 \myparagraph{Probability distribution Eq.~(\ref{eq:probHyperbolic})
   and quantum statistics} Eq.~(\ref{eq:probHyperbolic}) resembles two
 known probability distributions from physics.

First, we consider states with
$\lambda > \lambda_{d}$ (positive proliferation rate). These are
states that proliferate by themselves and whose proliferation is
limited by switches to other states. We can write
Eq.~(\ref{eq:probHyperbolic}) as
\begin{eqnarray*}
  \rho(\lambda)= 1 + \frac{(\lambda - \lambda_{d})/\xi}{1- (\lambda - \lambda_{d})/\xi} = 1 + \frac{1}{\frac{1}{(\lambda - \lambda_{d})/\xi} - 1} = 
1 + \frac{1}{\e^{(- \ln (\lambda - \lambda_{d})) - (- \ln \xi) } - 1}.
\end{eqnarray*}
This is the formula for   state occupancies in Bose-Einstein statistics, that is, 
plus a constant term 1. In the analogy, the negative logarithmic
proliferation rate $-\ln (\lambda-\lambda_{d})$ corresponds to the
scaled energy $\beta \,E$ of the bosonic particles, and the negative
logarithmic switching rate $-\ln \xi$ corresponds to the scaled
chemical potential $\beta\,\mu$.  Now, cells with a very low
proliferation rate (much lower than the switching rate) have a
probability weight of 1 (arising from the extra term 1 in the formula,
and corresponding to bosonic particles with high energies). Cells with
a proliferation rate close to the switching rate have much higher
probability weights (given by the occupancies of bosonic low-energy
states, plus the term 1).

Now,  second, we consider states with negative proliferation rates, i.e.
 $\lambda < \lambda_{d}$. These cell states would normally die out and
 are only populated through switching from other states. We can write
 the probability weights (\ref{eq:probHyperbolic}) as
\begin{eqnarray*}
  \rho(\lambda) =  \frac{1}{1+ (\lambda_{d} -\lambda)/\xi} = \frac{1}{\e^{\ln (\lambda_{d} -\lambda) -\ln \xi}+1}.
\end{eqnarray*}
The same formula describes state occupancies in Fermi-Dirac
statistics, that is, the energy distribution of fermionic particles.
In the analogy, the logarithm of the negative proliferation rate,
$\ln (\lambda_{d} -\lambda)$, corresponds to the scaled energy
$\beta \,E$, and the logarithmic switching rate $\ln \xi$ corresponds
to the scaled chemical potential $\beta\,\mu$ of the fermionic
particles. In a cell population, cells with negative proliferation
rates much larger than the switching rate will be rare (just like
fermionic particle states with high energies), while cells with
negative proliferation rates much smaller than the switching rate have
probability weights close to 1 (just like fermionic particle states
with low energies).

\end{appendix}

\end{document}